\documentclass[a4paper,11pt]{article}
\usepackage{amsmath, amsfonts, amscd, amssymb}
\usepackage{amsthm}
\usepackage{latexsym}       \usepackage{amsmath}
\usepackage{amsfonts}       \usepackage{amssymb}
\usepackage{amscd}          \usepackage[greek,english]{babel}
\usepackage[iso-8859-7]{inputenc}
\usepackage[dvipsnames,usenames]{color}
\usepackage{graphics}
\usepackage{graphicx}
\usepackage{longtable}
\usepackage{yhmath}

\numberwithin{equation}{section}

\theoremstyle{plain}

\theoremstyle{definition}
\newtheorem{definition}{Definition}[section]

\theoremstyle{proposition}
\newtheorem{proposition}{Proposition}[section]

\theoremstyle{theorem}
\newtheorem{theorem}{Theorem}[section]

\theoremstyle{corollary}
\newtheorem{corollary}{Corollary}[section]

\catcode`\@=11

\newcount\printerresolution
\global\printerresolution=300

\newcount\headsize
\global\headsize=450

\font\dotfont = lcircle10 at 3pt

\newcount\epihead
\global\epihead=200

\newcount\defaultscale
\def\setdefaultscale#1{\global\defaultscale=#1}
\setdefaultscale{100}

\newcount\actualtextarrowspace
\newcount\actualtextarrowlength

\newcommand{\computetextparameters}%
{\global\actualtextarrowspace=\textarrowlength%
\global\advance\actualtextarrowspace by 3%
\global\actualtextarrowlength=\textarrowlength%
\global\multiply\actualtextarrowlength by 100}

\newcount\textarrowlength
\def\settextarrowlength#1{\global\textarrowlength=#1%
\computetextparameters} \settextarrowlength{20}

\newcount\actualdisplayarrowspace
\newcount\actualdisplayarrowlength

\newcommand{\computedisplayparameters}%
{\global\actualdisplayarrowspace=\displayarrowlength%
\global\advance\actualdisplayarrowspace by 3%
\global\actualdisplayarrowlength=\displayarrowlength%
\global\multiply\actualdisplayarrowlength by 100}

\newcount\displayarrowlength
\def\setdisplayarrowlength#1{\global\displayarrowlength=#1%
\computedisplayparameters} \setdisplayarrowlength{30}

\def\@ifnexttok#1#2#3{\let\@tempe #1\def\@tempa{#2}\def\@tempb{#3}%
\futurelet\@tempc\@ifntok}
\def\@ifntok{\ifx \@tempc \@tempe\let\@tempd\@tempa\else\let\@tempd\@tempb\fi%
\@tempd}

\def\@diagramerror#1#2{%
\edef\@@tempc{#2}\expandafter\errhelp\expandafter{\@@tempc}%
\typeout{Diagram error. \space See User's guide for
explanation.^^J
 \space\@spaces\@spaces\@spaces Type \space H <return> \space for
 immediate help.}\errmessage{#1}}

\newif\ifdiagram

\def\testtextmode{%
\ifdiagram\@diagramerror{Text arrows are not allowed in
diagrams}{Here you should use east or west diagram arrows, not
forward or backward text arrows. Try proceeding now, typeset could
succeed but with unpredictable output.}
\else\ifmmode\relax\else%
\@diagramerror{Missing \string$}{Text arrows should be introduced
in math mode. Try proceeding now, typeset could succeed but output
could not be what you expected.}\fi\fi}

\def\testdiagrammode{\ifdiagram\relax\else
\@diagramerror{Diagram arrows are not allowed in formulas}{Here
you should use forward or backward text arrows, not diagram
arrows. Proceeding could work with unpredictable output, but
overflow arithmetic could also occur.}\fi}

\def\checkmode{\ifmmode\@diagramerror{Wrong mode: no diagrams
allowed in math mode.}{You should leave math mode before
introducing your diagram. All items in the diagram will
automatically be processed in math
mode.}\else\relax\fi\global\diagramtrue}

\global\diagramfalse

\def\DOT{{\dotfont q}}

\newcount\basicstep
\newcount\numberofsteps
\numberofsteps=\headsize \global\divide\numberofsteps by
\basicstep

\newcommand{\makehead}[3]{%
\begin{picture}(0,0)%
\multiput(0,0)(#1,#2){#3}{\DOT}%
\multiput(0,0)(-#2,#1){#3}{\DOT}%
\end{picture}}

\newcount\xstep
\newcount\ystep

\newsavebox{\northhead}
\savebox{\northhead}{%
\xstep=-\basicstep%
\multiply\xstep by 7071%
\divide\xstep by 10000%
\ystep=\xstep%
\makehead{\xstep}{\ystep}{\numberofsteps}}
\newcommand{\nhead}{\usebox{\northhead}}

\newsavebox{\easthead}
\savebox{\easthead}{%
\xstep=-\basicstep%
\multiply\xstep by 7071%
\divide\xstep by 10000%
\ystep=-\xstep%
\makehead{\xstep}{\ystep}{\numberofsteps}}
\newcommand{\ehead}{\usebox{\easthead}}

\newsavebox{\southhead}
\savebox{\southhead}{%
\xstep=\basicstep%
\multiply\xstep by 7071%
\divide\xstep by 10000%
\ystep=\xstep%
\makehead{\xstep}{\ystep}{\numberofsteps}}
\newcommand{\shead}{\usebox{\southhead}}

\newsavebox{\westhead}
\savebox{\westhead}{%
\xstep=\basicstep%
\multiply\xstep by 7071%
\divide\xstep by 10000%
\ystep=-\xstep%
\makehead{\xstep}{\ystep}{\numberofsteps}}
\newcommand{\whead}{\usebox{\westhead}}

\newsavebox{\northwesthead}
\savebox{\northwesthead}{%
\makehead{0}{-\basicstep}{\numberofsteps}}
\newcommand{\nwhead}{\usebox{\northwesthead}}

\newsavebox{\northeasthead}
\savebox{\northeasthead}{%
\makehead{-\basicstep}{0}{\numberofsteps}}
\newcommand{\nehead}{\usebox{\northeasthead}}

\newsavebox{\southwesthead}
\savebox{\southwesthead}{%
\makehead{\basicstep}{0}{\numberofsteps}}
\newcommand{\swhead}{\usebox{\southwesthead}}

\newsavebox{\southeasthead}
\savebox{\southeasthead}{%
\makehead{0}{\basicstep}{\numberofsteps}}
\newcommand{\sehead}{\usebox{\southeasthead}}

\newsavebox{\eastnortheasthead}
\savebox{\eastnortheasthead}{%
\xstep=-\basicstep%
\multiply\xstep by 9486%
\divide\xstep by 10000%
\ystep=\xstep%
\divide\ystep by -3%
\makehead{\xstep}{\ystep}{\numberofsteps}}
\newcommand{\enehead}{\usebox{\eastnortheasthead}}

\newsavebox{\northnortheasthead}
\savebox{\northnortheasthead}{%
\xstep=-\basicstep%
\multiply\xstep by 9486%
\divide\xstep by 10000%
\ystep=\xstep%
\divide\ystep by 3%
\makehead{\xstep}{\ystep}{\numberofsteps}}
\newcommand{\nnehead}{\usebox{\northnortheasthead}}

\newsavebox{\southsouthwesthead}
\savebox{\southsouthwesthead}{%
\xstep=\basicstep%
\multiply\xstep by 9486%
\divide\xstep by 10000%
\ystep=\xstep%
\divide\ystep by 3%
\makehead{\xstep}{\ystep}{\numberofsteps}}
\newcommand{\sswhead}{\usebox{\southsouthwesthead}}

\newsavebox{\westsouthwesthead}
\savebox{\westsouthwesthead}{%
\xstep=\basicstep%
\multiply\xstep by 9486%
\divide\xstep by 10000%
\ystep=\xstep%
\divide\ystep by -3%
\makehead{\xstep}{\ystep}{\numberofsteps}}
\newcommand{\wswhead}{\usebox{\westsouthwesthead}}

\newsavebox{\westnorthwesthead}
\savebox{\westnorthwesthead}{%
\xstep=\basicstep%
\multiply\xstep by 3162%
\divide\xstep by 10000%
\ystep=\xstep%
\multiply\ystep by -3%
\makehead{\xstep}{\ystep}{\numberofsteps}}
\newcommand{\wnwhead}{\usebox{\westnorthwesthead}}

\newsavebox{\eastsoutheasthead}
\savebox{\eastsoutheasthead}{%
\xstep=-\basicstep%
\multiply\xstep by 3162%
\divide\xstep by 10000%
\ystep=\xstep%
\multiply\ystep by -3%
\makehead{\xstep}{\ystep}{\numberofsteps}}
\newcommand{\esehead}{\usebox{\eastsoutheasthead}}

\newsavebox{\northnorthwesthead}
\savebox{\northnorthwesthead}{%
\xstep=-\basicstep%
\multiply\xstep by 3162%
\divide\xstep by 10000%
\ystep=\xstep%
\multiply\ystep by 3%
\makehead{\xstep}{\ystep}{\numberofsteps}}
\newcommand{\nnwhead}{\usebox{\northnorthwesthead}}

\newsavebox{\southsoutheasthead}
\savebox{\southsoutheasthead}{%
\xstep=\basicstep%
\multiply\xstep by 3162%
\divide\xstep by 10000%
\ystep=\xstep%
\multiply\ystep by 3%
\makehead{\xstep}{\ystep}{\numberofsteps}}
\newcommand{\ssehead}{\usebox{\southsoutheasthead}}

\newsavebox{\easteastnortheasthead}
\savebox{\easteastnortheasthead}{%
\xstep=-\basicstep%
\multiply\xstep by 8944%
\divide\xstep by 10000%
\ystep=\xstep%
\divide\ystep by -2%
\makehead{\xstep}{\ystep}{\numberofsteps}}
\newcommand{\eenehead}{\usebox{\easteastnortheasthead}}

\newsavebox{\northnorthnortheasthead}
\savebox{\northnorthnortheasthead}{%
\xstep=-\basicstep%
\multiply\xstep by 8944%
\divide\xstep by 10000%
\ystep=\xstep%
\divide\ystep by 2%
\makehead{\xstep}{\ystep}{\numberofsteps}}
\newcommand{\nnnehead}{\usebox{\northnorthnortheasthead}}

\newsavebox{\southsouthsouthwesthead}
\savebox{\southsouthsouthwesthead}{%
\xstep=\basicstep%
\multiply\xstep by 8944%
\divide\xstep by 10000%
\ystep=\xstep%
\divide\ystep by 2%
\makehead{\xstep}{\ystep}{\numberofsteps}}
\newcommand{\ssswhead}{\usebox{\southsouthsouthwesthead}}

\newsavebox{\westwestsouthwesthead}
\savebox{\westwestsouthwesthead}{%
\xstep=\basicstep%
\multiply\xstep by 8944%
\divide\xstep by 10000%
\ystep=\xstep%
\divide\ystep by -2%
\makehead{\xstep}{\ystep}{\numberofsteps}}
\newcommand{\wwswhead}{\usebox{\westwestsouthwesthead}}

\newsavebox{\westwestnorthwesthead}
\savebox{\westwestnorthwesthead}{%
\xstep=\basicstep%
\multiply\xstep by 4472%
\divide\xstep by 10000%
\ystep=\xstep%
\multiply\ystep by -2%
\makehead{\xstep}{\ystep}{\numberofsteps}}
\newcommand{\wwnwhead}{\usebox{\westwestnorthwesthead}}

\newsavebox{\easteastsoutheasthead}
\savebox{\easteastsoutheasthead}{%
\xstep=-\basicstep%
\multiply\xstep by 4472%
\divide\xstep by 10000%
\ystep=\xstep%
\multiply\ystep by -2%
\makehead{\xstep}{\ystep}{\numberofsteps}}
\newcommand{\eesehead}{\usebox{\easteastsoutheasthead}}

\newsavebox{\northnorthnorthwesthead}
\savebox{\northnorthnorthwesthead}{%
\xstep=-\basicstep%
\multiply\xstep by 4472%
\divide\xstep by 10000%
\ystep=\xstep%
\multiply\ystep by 2%
\makehead{\xstep}{\ystep}{\numberofsteps}}
\newcommand{\nnnwhead}{\usebox{\northnorthnorthwesthead}}

\newsavebox{\southsouthsoutheasthead}
\savebox{\southsouthsoutheasthead}{%
\xstep=\basicstep%
\multiply\xstep by 4472%
\divide\xstep by 10000%
\ystep=\xstep%
\multiply\ystep by 2%
\makehead{\xstep}{\ystep}{\numberofsteps}}
\newcommand{\sssehead}{\usebox{\southsouthsoutheasthead}}

\newsavebox{\northeasteastnortheasthead}
\savebox{\northeasteastnortheasthead}{%
\xstep=-\basicstep%
\multiply\xstep by 9806%
\divide\xstep by 10000%
\ystep=\xstep%
\divide\ystep by -5%
\makehead{\xstep}{\ystep}{\numberofsteps}}
\newcommand{\neenehead}{\usebox{\northeasteastnortheasthead}}

\newsavebox{\northeastnorthnortheasthead}
\savebox{\northeastnorthnortheasthead}{%
\xstep=-\basicstep%
\multiply\xstep by 9806%
\divide\xstep by 10000%
\ystep=\xstep%
\divide\ystep by 5%
\makehead{\xstep}{\ystep}{\numberofsteps}}
\newcommand{\nennehead}{\usebox{\northeastnorthnortheasthead}}

\newsavebox{\southwestsouthsouthwesthead}
\savebox{\southwestsouthsouthwesthead}{%
\xstep=\basicstep%
\multiply\xstep by 9806%
\divide\xstep by 10000%
\ystep=\xstep%
\divide\ystep by 5%
\makehead{\xstep}{\ystep}{\numberofsteps}}
\newcommand{\swsswhead}{\usebox{\southwestsouthsouthwesthead}}

\newsavebox{\southwestwestsouthwesthead}
\savebox{\southwestwestsouthwesthead}{%
\xstep=\basicstep%
\multiply\xstep by 9806%
\divide\xstep by 10000%
\ystep=\xstep%
\divide\ystep by -5%
\makehead{\xstep}{\ystep}{\numberofsteps}}
\newcommand{\swwswhead}{\usebox{\southwestwestsouthwesthead}}

\newsavebox{\northwestwestnorthwesthead}
\savebox{\northwestwestnorthwesthead}{%
\xstep=\basicstep%
\multiply\xstep by 1961%
\divide\xstep by 10000%
\ystep=\xstep%
\multiply\ystep by -5%
\makehead{\xstep}{\ystep}{\numberofsteps}}
\newcommand{\nwwnwhead}{\usebox{\northwestwestnorthwesthead}}

\newsavebox{\southeasteastsoutheasthead}
\savebox{\southeasteastsoutheasthead}{%
\xstep=-\basicstep%
\multiply\xstep by 1961%
\divide\xstep by 10000%
\ystep=\xstep%
\multiply\ystep by -5%
\makehead{\xstep}{\ystep}{\numberofsteps}}
\newcommand{\seesehead}{\usebox{\southeasteastsoutheasthead}}

\newsavebox{\northwestnorthnorthwesthead}
\savebox{\northwestnorthnorthwesthead}{%
\xstep=-\basicstep%
\multiply\xstep by 1961%
\divide\xstep by 10000%
\ystep=\xstep%
\multiply\ystep by 5%
\makehead{\xstep}{\ystep}{\numberofsteps}}
\newcommand{\nwnnwhead}{\usebox{\northwestnorthnorthwesthead}}

\newsavebox{\southeastsouthsoutheasthead}
\savebox{\southeastsouthsoutheasthead}{%
\xstep=\basicstep%
\multiply\xstep by 1961%
\divide\xstep by 10000%
\ystep=\xstep%
\multiply\ystep by 5%
\makehead{\xstep}{\ystep}{\numberofsteps}}
\newcommand{\sessehead}{\usebox{\southeastsouthsoutheasthead}}

\newsavebox{\isomorphismmark}
\newcommand{\isomark}[1]{\savebox{\isomorphismmark}{#1}}
\isomark{$\cong$}

\newif\ifuserdist
\newsavebox{\distributormark}
\newcommand{\distmark}[1]{\ifx#1\distcircle\userdistfalse\else%
\userdisttrue\savebox{\distributormark}{#1}\fi}
\newsavebox{\distributorcircle}
\savebox{\distributorcircle}{\begin{picture}(0,0)%
\put(0,0){\circle{4}}\end{picture}}

\userdistfalse

\newcount\monolength
\newcount\epilength
\newcount\bimolength
\newcount\secondmonolength
\newcount\secondepilength

\newcount\monotail%
\global\monotail=\headsize%
\global\multiply\monotail by 7070%
\global\divide\monotail by 10000%

\newcount\truemonotail%
\newcount\trueepihead%

\def\truetail{\truemonotail=\monotail%
\multiply\truemonotail by 100%
\divide\truemonotail by \SCALE}

\def\truehead{\trueepihead=\epihead%
\multiply\trueepihead by 100%
\divide\trueepihead by \SCALE}

\newcount\Monotail%
\global\Monotail=\headsize%
\global\divide\Monotail by 2%

\newcount\Epihead%
\global\Epihead=\epihead%
\global\multiply\Epihead by 7070%
\global\divide\Epihead by 10000%

\newcount\Truemonotail%
\newcount\Trueepihead%

\def\Truetail{\Truemonotail=\Monotail%
\multiply\Truemonotail by 100%
\divide\Truemonotail by \SCALE}%

\def\Truehead{\Trueepihead=\Epihead%
\multiply\Trueepihead by 100%
\divide\Trueepihead by \SCALE}

\newcount\MonoTail%
\global\MonoTail=\headsize%
\global\multiply\MonoTail by 6324%
\global\divide\MonoTail by 10000%

\newcount\EpiHead%
\global\EpiHead=\epihead%
\global\multiply\EpiHead by 8944%
\global\divide\EpiHead by 10000%

\newcount\TrueMonoTail%
\newcount\TrueEpiHead%

\def\TrueTail{\TrueMonoTail=\MonoTail%
\multiply\TrueMonoTail by 100%
\divide\TrueMonoTail by \SCALE}%

\def\TrueHead{\TrueEpiHead=\EpiHead%
\multiply\TrueEpiHead by 100%
\divide\TrueEpiHead by \SCALE}

\newcount\monotaiL%
\global\monotaiL=\headsize%
\global\multiply\monotaiL by 3162%
\global\divide\monotaiL by 10000%

\newcount\epiheaD%
\global\epiheaD=\epihead%
\global\multiply\epiheaD by 4472%
\global\divide\epiheaD by 10000%

\newcount\truemonotaiL%
\newcount\trueepiheaD%

\def\truetaiL{\truemonotaiL=\monotaiL%
\multiply\truemonotaiL by 100%
\divide\truemonotaiL by \SCALE}%

\def\trueheaD{\trueepiheaD=\epiheaD%
\multiply\trueepiheaD by 100%
\divide\trueepiheaD by \SCALE}

\newcount\SCALE%

\newcount\NUMBER%

\newcount\LINE%

\newcount\COLUMN%

\newcount\WIDTH%

\newcount\SOURCE%

\newcount\ARROW%

\newcount\TARGET%

\newcount\ARROWLENGTH%

\newcount\NUMBEROFDOTS%

\newcounter{x}%

\newcounter{y}%

\newcounter{z}%

\newcounter{horizontal}%

\newcounter{vertical}%

\newskip\itemlength%

\newskip\firstitem%

\newskip\seconditem%

\newcommand{\printarrow}{}%

\newcount\X%

\newcount\Y%

\newcount\Z%

\newcommand{\truex}[1]{%
\NUMBER=#1%
\multiply\NUMBER by 100%
\divide\NUMBER by \SCALE%
\setcounter{x}{\NUMBER}}%

\newcommand{\truey}[1]{%
\NUMBER=#1%
\multiply\NUMBER by 100%
\divide\NUMBER by \SCALE%
\setcounter{y}{\NUMBER}}%

\newcommand{\truez}[1]{%
\NUMBER=#1%
\multiply\NUMBER by 100%
\divide\NUMBER by \SCALE%
\setcounter{z}{\NUMBER}}%

\newcommand{\changecounters}[1]{%
\SOURCE=\ARROW%
\ARROW=\TARGET%
\settowidth{\itemlength}{#1}%
\ifdim \itemlength > 2800\unitlength%
\addtolength{\itemlength}{-2800\unitlength}%
\TARGET=\itemlength%
\divide\TARGET by 1310%
\multiply\TARGET by 100%
\divide\TARGET by \SCALE%
\else%
\TARGET=0%
\fi%
\ARROWLENGTH=5000%
\advance\ARROWLENGTH by -\SOURCE%
\advance\ARROWLENGTH by -\TARGET%
\divide\ARROWLENGTH by 100%
\advance\SOURCE by -\TARGET}%

\newcommand{\initialize}[1]{%
\LINE=0%
\COLUMN=0%
\WIDTH=0%
\ARROW=0%
\TARGET=0%
\changecounters{#1}%
\renewcommand{\printarrow}{#1}%
\begin{center}%
\vspace{2pt}%
\begin{picture}(0,0)}%

\newcommand{\DIAGV}[2]{%
\checkmode%
\SCALE=#1%
\setlength{\unitlength}{655sp}%
\multiply\unitlength by \SCALE%
\divide\unitlength by 100%
\initialize{\mbox{$#2$}}}%

\newcommand{\n}[1]{%
\changecounters{\mbox{$#1$}}%
\put(\COLUMN,\LINE){\makebox(0,0){\printarrow}}%
\thinlines%
\renewcommand{\printarrow}{\mbox{$#1$}}%
\advance\COLUMN by 4000}%

\newcommand{\nn}[1]{%
\put(\COLUMN,\LINE){\makebox(0,0){\printarrow}}%
\thinlines%
\ifnum \WIDTH < \COLUMN%
\WIDTH=\COLUMN%
\else%
\fi%
\advance\LINE by -4000%
\COLUMN=0%
\ARROW=0%
\TARGET=0%
\changecounters{\mbox{$#1$}}%
\renewcommand{\printarrow}{\mbox{$#1$}}}%

\newcommand{\conclude}{%
\put(\COLUMN,\LINE){\makebox(0,0){\printarrow}}%
\thinlines%
\ifnum \WIDTH < \COLUMN%
\WIDTH=\COLUMN%
\else%
\fi%
\setcounter{horizontal}{\WIDTH}%
\setcounter{vertical}{-\LINE}%
\end{picture}}%

\newcommand{\diag}{%
\conclude%
\raisebox{0pt}[0pt][\value{vertical}\unitlength]{}%
\hspace*{\value{horizontal}\unitlength}%
\vspace{12pt}%
\end{center}%
\setlength{\unitlength}{1pt}%
\global\diagramfalse}%

\newcommand{\diagv}[3]{%
\conclude%
\NUMBER=#1%
\rule{0pt}{\NUMBER pt}%
\hspace*{-#2pt}%
\raisebox{0pt}[0pt][\value{vertical}\unitlength]{}%
\hspace*{\value{horizontal}\unitlength}
\NUMBER=#3%
\advance\NUMBER by 12%
\vspace*{\NUMBER pt}%
\end{center}%
\setlength{\unitlength}{1pt}%
\global\diagramfalse}%

\def\movename(#1,#2)#3{%
\hspace{#1pt}%
\raisebox{#2pt}[5pt][2pt]{\raisebox{#2pt}{$#3$}}%
\hspace{-#1pt}}%

\def\movearrow(#1,#2)#3{%
\makebox[0pt]{%
\hspace{#1pt}\hspace{#1pt}%
\raisebox{#2pt}[0pt][0pt]{\raisebox{#2pt}{$#3$}}}}%

\def\movevertex(#1,#2)#3{%
\mbox{\hspace{#1pt}%
\raisebox{#2pt}{\raisebox{#2pt}{$#3$}}%
\hspace{-#1pt}}}%

\newcommand{\crosslength}[2]{%
\settowidth{\firstitem}{#1}%
\settowidth{\seconditem}{#2}%
\ifdim\firstitem < \seconditem%
\itemlength=\seconditem%
\else%
\itemlength=\firstitem%
\fi%
\divide\itemlength by 2%
\hspace{\itemlength}}%

\def\basicDIAG#1€{\DIAGV{\defaultscale}{#1}\@ifnexttok€{\finishline}{\basicn}}
\def\basicDIAGV[#1]#2€{\DIAGV{#1}{#2}\@ifnexttok€{\finishline}{\basicn}}
\def\basicn#1€{\n{#1}\@ifnexttok€{\finishline}{\basicn}}
\def\basicnn#1€{\nn{#1}\@ifnexttok€{\finishline}{\basicn}}
\def\finishline#1{\@ifnextchar\end{\diag}%
{\@ifnextchar\spacing{\relax}{\basicnn}}}
\def\spacing(#1,#2,#3){\diagv{#1}{#2}{#3}}

\newif\ifcaption%

\newenvironment{diagram}{%
\iffloatdiag\relax\else
\global\def\diagramcaption##1{%
\global\captiontrue%
\global\def\@diagcaption{##1}}%
\global\def\@diagcaption{}\fi%
\@ifnextchar[{\basicDIAGV}{\basicDIAG}}%
{\iffloatdiag\relax\else%
\ifcaption
\begin{center}\mbox{}\@diagcaption\end{center}%
\else\relax\fi\fi\global\captionfalse}

\global\captionfalse

\gdef\@diaglabel{Diagram}
\gdef\diagramlabel#1{\gdef\@diaglabel{#1}}

\newcounter{Diagram}
\def\theDiagram{\@arabic\c@Diagram}
\def\fps@Diagram{tbp}
\def\ftype@Diagram{1}
\def\ext@Diagram{lof}
\def\fnum@Diagram{\@diaglabel\ \theDiagram}
\def\Diagram{\@float{Diagram}}
\let\endDiagram\end@float
\@namedef{Diagram*}{\@dblfloat{Diagram}}
\@namedef{endDiagram*}{\end@dblfloat}

\def\setdiagramcounter#1{\@addtoreset{Diagram}{#1}%
\def\theDiagram{\arabic{#1}.\@arabic\c@Diagram}}

\newif\iffloatdiag

{\end{diagram}\caption{\@diagcaption}\end{Diagram}%
\global\floatdiagfalse}

\global\floatdiagfalse

\newcommand{\TUP}[1]{\raisebox{0pt}[0pt][3pt]{}#1}

\newcommand{\TDOWN}[1]{\raisebox{0pt}[6pt][0pt]{}#1}

\newcommand{\tlowername}[2]%
{$\stackrel{\makebox[1pt]{#1}}%
{\begin{picture}(0,0)%
\put(0,0){\makebox(0,6)[t]{\makebox[1pt]{$\scriptstyle#2$}}}%
\end{picture}}$}%

\newcommand{\tcase}[1]{%
\testtextmode%
\setlength{\unitlength}{0.01pt}%
\makebox[\actualtextarrowspace pt]%
{\raisebox{2.5pt}{#1{\actualtextarrowlength}}}%
\setlength{\unitlength}{1pt}}%

\newcommand{\Tcase}[2]{%
\testtextmode%
\setlength{\unitlength}{0.01pt}%
\makebox[\actualtextarrowspace pt]%
{\raisebox{2.5pt}{$\stackrel{\scriptstyle #2}{#1{\actualtextarrowlength}}$}}%
\setlength{\unitlength}{1pt}}%

\newcommand{\tbicase}[1]{%
\testtextmode%
\setlength{\unitlength}{0.01pt}%
\makebox[\actualtextarrowspace pt]%
{\raisebox{1pt}{#1{\actualtextarrowlength}}}%
\setlength{\unitlength}{1pt}}%

\newcommand{\Tbicase}[3]{%
\testtextmode%
\setlength{\unitlength}{0.01pt}%
\makebox[\actualtextarrowspace pt]%
{\raisebox{-1pt}%
{$\stackrel{\scriptstyle #2}%
{\mbox{\tlowername{#1{\actualtextarrowlength}}%
{\scriptstyle #3}}}$}}%
\setlength{\unitlength}{1pt}}%

\newcommand{\DUP}[1]{\raisebox{0pt}[0pt][4pt]{}#1}

\newcommand{\DDOWN}[1]{\raisebox{0pt}[9pt][0pt]{}#1}

\newcommand{\dlowername}[2]%
{$\stackrel{\makebox[1pt]{#1}}%
{\begin{picture}(0,0)%
\put(0,0){\makebox(0,6)[t]{\makebox[1pt]{$\textstyle#2$}}}%
\end{picture}}$}%

\newcommand{\dcase}[1]{%
\testtextmode%
\setlength{\unitlength}{0.01pt}%
\makebox[\actualdisplayarrowspace pt]%
{\raisebox{2.5pt}{#1{\actualdisplayarrowlength}}}%
\setlength{\unitlength}{1pt}}%

\newcommand{\Dcase}[2]{%
\testtextmode%
\setlength{\unitlength}{0.01pt}%
\makebox[\actualdisplayarrowspace pt]%
{\raisebox{2.5pt}{$\stackrel{\textstyle #2}{#1{\actualdisplayarrowlength}}$}}%
\setlength{\unitlength}{1pt}}%

\newcommand{\dbicase}[1]{%
\testtextmode%
\setlength{\unitlength}{0.01pt}%
\makebox[\actualdisplayarrowspace pt]%
{\raisebox{1pt}{#1{\actualdisplayarrowlength}}}%
\setlength{\unitlength}{1pt}}%

\newcommand{\Dbicase}[3]{%
\testtextmode%
\setlength{\unitlength}{0.01pt}%
\makebox[\actualdisplayarrowspace pt]%
{\raisebox{-1pt}%
{$\stackrel{\textstyle #2}%
{\mbox{\tlowername{#1{\actualdisplayarrowlength}}%
{\textstyle #3}}}$}}%
\setlength{\unitlength}{1pt}}%

\newcommand{\AR}[1]%
{\begin{picture}(#1,0)%
\put(0,0){\line(1,0){#1}}%
\put(#1,0){\ehead}%
\end{picture}}%

\newcommand{\DIST}[1]%
{\begin{picture}(#1,0)%
\put(0,0){\line(1,0){#1}}%
\put(#1,0){\ehead}%
\NUMBER=#1%
\divide\NUMBER by 2%
\put(\NUMBER,0){\circle{400}}%
\end{picture}}%

\newcommand{\DOTAR}[1]%
{\NUMBEROFDOTS=#1%
\divide\NUMBEROFDOTS by 300%
\advance\NUMBEROFDOTS by 1%
\begin{picture}(#1,0)%
\multiput(0,0)(300,0){\NUMBEROFDOTS}{\circle*{100}}%
\put(#1,0){\ehead}%
\end{picture}}%

\newcommand{\MONO}[1]%
{\monolength=#1%
\advance\monolength by -\monotail%
\begin{picture}(#1,0)%
\put(\monotail,0){\line(1,0){\monolength}}%
\put(#1,0){\ehead}%
\put(\monotail,0){\ehead}%
\end{picture}}%

\newcommand{\EPI}[1]%
{\epilength=#1%
\advance\epilength by -\epihead%
\begin{picture}(#1,0)(-#1,0)%
\put(-#1,0){\line(1,0){\epilength}}%
\put(-\epihead,0){\ehead}%
\put(0,0){\ehead}%
\end{picture}}%

\newcommand{\BIMO}[1]%
{\monolength=#1%
\advance\monolength by -\monotail%
\epilength=\monolength%
\advance\epilength by -\epihead%
\begin{picture}(#1,0)(-#1,0)%
\put(-\monolength,0){\line(1,0){\epilength}}%
\put(-\monolength,0){\ehead}%
\put(-\epihead,0){\ehead}%
\put(0,0){\ehead}%
\end{picture}}%

\newcommand{\BIAR}[1]%
{\begin{picture}(#1,700)%
\put(0,0){\line(1,0){#1}}%
\put(#1,0){\ehead}%
\put(0,700){\line(1,0){#1}}%
\put(#1,700){\ehead}%
\end{picture}}%

\newcommand{\BIDIST}[1]%
{\begin{picture}(#1,700)%
\put(0,0){\line(1,0){#1}}%
\put(#1,0){\ehead}%
\put(0,700){\line(1,0){#1}}%
\put(#1,700){\ehead}%
\NUMBER=#1%
\divide\NUMBER by 2%
\put(\NUMBER,0){\circle{400}}%
\put(\NUMBER,700){\circle{400}}%
\end{picture}}%

\newcommand{\EQL}[1]%
{\begin{picture}(#1,0)%
\put(0,100){\line(1,0){#1}}%
\put(0,-100){\line(1,0){#1}}%
\end{picture}}%

\newcommand{\ADJAR}[1]%
{\begin{picture}(#1,700)%
\put(0,0){\line(1,0){#1}}%
\put(#1,0){\ehead}%
\put(#1,700){\line(-1,0){#1}}%
\put(0,700){\whead}
\end{picture}}%

\newcommand{\ADJDIST}[1]%
{\begin{picture}(#1,700)%
\put(0,0){\line(1,0){#1}}%
\put(#1,0){\ehead}%
\put(#1,700){\line(-1,0){#1}}%
\put(0,700){\whead}
\NUMBER=#1%
\divide\NUMBER by 2%
\put(\NUMBER,0){\circle{400}}%
\put(\NUMBER,700){\circle{400}}%
\end{picture}}%

\newcommand{\ar}{\ifinner\tcase{\AR}\else\dcase{\AR}\fi}%

\newcommand{\Ar}[1]{\ifinner\Tcase{\AR}{#1}\else\Dcase{\AR}{#1}\fi}%

\newcommand{\dist}{\ifinner\tcase{\DIST}\else\dcase{\DIST}\fi}%

\newcommand{\Dist}[1]{\ifinner\Tcase{\DIST}{\TUP{#1}}%
\else\Dcase{\DIST}{\TUP{#1}}\fi}%

\newcommand{\dotar}{\ifinner\tcase{\DOTAR}\else\dcase{\DOTAR}\fi}%

\newcommand{\Dotar}[1]{\ifinner\Tcase{\DOTAR}{#1}%
\else\Dcase{\DOTAR}{#1}\fi}%

\newcommand{\mono}{\ifinner\tcase{\MONO}\else\dcase{\MONO}\fi}%

\newcommand{\Mono}[1]{\ifinner\Tcase{\MONO}{#1}\else\Dcase{\MONO}{#1}\fi}%

\newcommand{\epi}{\ifinner\tcase{\EPI}\else\dcase{\EPI}\fi}%

\newcommand{\Epi}[1]{\ifinner\Tcase{\EPI}{#1}\else\Dcase{\EPI}{#1}\fi}%

\newcommand{\bimo}{\ifinner\tcase{\BIMO}\else\dcase{\BIMO}\fi}%

\newcommand{\Bimo}[1]{\ifinner\Tcase{\BIMO}{#1}%
\else\Dcase{\BIMO}{#1}\fi}%

\newcommand{\iso}{\ifinner\Tcase{\AR}{\cong}\else\Dcase{\AR}{\cong}\fi}%

\newcommand{\Iso}[1]{\ifinner\Tcase{\AR}{\cong{#1}}%
\else\Dcase{\AR}{\cong{#1}}\fi}%

\newcommand{\biar}{\ifinner\tbicase{\BIAR}\else\dbicase{\BIAR}\fi}%

\newcommand{\Biar}[2]{\ifinner\Tbicase{\BIAR}{#1}{#2}%
\else\Dbicase{\BIAR}{#1}{#2}\fi}%

\newcommand{\bidist}{\ifinner\tbicase{\BIDIST}\else\dbicase{\BIDIST}\fi}%

\newcommand{\Bidist}[2]{\ifinner\Tbicase{\BIDIST}{\TUP{#1}}{\TDOWN{#2}}%
\else\Dbicase{\BIDIST}{\DUP{#1}}{\DDOWN{#2}}\fi}%

\newcommand{\eql}{\ifinner\tcase{\EQL}\else\dcase{\EQL}\fi}%

\newcommand{\Eql}[1]{\ifinner\Tcase{\EQL}{\TUP{#1}}%
\else\Dcase{\EQL}{\DUP{#1}}\fi}%

\newcommand{\adjar}{\ifinner\tbicase{\ADJAR}\else\dbicase{\ADJAR}\fi}%

\newcommand{\Adjar}[2]{\ifinner\Tbicase{\ADJAR}{#1}{#2}%
\else\Dbicase{\ADJAR}{#1}{#2}\fi}%

\newcommand{\adjdist}{\ifinner\tbicase{\ADJDIST}\else\dbicase{\ADJDIST}\fi}%

\newcommand{\Adjdist}[2]{\ifinner\Tbicase{\ADJDIST}{\TUP{#1}}{\TDOWN{#2}}%
\else\Dbicase{\ADJDIST}{\DUP{#1}}{\DDOWN{#2}}\fi}%

\newcommand{\BKAR}[1]%
{\begin{picture}(#1,0)%
\put(#1,0){\line(-1,0){#1}}%
\put(0,0){\whead}%
\end{picture}}%

\newcommand{\BKDIST}[1]%
{\begin{picture}(#1,0)%
\put(#1,0){\line(-1,0){#1}}%
\put(0,0){\whead}%
\NUMBER=#1%
\divide\NUMBER by 2%
\put(\NUMBER,0){\circle{400}}%
\end{picture}}%

\newcommand{\BKDOTAR}[1]%
{\NUMBEROFDOTS=#1%
\divide\NUMBEROFDOTS by 300%
\advance\NUMBEROFDOTS by 1%
\begin{picture}(#1,0)%
\multiput(#1,0)(-300,0){\NUMBEROFDOTS}{\circle*{100}}%
\put(0,0){\whead}%
\end{picture}}%

\newcommand{\BKMONO}[1]%
{\monolength=#1%
\advance\monolength by -\monotail%
\begin{picture}(#1,0)(-#1,0)%
\put(-\monotail,0){\line(-1,0){\monolength}}%
\put(-\monotail,0){\whead}%
\put(-#1,0){\whead}%
\end{picture}}%

\newcommand{\BKEPI}[1]%
{\epilength=#1%
\advance\epilength by -\epihead%
\begin{picture}(#1,0)%
\put(#1,0){\line(-1,0){\epilength}}%
\put(\epihead,0){\whead}%
\put(0,0){\whead}%
\end{picture}}%

\newcommand{\BKBIMO}[1]%
{\monolength=#1%
\advance\monolength by -\monotail%
\epilength=\monolength%
\advance\epilength by -\epihead%
\begin{picture}(#1,0)%
\put(\monolength,0){\line(-1,0){\epilength}}%
\put(\monolength,0){\whead}%
\put(\epihead,0){\whead}%
\put(0,0){\whead}%
\end{picture}}%

\newcommand{\BKBIAR}[1]%
{\begin{picture}(#1,700)%
\put(#1,0){\line(-1,0){#1}}%
\put(0,0){\whead}%
\put(#1,700){\line(-1,0){#1}}%
\put(0,700){\whead}%
\end{picture}}%

\newcommand{\BKBIDIST}[1]%
{\begin{picture}(#1,700)%
\put(#1,0){\line(-1,0){#1}}%
\put(0,0){\whead}%
\put(#1,700){\line(-1,0){#1}}%
\put(0,700){\whead}%
\NUMBER=#1%
\divide\NUMBER by 2%
\put(\NUMBER,0){\circle{400}}%
\put(\NUMBER,700){\circle{400}}%
\end{picture}}%

\newcommand{\BKADJAR}[1]%
{\begin{picture}(#1,700)%
\put(0,700){\line(1,0){#1}}%
\put(#1,700){\ehead}%
\put(#1,0){\line(-1,0){#1}}%
\put(0,0){\whead}%
\end{picture}}%

\newcommand{\BKADJDIST}[1]%
{\begin{picture}(#1,700)%
\put(0,700){\line(1,0){#1}}%
\put(#1,700){\ehead}%
\put(#1,0){\line(-1,0){#1}}%
\put(0,0){\whead}%
\NUMBER=#1%
\divide\NUMBER by 2%
\put(\NUMBER,0){\circle{400}}%
\put(\NUMBER,700){\circle{400}}%
\end{picture}}%

\newcommand{\bkar}{\ifinner\tcase{\BKAR}\else\dcase{\BKAR}\fi}%

\newcommand{\Bkar}[1]{\ifinner\Tcase{\BKAR}{#1}\else\Dcase{\BKAR}{#1}\fi}%

\newcommand{\bkdist}{\ifinner\tcase{\BKDIST}\else\dcase{\BKDIST}\fi}%

\newcommand{\Bkdist}[1]{\ifinner\Tcase{\BKDIST}{\TUP{#1}}%
\else\Dcase{\BKDIST}{\TUP{#1}}\fi}%

\newcommand{\bkdotar}{\ifinner\tcase{\BKDOTAR}\else\dcase{\BKDOTAR}\fi}%

\newcommand{\Bkdotar}[1]{\ifinner\Tcase{\BKDOTAR}{#1}%
\else\Dcase{\BKDOTAR}{#1}\fi}%

\newcommand{\bkmono}{\ifinner\tcase{\BKMONO}\else\dcase{\BKMONO}\fi}%

\newcommand{\Bkmono}[1]{\ifinner\Tcase{\BKMONO}{#1}%
\else\Dcase{\BKMONO}{#1}\fi}%

\newcommand{\bkepi}{\ifinner\tcase{\BKEPI}\else\dcase{\BKEPI}\fi}%

\newcommand{\Bkepi}[1]{\ifinner\Tcase{\BKEPI}{#1}%
\else\Dcase{\BKEPI}{#1}\fi}%

\newcommand{\bkbimo}{\ifinner\tcase{\BKBIMO}\else\dcase{\BKBIMO}\fi}%

\newcommand{\Bkbimo}[1]{\ifinner\Tcase{\BKBIMO}{\hspace{9pt}#1}%
\else\Dcase{\BKBIMO}{\hspace{9pt}#1}\fi}%

\newcommand{\bkiso}{\ifinner\Tcase{\BKAR}{\cong}%
\else\Dcase{\BKAR}{\cong}\fi}%

\newcommand{\Bkiso}[1]{\ifinner\Tcase{\BKAR}{\cong{#1}}%
\else\Dcase{\BKAR}{\cong{#1}}\fi}%

\newcommand{\bkbiar}{\ifinner\tbicase{\BKBIAR}\else\dbicase{\BKBIAR}\fi}%

\newcommand{\Bkbiar}[2]{\ifinner\Tbicase{\BKBIAR}{#1}{#2}%
\else\Dbicase{\BKBIAR}{#1}{#2}\fi}%

\newcommand{\bkbidist}{\ifinner\tbicase{\BKBIDIST}%
\else\dbicase{\BKBIDIST}\fi}%

\newcommand{\Bkbidist}[2]{\ifinner\Tbicase{\BKBIDIST}{\TUP{#1}}{\TDOWN{#2}}%
\else\Tbicase{\BKBIDIST}{\DUP{#1}}{\DDOWN{#2}}\fi}%

\newcommand{\bkadjar}{\ifinner\tbicase{\BKADJAR}%
\else\dbicase{\BKADJAR}\fi}%

\newcommand{\Bkadjar}[2]{\ifinner\Tbicase{\BKADJAR}{#1}{#2}%
\else\Dbicase{\BKADJAR}{#1}{#2}\fi}%

\newcommand{\bkadjdist}{\ifinner\tbicase{\BKADJDIST}%
\else\dbicase{\BKADJDIST}\fi}%

\newcommand{\Bkadjdist}[2]{\ifinner\Tbicase{\BKADJDIST}{\TUP{#1}}{\TDOWN{#2}}%
\else\Dbicase{\BKADJDIST}{\TUP{#1}}{\TDOWN{#2}}\fi}%

\newcommand{\lowername}[2]%
{$\stackrel{\makebox[1pt]{#1}}%
{\begin{picture}(0,0)%
\truex{600}%
\put(0,0){\makebox(0,\value{x})[t]{\makebox[1pt]{$#2$}}}%
\end{picture}}$}%

\newcommand{\hcase}[2]%
{\testdiagrammode\makebox[0pt]%
{\raisebox{0pt}[0pt][0pt]{#1{#2}}}}%

\newcommand{\Hcase}[3]%
{\testdiagrammode\makebox[0pt]
{\raisebox{0pt}[0pt][0pt]%
{$\stackrel{\makebox[0pt]{$\textstyle{#2}$}}{#1{#3}}$}}}%

\newcommand{\hcasE}[3]%
{\testdiagrammode\makebox[0pt]%
{\raisebox{-8pt}[0pt][0pt]%
{\lowername{#1{#3}}{#2}}}}%

\newcommand{\Hisocase}[4]%
{\testdiagrammode\makebox[0pt]
{\raisebox{-8pt}[0pt][0pt]%
{$\stackrel{\makebox[0pt]{$\textstyle{#2}$}}%
{\mbox{\lowername{#1{#4}}{#3}}}$}}}%

\newcommand{\hbicase}[2]%
{\testdiagrammode\makebox[0pt]%
{\raisebox{-2.4pt}[0pt][0pt]{#1{#2}}}}%

\newcommand{\Hbicase}[4]%
{\testdiagrammode\makebox[0pt]
{\raisebox{-10.4pt}[0pt][0pt]%
{$\stackrel{\makebox[0pt]{$\textstyle{#2}$}}%
{\mbox{\lowername{#1{#4}}{#3}}}$}}}%

\newcommand{\EAR}[1]%
{\begin{picture}(#1,0)%
\put(0,0){\line(1,0){#1}}%
\put(#1,0){\ehead}%
\end{picture}}%

\newcommand{\EDIST}[1]%
{\begin{picture}(#1,0)%
\put(0,0){\line(1,0){#1}}%
\put(#1,0){\ehead}%
\truex{400}
\NUMBER=#1%
\divide\NUMBER by 2%
\put(\NUMBER,0){\circle{\value{x}}}
\end{picture}}%

\newcommand{\EDOTAR}[1]%
{\truex{100}\truey{300}%
\NUMBEROFDOTS=#1%
\divide\NUMBEROFDOTS by \value{y}%
\advance\NUMBEROFDOTS by 1%
\begin{picture}(#1,0)%
\multiput(0,0)(\value{y},0){\NUMBEROFDOTS}%
{\circle*{\value{x}}}%
\put(#1,0){\ehead}%
\end{picture}}%

\newcommand{\EMONO}[1]%
{\truetail
\monolength=#1%
\advance\monolength by -\truemonotail%
\begin{picture}(#1,0)%
\put(\truemonotail,0){\line(1,0){\monolength}}%
\put(#1,0){\ehead}%
\put(\truemonotail,0){\ehead}%
\end{picture}}%

\newcommand{\EEPI}[1]%
{\truehead%
\epilength=#1%
\advance\epilength by -\trueepihead%
\begin{picture}(#1,0)(-#1,0)%
\put(-#1,0){\line(1,0){\epilength}}%
\put(-\trueepihead,0){\ehead}%
\put(0,0){\ehead}%
\end{picture}}%

\newcommand{\EBIMO}[1]%
{\truehead\truetail%
\monolength=#1%
\advance\monolength by -\truemonotail%
\epilength=\monolength%
\advance\epilength by -\trueepihead%
\begin{picture}(#1,0)(-#1,0)%
\put(-\monolength,0){\line(1,0){\epilength}}%
\put(-\monolength,0){\ehead}%
\put(-\trueepihead,0){\ehead}%
\put(0,0){\ehead}%
\end{picture}}%

\newcommand{\EBIAR}[1]%
{\truex{700}%
\begin{picture}(#1,\value{x})%
\put(0,0){\line(1,0){#1}}%
\put(#1,0){\ehead}%
\put(0,\value{x}){\line(1,0){#1}}%
\put(#1,\value{x}){\ehead}%
\end{picture}}%

\newcommand{\EBIDIST}[1]%
{\truex{700}%
\begin{picture}(#1,\value{x})%
\put(0,0){\line(1,0){#1}}%
\put(#1,0){\ehead}%
\put(0,\value{x}){\line(1,0){#1}}%
\put(#1,\value{x}){\ehead}%
\truey{400}%
\NUMBER=#1%
\divide\NUMBER by 2%
\put(\NUMBER,0){\circle{\value{y}}}
\put(\NUMBER,\value{x}){\circle{\value{y}}}%
\end{picture}}%

\newcommand{\EEQL}[1]%
{\begin{picture}(#1,0)%
\truex{200}%
\put(0,\value{x}){\line(1,0){#1}}%
\put(0,0){\line(1,0){#1}}%
\end{picture}}%

\newcommand{\EADJAR}[1]%
{\truex{700}%
\begin{picture}(#1,\value{x})%
\put(0,0){\line(1,0){#1}}%
\put(#1,0){\ehead}%
\put(#1,\value{x}){\line(-1,0){#1}}%
\put(0,\value{x}){\whead}%
\end{picture}}%

\newcommand{\EADJDIST}[1]%
{\truex{700}%
\begin{picture}(#1,\value{x})%
\put(0,0){\line(1,0){#1}}%
\put(#1,0){\ehead}%
\put(#1,\value{x}){\line(-1,0){#1}}%
\put(0,\value{x}){\whead}%
\truey{400}%
\NUMBER=#1%
\divide\NUMBER by 2%
\put(\NUMBER,0){\circle{\value{y}}}
\put(\NUMBER,\value{x}){\circle{\value{y}}}%
\end{picture}}%

\def\basicear[#1]{%
\Z=#1%
\multiply \Z by 100%
\hcase{\EAR}{\Z}}%

\newcommand{\ear}{\@ifnextchar[{\basicear}%
{\hspace{\SOURCE\unitlength}\basicear[\ARROWLENGTH]}}%

\def\basicEar[#1]#2{%
\Z=#1%
\multiply \Z by 100%
\Hcase{\EAR}{#2}{\Z}}%

\newcommand{\Ear}{\@ifnextchar[{\basicEar}%
{\hspace{\SOURCE\unitlength}\basicEar[\ARROWLENGTH]}}%

\def\basiceaR[#1]#2{%
\Z=#1%
\multiply \Z by 100%
\hcasE{\EAR}{#2}{\Z}}%

\newcommand{\eaR}{\@ifnextchar[{\basiceaR}%
{\hspace{\SOURCE\unitlength}\basiceaR[\ARROWLENGTH]}}%

\def\basicedist[#1]{%
\Z=#1%
\multiply \Z by 100%
\hcase{\EDIST}{\Z}}%

\newcommand{\edist}{\@ifnextchar[{\basicedist}%
{\hspace{\SOURCE\unitlength}\basicedist[\ARROWLENGTH]}}%

\def\basicEdist[#1]#2{%
\Z=#1%
\multiply \Z by 100%
\Hcase{\EDIST}{\DUP{#2}}{\Z}}%

\newcommand{\Edist}{\@ifnextchar[{\basicEdist}%
{\hspace{\SOURCE\unitlength}\basicEdist[\ARROWLENGTH]}}%

\def\basicedisT[#1]#2{%
\Z=#1%
\multiply \Z by 100%
\hcasE{\EDIST}{\DDOWN{#2}}{\Z}}%

\newcommand{\edisT}{\@ifnextchar[{\basicedisT}%
{\hspace{\SOURCE\unitlength}\basicedisT[\ARROWLENGTH]}}%

\def\basicedotar[#1]{%
\Z=#1%
\multiply \Z by 100%
\hcase{\EDOTAR}{\Z}}%

\newcommand{\edotar}{\@ifnextchar[{\basicedotar}%
{\hspace{\SOURCE\unitlength}\basicedotar[\ARROWLENGTH]}}%

\def\basicEdotar[#1]#2{%
\Z=#1%
\multiply \Z by 100%
\Hcase{\EDOTAR}{#2}{\Z}}%

\newcommand{\Edotar}{\@ifnextchar[{\basicEdotar}%
{\hspace{\SOURCE\unitlength}\basicEdotar[\ARROWLENGTH]}}%

\def\basicedotaR[#1]#2{%
\Z=#1%
\multiply \Z by 100%
\hcasE{\EDOTAR}{#2}{\Z}}%

\newcommand{\edotaR}{\@ifnextchar[{\basicedotaR}%
{\hspace{\SOURCE\unitlength}\basicedotaR[\ARROWLENGTH]}}%

\def\basicemono[#1]{%
\Z=#1%
\multiply \Z by 100%
\hcase{\EMONO}{\Z}}%

\newcommand{\emono}{\@ifnextchar[{\basicemono}%
{\hspace{\SOURCE\unitlength}\basicemono[\ARROWLENGTH]}}%

\def\basicEmono[#1]#2{%
\Z=#1%
\multiply \Z by 100%
\Hcase{\EMONO}{#2}{\Z}}%

\newcommand{\Emono}{\@ifnextchar[{\basicEmono}%
{\hspace{\SOURCE\unitlength}\basicEmono[\ARROWLENGTH]}}%

\def\basicemonO[#1]#2{%
\Z=#1%
\multiply \Z by 100%
\hcasE{\EMONO}{#2}{\Z}}%

\newcommand{\emonO}{\@ifnextchar[{\basicemonO}%
{\hspace{\SOURCE\unitlength}\basicemonO[\ARROWLENGTH]}}%

\def\basiceepi[#1]{%
\Z=#1%
\multiply \Z by 100%
\hcase{\EEPI}{\Z}}%

\newcommand{\eepi}{\@ifnextchar[{\basiceepi}%
{\hspace{\SOURCE\unitlength}\basiceepi[\ARROWLENGTH]}}%

\def\basicEepi[#1]#2{%
\Z=#1%
\multiply \Z by 100%
\Hcase{\EEPI}{#2}{\Z}}%

\newcommand{\Eepi}{\@ifnextchar[{\basicEepi}%
{\hspace{\SOURCE\unitlength}\basicEepi[\ARROWLENGTH]}}%

\def\basiceepI[#1]#2{%
\Z=#1%
\multiply \Z by 100%
\hcasE{\EEPI}{#2}{\Z}}%

\newcommand{\eepI}{\@ifnextchar[{\basiceepI}%
{\hspace{\SOURCE\unitlength}\basiceepI[\ARROWLENGTH]}}%

\def\basicebimo[#1]{%
\Z=#1%
\multiply \Z by 100%
\hcase{\EBIMO}{\Z}}%

\newcommand{\ebimo}{\@ifnextchar[{\basicebimo}%
{\hspace{\SOURCE\unitlength}\basicebimo[\ARROWLENGTH]}}%

\def\basicEbimo[#1]#2{%
\Z=#1%
\multiply \Z by 100%
\Hcase{\EBIMO}{#2}{\Z}}%

\newcommand{\Ebimo}{\@ifnextchar[{\basicEbimo}%
{\hspace{\SOURCE\unitlength}\basicEbimo[\ARROWLENGTH]}}%

\def\basicebimO[#1]#2{%
\Z=#1%
\multiply \Z by 100%
\hcasE{\EBIMO}{#2}{\Z}}%

\newcommand{\ebimO}{\@ifnextchar[{\basicebimO}%
{\hspace{\SOURCE\unitlength}\basicebimO[\ARROWLENGTH]}}%

\def\basiceiso[#1]{%
\Z=#1%
\multiply \Z by 100%
\Hisocase{\EAR}{\cong}{}{\Z}}%

\newcommand{\eiso}{\@ifnextchar[{\basiceiso}%
{\hspace{\SOURCE\unitlength}\basiceiso[\ARROWLENGTH]}}%

\def\basicEiso[#1]#2{%
\Z=#1%
\multiply \Z by 100%
\Hisocase{\EAR}{#2}{\cong}{\Z}}%

\newcommand{\Eiso}{\@ifnextchar[{\basicEiso}%
{\hspace{\SOURCE\unitlength}\basicEiso[\ARROWLENGTH]}}%

\def\basiceisO[#1]#2{%
\Z=#1%
\multiply \Z by 100%
\Hisocase{\EAR}{\cong}{#2}{\Z}}%

\newcommand{\eisO}{\@ifnextchar[{\basiceisO}%
{\hspace{\SOURCE\unitlength}\basiceisO[\ARROWLENGTH]}}%

\def\basiceeql[#1]{%
\Z=#1%
\multiply \Z by 100%
\hcase{\EEQL}{\Z}}%

\newcommand{\eeql}{\@ifnextchar[{\basiceeql}%
{\hspace{\SOURCE\unitlength}\basiceeql[\ARROWLENGTH]}}%

\def\basicEeql[#1]#2{%
\Z=#1%
\multiply \Z by 100%
\Hcase{\EEQL}{\DUP{#2}}{\Z}}%

\newcommand{\Eeql}{\@ifnextchar[{\basicEeql}%
{\hspace{\SOURCE\unitlength}\basicEeql[\ARROWLENGTH]}}%

\def\basiceeqL[#1]#2{%
\Z=#1%
\multiply \Z by 100%
\hcasE{\EEQL}{#2}{\Z}}%

\newcommand{\eeqL}{\@ifnextchar[{\basiceeqL}%
{\hspace{\SOURCE\unitlength}\basiceeqL[\ARROWLENGTH]}}%

\def\basicebiar[#1]{%
\Z=#1%
\multiply \Z by 100%
\hbicase{\EBIAR}{\Z}}%

\newcommand{\ebiar}{\@ifnextchar[{\basicebiar}%
{\hspace{\SOURCE\unitlength}\basicebiar[\ARROWLENGTH]}}%

\def\basicEbiar[#1]#2#3{%
\Z=#1%
\multiply \Z by 100%
\Hbicase{\EBIAR}{#2}{#3}{\Z}}%

\newcommand{\Ebiar}{\@ifnextchar[{\basicEbiar}%
{\hspace{\SOURCE\unitlength}\basicEbiar[\ARROWLENGTH]}}%

\def\basicebidist[#1]{%
\Z=#1%
\multiply \Z by 100%
\hbicase{\EBIDIST}{\Z}}%

\newcommand{\ebidist}{\@ifnextchar[{\basicebidist}%
{\hspace{\SOURCE\unitlength}\basicebidist[\ARROWLENGTH]}}%

\def\basicEbidist[#1]#2#3{%
\Z=#1%
\multiply \Z by 100%
\Hbicase{\EBIDIST}{\DUP{#2}}{\DDOWN{#3}}{\Z}}%

\newcommand{\Ebidist}{\@ifnextchar[{\basicEbidist}%
{\hspace{\SOURCE\unitlength}\basicEbidist[\ARROWLENGTH]}}%

\def\basiceadjar[#1]{%
\Z=#1%
\multiply \Z by 100%
\hbicase{\EADJAR}{\Z}}%

\newcommand{\eadjar}{\@ifnextchar[{\basiceadjar}%
{\hspace{\SOURCE\unitlength}\basiceadjar[\ARROWLENGTH]}}%

\def\basicEadjar[#1]#2#3{%
\Z=#1%
\multiply \Z by 100%
\Hbicase{\EADJAR}{#2}{#3}{\Z}}%

\newcommand{\Eadjar}{\@ifnextchar[{\basicEadjar}%
{\hspace{\SOURCE\unitlength}\basicEadjar[\ARROWLENGTH]}}%

\def\basiceadjdist[#1]{%
\Z=#1%
\multiply \Z by 100%
\hbicase{\EADJDIST}{\Z}}%

\newcommand{\eadjdist}{\@ifnextchar[{\basiceadjdist}%
{\hspace{\SOURCE\unitlength}\basiceadjdist[\ARROWLENGTH]}}%

\def\basicEadjdist[#1]#2#3{%
\Z=#1%
\multiply \Z by 100%
\Hbicase{\EADJDIST}{\DUP{#2}}{\DDOWN{#3}}{\Z}}%

\newcommand{\Eadjdist}{\@ifnextchar[{\basicEadjdist}%
{\hspace{\SOURCE\unitlength}\basicEadjdist[\ARROWLENGTH]}}%

\newcommand{\WAR}[1]%
{\begin{picture}(#1,0)%
\put(#1,0){\line(-1,0){#1}}%
\put(0,0){\whead}%
\end{picture}}%

\newcommand{\WDIST}[1]%
{\begin{picture}(#1,0)%
\put(#1,0){\line(-1,0){#1}}%
\put(0,0){\whead}%
\truex{400}%
\NUMBER=#1%
\divide\NUMBER by 2%
\put(\NUMBER,0){\circle{\value{x}}}%
\end{picture}}%

\newcommand{\WDOTAR}[1]%
{\truex{100}\truey{300}%
\NUMBEROFDOTS=#1%
\divide\NUMBEROFDOTS by \value{y}%
\advance\NUMBEROFDOTS by 1%
\begin{picture}(#1,0)%
\multiput(#1,0)(-\value{y},0){\NUMBEROFDOTS}%
{\circle*{\value{x}}}%
\put(0,0){\whead}%
\end{picture}}%

\newcommand{\WMONO}[1]%
{\truetail%
\monolength=#1%
\advance\monolength by -\truemonotail%
\begin{picture}(#1,0)(-#1,0)%
\put(-\truemonotail,0){\line(-1,0){\monolength}}%
\put(-\truemonotail,0){\whead}%
\put(-#1,0){\whead}%
\end{picture}}%

\newcommand{\WEPI}[1]%
{\truehead%
\epilength=#1%
\advance\epilength by -\trueepihead%
\begin{picture}(#1,0)%
\put(#1,0){\line(-1,0){\epilength}}%
\put(\trueepihead,0){\whead}%
\put(0,0){\whead}%
\end{picture}}%

\newcommand{\WBIMO}[1]%
{\truehead\truetail%
\monolength=#1
\advance\monolength by -\truemonotail%
\epilength=\monolength%
\advance\epilength by -\trueepihead%
\begin{picture}(#1,0)%
\put(\monolength,0){\line(-1,0){\epilength}}%
\put(\monolength,0){\whead}%
\put(\trueepihead,0){\whead}%
\put(0,0){\whead}%
\end{picture}}%

\newcommand{\WBIAR}[1]%
{\truex{700}%
\begin{picture}(#1,\value{x})%
\put(#1,0){\line(-1,0){#1}}%
\put(0,0){\whead}%
\put(#1,\value{x}){\line(-1,0){#1}}%
\put(0,\value{x}){\whead}%
\end{picture}}%

\newcommand{\WBIDIST}[1]%
{\truex{700}%
\begin{picture}(#1,\value{x})%
\put(#1,0){\line(-1,0){#1}}%
\put(0,0){\whead}%
\put(#1,\value{x}){\line(-1,0){#1}}%
\put(0,\value{x}){\whead}%
\truey{400}%
\NUMBER=#1%
\divide\NUMBER by 2%
\put(\NUMBER,0){\circle{\value{y}}}%
\put(\NUMBER,\value{x}){\circle{\value{y}}}%
\end{picture}}%

\newcommand{\WADJAR}[1]%
{\truex{700}%
\begin{picture}(#1,\value{x})%
\put(0,\value{x}){\line(1,0){#1}}%
\put(#1,\value{x}){\ehead}%
\put(#1,0){\line(-1,0){#1}}%
\put(0,0){\whead}%
\end{picture}}%

\newcommand{\WADJDIST}[1]%
{\truex{700}%
\begin{picture}(#1,\value{x})%
\put(0,\value{x}){\line(1,0){#1}}%
\put(#1,\value{x}){\ehead}%
\put(#1,0){\line(-1,0){#1}}%
\put(0,0){\whead}%
\truey{400}%
\NUMBER=#1%
\divide\NUMBER by 2%
\put(\NUMBER,0){\circle{\value{y}}}%
\put(\NUMBER,\value{x}){\circle{\value{y}}}%
\end{picture}}%

\def\basicwar[#1]{%
\Z=#1%
\multiply \Z by 100%
\hcase{\WAR}{\Z}}%

\newcommand{\war}{\@ifnextchar[{\basicwar}%
{\hspace{\SOURCE\unitlength}\basicwar[\ARROWLENGTH]}}%

\def\basicWar[#1]#2{%
\Z=#1%
\multiply \Z by 100%
\Hcase{\WAR}{#2}{\Z}}%

\newcommand{\War}{\@ifnextchar[{\basicWar}%
{\hspace{\SOURCE\unitlength}\basicWar[\ARROWLENGTH]}}%

\def\basicwaR[#1]#2{%
\Z=#1%
\multiply \Z by 100%
\hcasE{\WAR}{#2}{\Z}}%

\newcommand{\waR}{\@ifnextchar[{\basicwaR}%
{\hspace{\SOURCE\unitlength}\basicwaR[\ARROWLENGTH]}}%

\def\basicwdist[#1]{%
\Z=#1%
\multiply \Z by 100%
\hcase{\WDIST}{\Z}}%

\newcommand{\wdist}{\@ifnextchar[{\basicwdist}%
{\hspace{\SOURCE\unitlength}\basicwdist[\ARROWLENGTH]}}%

\def\basicWdist[#1]#2{%
\Z=#1%
\multiply \Z by 100%
\Hcase{\WDIST}{\DUP{#2}}{\Z}}%

\newcommand{\Wdist}{\@ifnextchar[{\basicWdist}%
{\hspace{\SOURCE\unitlength}\basicWdist[\ARROWLENGTH]}}%

\def\basicwdisT[#1]#2{%
\Z=#1%
\multiply \Z by 100%
\hcasE{\WDIST}{\DDOWN{#2}}{\Z}}%

\newcommand{\wdisT}{\@ifnextchar[{\basicwdisT}%
{\hspace{\SOURCE\unitlength}\basicwdisT[\ARROWLENGTH]}}%

\def\basicwdotar[#1]{%
\Z=#1%
\multiply \Z by 100%
\hcase{\WDOTAR}{\Z}}%

\newcommand{\wdotar}{\@ifnextchar[{\basicwdotar}%
{\hspace{\SOURCE\unitlength}\basicwdotar[\ARROWLENGTH]}}%

\def\basicWdotar[#1]#2{%
\Z=#1%
\multiply \Z by 100%
\Hcase{\WDOTAR}{#2}{\Z}}%

\newcommand{\Wdotar}{\@ifnextchar[{\basicWdotar}%
{\hspace{\SOURCE\unitlength}\basicWdotar[\ARROWLENGTH]}}%

\def\basicwdotaR[#1]#2{%
\Z=#1%
\multiply \Z by 100%
\hcasE{\WDOTAR}{#2}{\Z}}%

\newcommand{\wdotaR}{\@ifnextchar[{\basicwdotaR}%
{\hspace{\SOURCE\unitlength}\basicwdotaR[\ARROWLENGTH]}}%

\def\basicwmono[#1]{%
\Z=#1%
\multiply \Z by 100%
\hcase{\WMONO}{\Z}}%

\newcommand{\wmono}{\@ifnextchar[{\basicwmono}%
{\hspace{\SOURCE\unitlength}\basicwmono[\ARROWLENGTH]}}%

\def\basicWmono[#1]#2{%
\Z=#1%
\multiply \Z by 100%
\Hcase{\WMONO}{#2}{\Z}}%

\newcommand{\Wmono}{\@ifnextchar[{\basicWmono}%
{\hspace{\SOURCE\unitlength}\basicWmono[\ARROWLENGTH]}}%

\def\basicwmonO[#1]#2{%
\Z=#1%
\multiply \Z by 100%
\hcasE{\WMONO}{#2}{\Z}}%

\newcommand{\wmonO}{\@ifnextchar[{\basicwmonO}%
{\hspace{\SOURCE\unitlength}\basicwmonO[\ARROWLENGTH]}}%

\def\basicwepi[#1]{%
\Z=#1%
\multiply \Z by 100%
\hcase{\WEPI}{\Z}}%

\newcommand{\wepi}{\@ifnextchar[{\basicwepi}%
{\hspace{\SOURCE\unitlength}\basicwepi[\ARROWLENGTH]}}%

\def\basicWepi[#1]#2{%
\Z=#1%
\multiply \Z by 100%
\Hcase{\WEPI}{#2}{\Z}}%

\newcommand{\Wepi}{\@ifnextchar[{\basicWepi}%
{\hspace{\SOURCE\unitlength}\basicWepi[\ARROWLENGTH]}}%

\def\basicwepI[#1]#2{%
\Z=#1%
\multiply \Z by 100%
\hcasE{\WEPI}{#2}{\Z}}%

\newcommand{\wepI}{\@ifnextchar[{\basicwepI}%
{\hspace{\SOURCE\unitlength}\basicwepI[\ARROWLENGTH]}}%

\def\basicwbimo[#1]{%
\Z=#1%
\multiply \Z by 100%
\hcase{\WBIMO}{\Z}}%

\newcommand{\wbimo}{\@ifnextchar[{\basicwbimo}%
{\hspace{\SOURCE\unitlength}\basicwbimo[\ARROWLENGTH]}}%

\def\basicWbimo[#1]#2{%
\Z=#1%
\multiply \Z by 100%
\Hcase{\WBIMO}{#2}{\Z}}%

\newcommand{\Wbimo}{\@ifnextchar[{\basicWbimo}%
{\hspace{\SOURCE\unitlength}\basicWbimo[\ARROWLENGTH]}}%

\def\basicwbimO[#1]#2{%
\Z=#1%
\multiply \Z by 100%
\hcasE{\WBIMO}{#2}{\Z}}%

\newcommand{\wbimO}{\@ifnextchar[{\basicwbimO}%
{\hspace{\SOURCE\unitlength}\basicwbimO[\ARROWLENGTH]}}%

\def\basicwiso[#1]{%
\Z=#1%
\multiply \Z by 100%
\Hisocase{\WAR}{\cong}{}{\Z}}%

\newcommand{\wiso}{\@ifnextchar[{\basicwiso}%
{\hspace{\SOURCE\unitlength}\basicwiso[\ARROWLENGTH]}}%

\def\basicWiso[#1]#2{%
\Z=#1%
\multiply \Z by 100%
\Hisocase{\WAR}{#2}{\cong}{\Z}}%

\newcommand{\Wiso}{\@ifnextchar[{\basicWiso}%
{\hspace{\SOURCE\unitlength}\basicWiso[\ARROWLENGTH]}}%

\def\basicwisO[#1]#2{%
\Z=#1%
\multiply \Z by 100%
\Hisocase{\WAR}{\cong}{#2}{\Z}}%

\newcommand{\wisO}{\@ifnextchar[{\basicwisO}%
{\hspace{\SOURCE\unitlength}\basicwisO[\ARROWLENGTH]}}%

\def\basicwbiar[#1]{%
\Z=#1%
\multiply \Z by 100%
\hbicase{\WBIAR}{\Z}}%

\newcommand{\wbiar}{\@ifnextchar[{\basicwbiar}%
{\hspace{\SOURCE\unitlength}\basicwbiar[\ARROWLENGTH]}}%

\def\basicWbiar[#1]#2#3{%
\Z=#1%
\multiply \Z by 100%
\Hbicase{\WBIAR}{#2}{#3}{\Z}}%

\newcommand{\Wbiar}{\@ifnextchar[{\basicWbiar}%
{\hspace{\SOURCE\unitlength}\basicWbiar[\ARROWLENGTH]}}%

\def\basicwbidist[#1]{%
\Z=#1%
\multiply \Z by 100%
\hbicase{\WBIDIST}{\Z}}%

\newcommand{\wbidist}{\@ifnextchar[{\basicwbidist}%
{\hspace{\SOURCE\unitlength}\basicwbidist[\ARROWLENGTH]}}%

\def\basicWbidist[#1]#2#3{%
\Z=#1%
\multiply \Z by 100%
\Hbicase{\WBIDIST}{\DUP{#2}}{\DDOWN{#3}}{\Z}}%

\newcommand{\Wbidist}{\@ifnextchar[{\basicWbidist}%
{\hspace{\SOURCE\unitlength}\basicWbidist[\ARROWLENGTH]}}%

\def\basicwadjar[#1]{%
\Z=#1%
\multiply \Z by 100%
\hbicase{\WADJAR}{\Z}}%

\newcommand{\wadjar}{\@ifnextchar[{\basicwadjar}%
{\hspace{\SOURCE\unitlength}\basicwadjar[\ARROWLENGTH]}}%

\def\basicWadjar[#1]#2#3{%
\Z=#1%
\multiply \Z by 100%
\Hbicase{\WADJAR}{#2}{#3}{\Z}}%

\newcommand{\Wadjar}{\@ifnextchar[{\basicWadjar}%
{\hspace{\SOURCE\unitlength}\basicWadjar[\ARROWLENGTH]}}%

\def\basicwadjdist[#1]{%
\Z=#1%
\multiply \Z by 100%
\hbicase{\WADJDIST}{\Z}}%

\newcommand{\wadjdist}{\@ifnextchar[{\basicwadjdist}%
{\hspace{\SOURCE\unitlength}\basicwadjdist[\ARROWLENGTH]}}%

\def\basicWadjdist[#1]#2#3{%
\Z=#1%
\multiply \Z by 100%
\Hbicase{\WADJDIST}{\DUP{#2}}{\DDOWN{#3}}{\Z}}%

\newcommand{\Wadjdist}{\@ifnextchar[{\basicWadjdist}%
{\hspace{\SOURCE\unitlength}\basicWadjdist[\ARROWLENGTH]}}%

\newcommand{\vcase}[2]{\testdiagrammode#1{#2}}%

\newcommand{\Vcase}[3]{\testdiagrammode\makebox[0pt]%
{\makebox[0pt][r]{\raisebox{0pt}[0pt][0pt]{${#2}\hspace{2pt}$}}}#1{#3}}%

\newcommand{\vcasE}[3]{\testdiagrammode\makebox[0pt]%
{#1{#3}\makebox[0pt][l]{\raisebox{0pt}[0pt][0pt]{\hspace{2pt}$#2$}}}}%

\newcommand{\Visocase}[4]{\testdiagrammode\makebox[0pt]%
{\makebox[0pt][r]{\raisebox{0pt}[0pt][0pt]{$#2$\hspace{2pt}}}#1{#4}%
\makebox[0pt][l]{\raisebox{0pt}[0pt][0pt]{\hspace{2pt}$#3$}}}}%
\newcommand{\vbicase}[2]{\testdiagrammode\makebox[0pt]{{#1{#2}}}}%

\newcommand{\Vbicase}[4]{\testdiagrammode\makebox[0pt]%
{\makebox[0pt][r]{\raisebox{0pt}[0pt][0pt]{$#2$\hspace{5.5pt}}}#1{#4}%
\makebox[0pt][l]{\raisebox{0pt}[0pt][0pt]{\hspace{6.5pt}$#3$}}}}%

\newcommand{\SAR}[1]%
{\begin{picture}(0,0)%
\put(0,0){\makebox(0,0)%
{\begin{picture}(0,#1)%
\put(0,#1){\line(0,-1){#1}}%
\put(0,0){\shead}%
\end{picture}}}\end{picture}}%

\newcommand{\SDIST}[1]%
{\begin{picture}(0,0)%
\put(0,0){\makebox(0,0)%
{\begin{picture}(0,#1)%
\put(0,#1){\line(0,-1){#1}}%
\put(0,0){\shead}%
\end{picture}}}%
\truex{400}%
\put(0,0){\circle{\value{x}}}%
\end{picture}}%

\newcommand{\SDOTAR}[1]%
{\truex{100}\truey{300}%
\NUMBEROFDOTS=#1%
\divide\NUMBEROFDOTS by \value{y}%
\advance\NUMBEROFDOTS by 1%
\begin{picture}(0,0)%
\put(0,0){\makebox(0,0)%
{\begin{picture}(0,#1)%
\multiput(0,#1)(0,-\value{y}){\NUMBEROFDOTS}%
{\circle*{\value{x}}}%
\put(0,0){\shead}%
\end{picture}}}\end{picture}}%

\newcommand{\SMONO}[1]%
{\truetail%
\monolength=#1%
\advance\monolength by -\truemonotail%
\begin{picture}(0,0)%
\put(0,0){\makebox(0,0)%
{\begin{picture}(0,#1)%
\put(0,\monolength){\line(0,-1){\monolength}}%
\put(0,\monolength){\shead}%
\put(0,0){\shead}%
\end{picture}}}\end{picture}}%

\newcommand{\SEPI}[1]%
{\truehead%
\epilength=#1%
\advance\epilength by -\trueepihead%
\begin{picture}(0,0)%
\put(0,0){\makebox(0,0)%
{\begin{picture}(0,#1)%
\put(0,#1){\line(0,-1){\epilength}}%
\put(0,\trueepihead){\shead}%
\put(0,0){\shead}%
\end{picture}}}\end{picture}}%

\newcommand{\SBIMO}[1]%
{\truehead\truetail%
\monolength=#1%
\advance\monolength by -\truemonotail%
\epilength=\monolength%
\advance\epilength by -\trueepihead%
\begin{picture}(0,0)%
\put(0,0){\makebox(0,0)%
{\begin{picture}(0,#1)%
\put(0,\monolength){\line(0,-1){\epilength}}%
\put(0,\monolength){\shead}%
\put(0,\trueepihead){\shead}%
\put(0,0){\shead}%
\end{picture}}}\end{picture}}%

\newcommand{\SBIAR}[1]%
{\begin{picture}(0,0)%
\truex{350}%
\put(0,0){\makebox(0,0)%
{\begin{picture}(0,#1)%
\put(-\value{x},#1){\line(0,-1){#1}}%
\put(-\value{x},0){\shead}%
\put(\value{x},#1){\line(0,-1){#1}}%
\put(\value{x},0){\shead}%
\end{picture}}}\end{picture}}%

\newcommand{\SBIDIST}[1]%
{\begin{picture}(0,0)%
\truex{350}%
\put(0,0){\makebox(0,0)%
{\begin{picture}(0,#1)%
\put(-\value{x},#1){\line(0,-1){#1}}%
\put(-\value{x},0){\shead}%
\put(\value{x},#1){\line(0,-1){#1}}%
\put(\value{x},0){\shead}%
\end{picture}}}%
\truey{400}%
\put(-\value{x},0){\circle{\value{y}}}%
\put(\value{x},0){\circle{\value{y}}}%
\end{picture}}%

\newcommand{\SEQL}[1]%
{\begin{picture}(0,0)%
\truex{100}%
\put(0,0){\makebox(0,0)%
{\begin{picture}(0,#1)\put(-\value{x},#1){\line(0,-1){#1}}%
\put(\value{x},#1){\line(0,-1){#1}}%
\end{picture}}}\end{picture}}%

\newcommand{\SADJAR}[1]{\begin{picture}(0,0)%
\truex{350}%
\put(0,0){\makebox(0,0)%
{\begin{picture}(0,#1)%
\put(-\value{x},#1){\line(0,-1){#1}}%
\put(-\value{x},0){\shead}%
\put(\value{x},0){\line(0,1){#1}}%
\put(\value{x},#1){\nhead}%
\end{picture}}}\end{picture}}%

\newcommand{\SADJDIST}[1]{\begin{picture}(0,0)%
\truex{350}%
\put(0,0){\makebox(0,0)%
{\begin{picture}(0,#1)%
\put(-\value{x},#1){\line(0,-1){#1}}%
\put(-\value{x},0){\shead}%
\put(\value{x},0){\line(0,1){#1}}%
\put(\value{x},#1){\nhead}%
\end{picture}}}%
\truey{400}%
\put(-\value{x},0){\circle{\value{y}}}%
\put(\value{x},0){\circle{\value{y}}}%
\end{picture}}%

\def\basicsar[#1]{\vcase{\SAR}{#100}}%

\newcommand{\sar}{\@ifnextchar[{\basicsar}{\basicsar[50]}}%

\def\basicSar[#1]#2{\Vcase{\SAR}{#2}{#100}}%

\newcommand{\Sar}{\@ifnextchar[{\basicSar}{\basicSar[50]}}%

\def\basicsaR[#1]#2{\vcasE{\SAR}{#2}{#100}}%

\newcommand{\saR}{\@ifnextchar[{\basicsaR}{\basicsaR[50]}}%

\def\basicsdist[#1]{\vcase{\SDIST}{#100}}%

\newcommand{\sdist}{\@ifnextchar[{\basicsdist}{\basicsdist[50]}}%

\def\basicSdist[#1]#2{\Vcase{\SDIST}{#2\hspace*{2pt}}{#100}}%

\newcommand{\Sdist}{\@ifnextchar[{\basicSdist}{\basicSdist[50]}}%

\def\basicsdisT[#1]#2{\vcasE{\SDIST}{\hspace*{2pt}#2}{#100}}%

\newcommand{\sdisT}{\@ifnextchar[{\basicsdisT}{\basicsdisT[50]}}%

\def\basicsdotar[#1]{\vcase{\SDOTAR}{#100}}%

\newcommand{\sdotar}{\@ifnextchar[{\basicsdotar}{\basicsdotar[50]}}%

\def\basicSdotar[#1]#2{\Vcase{\SDOTAR}{#2}{#100}}%

\newcommand{\Sdotar}{\@ifnextchar[{\basicSdotar}{\basicSdotar[50]}}%

\def\basicsdotaR[#1]#2{\vcasE{\SDOTAR}{#2}{#100}}%

\newcommand{\sdotaR}{\@ifnextchar[{\basicsdotaR}{\basicsdotaR[50]}}%

\def\basicsmono[#1]{\vcase{\SMONO}{#100}}%

\newcommand{\smono}{\@ifnextchar[{\basicsmono}{\basicsmono[50]}}%

\def\basicSmono[#1]#2{\Vcase{\SMONO}{#2}{#100}}%

\newcommand{\Smono}{\@ifnextchar[{\basicSmono}{\basicSmono[50]}}%

\def\basicsmonO[#1]#2{\vcasE{\SMONO}{#2}{#100}}%

\newcommand{\smonO}{\@ifnextchar[{\basicsmonO}{\basicsmonO[50]}}%

\def\basicsepi[#1]{\vcase{\SEPI}{#100}}%

\newcommand{\sepi}{\@ifnextchar[{\basicsepi}{\basicsepi[50]}}%

\def\basicSepi[#1]#2{\Vcase{\SEPI}{#2}{#100}}%

\newcommand{\Sepi}{\@ifnextchar[{\basicSepi}{\basicSepi[50]}}%

\def\basicsepI[#1]#2{\vcasE{\SEPI}{#2}{#100}}%

\newcommand{\sepI}{\@ifnextchar[{\basicsepI}{\basicsepI[50]}}%

\def\basicsbimo[#1]{\vcase{\SBIMO}{#100}}%

\newcommand{\sbimo}{\@ifnextchar[{\basicsbimo}{\basicsbimo[50]}}%

\def\basicSbimo[#1]#2{\Vcase{\SBIMO}{#2}{#100}}%

\newcommand{\Sbimo}{\@ifnextchar[{\basicSbimo}{\basicSbimo[50]}}%

\def\basicsbimO[#1]#2{\vcasE{\SBIMO}{#2}{#100}}%

\newcommand{\sbimO}{\@ifnextchar[{\basicsbimO}{\basicsbimO[50]}}%

\def\basicsiso[#1]{\Visocase{\SAR}{\cong}{}{#100}}%

\newcommand{\siso}{\@ifnextchar[{\basicsiso}{\basicsiso[50]}}%

\def\basicSiso[#1]#2{\Visocase{\SAR}{#2}{\cong}{#100}}%

\newcommand{\Siso}{\@ifnextchar[{\basicSiso}{\basicSiso[50]}}%

\def\basicsisO[#1]#2{\Visocase{\SAR}{\cong}{#2}{#100}}%

\newcommand{\sisO}{\@ifnextchar[{\basicsisO}{\basicsisO[50]}}%

\def\basicseql[#1]{\vcase{\SEQL}{#100}}%

\newcommand{\seql}{\@ifnextchar[{\basicseql}{\basicseql[50]}}%

\def\basicSeql[#1]#2{\Vcase{\SEQL}{#2\hspace*{2pt}}{#100}}%

\newcommand{\Seql}{\@ifnextchar[{\basicSeql}{\basicSeql[50]}}%

\def\basicseqL[#1]#2{\vcasE{\SEQL}{\hspace*{2pt}#2}{#100}}%

\newcommand{\seqL}{\@ifnextchar[{\basicseqL}{\basicseqL[50]}}%

\def\basicsbiar[#1]{\vbicase{\SBIAR}{#100}}%

\newcommand{\sbiar}{\@ifnextchar[{\basicsbiar}{\basicsbiar[50]}}%

\def\basicSbiar[#1]#2#3{\Vbicase{\SBIAR}{#2}{#3}{#100}}%

\newcommand{\Sbiar}{\@ifnextchar[{\basicSbiar}{\basicSbiar[50]}}%

\def\basicsbidist[#1]{\vbicase{\SBIDIST}{#100}}%

\newcommand{\sbidist}{\@ifnextchar[{\basicsbidist}{\basicsbidist[50]}}%

\def\basicSbidist[#1]#2#3%
{\Vbicase{\SBIDIST}{#2\hspace*{2pt}}{\hspace*{2pt}#3}{#100}}%

\newcommand{\Sbidist}{\@ifnextchar[{\basicSbidist}{\basicSbidist[50]}}%

\def\basicsadjar[#1]{\vbicase{\SADJAR}{#100}}%

\newcommand{\sadjar}{\@ifnextchar[{\basicsadjar}{\basicsadjar[50]}}%

\def\basicSadjar[#1]#2#3{\Vbicase{\SADJAR}{#2}{#3}{#100}}%

\newcommand{\Sadjar}{\@ifnextchar[{\basicSadjar}{\basicSadjar[50]}}%

\def\basicsadjdist[#1]{\vbicase{\SADJDIST}{#100}}%

\newcommand{\sadjdist}{\@ifnextchar[{\basicsadjdist}{\basicsadjdist[50]}}%

\def\basicSadjdist[#1]#2#3%
{\Vbicase{\SADJDIST}{#2\hspace*{2pt}}{\hspace*{2pt}#3}{#100}}%

\newcommand{\Sadjdist}{\@ifnextchar[{\basicSadjdist}{\basicSadjdist[50]}}%

\newcommand{\NAR}[1]%
{\begin{picture}(0,0)%
\put(0,0){\makebox(0,0)%
{\begin{picture}(0,#1)%
\put(0,0){\line(0,1){#1}}%
\put(0,#1){\nhead}%
\end{picture}}}\end{picture}}%

\newcommand{\NDIST}[1]%
{\begin{picture}(0,0)%
\put(0,0){\makebox(0,0)%
{\begin{picture}(0,#1)%
\put(0,0){\line(0,1){#1}}%
\put(0,#1){\nhead}%
\end{picture}}}
\truex{400}%
\put(0,0){\circle{\value{x}}}%
\end{picture}}%

\newcommand{\NDOTAR}[1]%
{\truex{100}\truey{300}%
\NUMBEROFDOTS=#1%
\divide\NUMBEROFDOTS by \value{y}%
\advance\NUMBEROFDOTS by 1%
\begin{picture}(0,0)%
\put(0,0){\makebox(0,0)%
{\begin{picture}(0,#1)%
\multiput(0,0)(0,\value{y}){\NUMBEROFDOTS}%
{\circle*{\value{x}}}%
\put(0,#1){\nhead}%
\end{picture}}}\end{picture}}%

\newcommand{\NMONO}[1]%
{\truetail%
\monolength=#1%
\advance\monolength by -\truemonotail%
\begin{picture}(0,0)%
\put(0,0){\makebox(0,0)%
{\begin{picture}(0,#1)%
\put(0,\truemonotail){\line(0,1){\monolength}}%
\put(0,#1){\nhead}%
\put(0,\truemonotail){\nhead}%
\end{picture}}}\end{picture}}%

\newcommand{\NEPI}[1]%
{\truehead%
\epilength=#1%
\advance\epilength by -\trueepihead%
\begin{picture}(0,0)%
\put(0,0){\makebox(0,0)%
{\begin{picture}(0,#1)%
\put(0,0){\line(0,1){\epilength}}%
\put(0,#1){\nhead}%
\put(0,\epilength){\nhead}%
\end{picture}}}\end{picture}}%

\newcommand{\NBIMO}[1]%
{\truehead\truetail%
\epilength=#1%
\advance\epilength by -\trueepihead%
\monolength=\epilength%
\advance\monolength by -\truemonotail%
\begin{picture}(0,0)%
\put(0,0){\makebox(0,0)%
{\begin{picture}(0,#1)%
\put(0,\truemonotail){\line(0,1){\monolength}}%
\put(0,#1){\nhead}%
\put(0,\truemonotail){\nhead}%
\put(0,\epilength){\nhead}%
\end{picture}}}\end{picture}}%

\newcommand{\NBIAR}[1]%
{\begin{picture}(0,0)%
\truex{350}%
\put(0,0){\makebox(0,0)%
{\begin{picture}(0,#1)%
\put(-\value{x},0){\line(0,1){#1}}%
\put(-\value{x},#1){\nhead}%
\put(\value{x},0){\line(0,1){#1}}%
\put(\value{x},#1){\nhead}%
\end{picture}}}\end{picture}}%

\newcommand{\NBIDIST}[1]%
{\begin{picture}(0,0)%
\truex{350}%
\put(0,0){\makebox(0,0)%
{\begin{picture}(0,#1)%
\put(-\value{x},0){\line(0,1){#1}}%
\put(-\value{x},#1){\nhead}%
\put(\value{x},0){\line(0,1){#1}}%
\put(\value{x},#1){\nhead}%
\end{picture}}}
\truey{400}%
\put(-\value{x},0){\circle{\value{y}}}%
\put(\value{x},0){\circle{\value{y}}}%
\end{picture}}%

\newcommand{\NADJAR}[1]{\begin{picture}(0,0)%
\truex{350}%
\put(0,0){\makebox(0,0)%
{\begin{picture}(0,#1)%
\put(\value{x},#1){\line(0,-1){#1}}%
\put(\value{x},0){\shead}%
\put(-\value{x},0){\line(0,1){#1}}%
\put(-\value{x},#1){\nhead}%
\end{picture}}}\end{picture}}%

\newcommand{\NADJDIST}[1]{\begin{picture}(0,0)%
\truex{350}%
\put(0,0){\makebox(0,0)%
{\begin{picture}(0,#1)%
\put(\value{x},#1){\line(0,-1){#1}}%
\put(\value{x},0){\shead}%
\put(-\value{x},0){\line(0,1){#1}}%
\put(-\value{x},#1){\nhead}%
\end{picture}}}
\truey{400}%
\put(-\value{x},0){\circle{\value{y}}}%
\put(\value{x},0){\circle{\value{y}}}%
\end{picture}}%

\def\basicnar[#1]{\vcase{\NAR}{#100}}%

\newcommand{\nar}{\@ifnextchar[{\basicnar}{\basicnar[50]}}%

\def\basicNar[#1]#2{\Vcase{\NAR}{#2}{#100}}%

\newcommand{\Nar}{\@ifnextchar[{\basicNar}{\basicNar[50]}}%

\def\basicnaR[#1]#2{\vcasE{\NAR}{#2}{#100}}%

\newcommand{\naR}{\@ifnextchar[{\basicnaR}{\basicnaR[50]}}%

\def\basicndist[#1]{\vcase{\NDIST}{#100}}%

\newcommand{\ndist}{\@ifnextchar[{\basicndist}{\basicndist[50]}}%

\def\basicNdist[#1]#2{\Vcase{\NDIST}{#2\hspace*{2pt}}{#100}}%

\newcommand{\Ndist}{\@ifnextchar[{\basicNdist}{\basicNdist[50]}}%

\def\basicndisT[#1]#2{\vcasE{\NDIST}{\hspace*{2pt}#2}{#100}}%

\newcommand{\ndisT}{\@ifnextchar[{\basicndisT}{\basicndisT[50]}}%

\def\basicndotar[#1]{\vcase{\NDOTAR}{#100}}%

\newcommand{\ndotar}{\@ifnextchar[{\basicndotar}{\basicndotar[50]}}%

\def\basicNdotar[#1]#2{\Vcase{\NDOTAR}{#2}{#100}}%

\newcommand{\Ndotar}{\@ifnextchar[{\basicNdotar}{\basicNdotar[50]}}%

\def\basicndotaR[#1]#2{\vcasE{\NDOTAR}{#2}{#100}}%

\newcommand{\ndotaR}{\@ifnextchar[{\basicndotaR}{\basicndotaR[50]}}%

\def\basicnmono[#1]{\vcase{\NMONO}{#100}}%

\newcommand{\nmono}{\@ifnextchar[{\basicnmono}%
{\basicnmono[50]}}%

\def\basicNmono[#1]#2{\Vcase{\NMONO}{#2}{#100}}%

\newcommand{\Nmono}{\@ifnextchar[{\basicNmono}{\basicNmono[50]}}%

\def\basicnmonO[#1]#2{\vcasE{\NMONO}{#2}{#100}}%

\newcommand{\nmonO}{\@ifnextchar[{\basicnmonO}{\basicnmonO[50]}}%

\def\basicnepi[#1]{\vcase{\NEPI}{#100}}%

\newcommand{\nepi}{\@ifnextchar[{\basicnepi}{\basicnepi[50]}}%

\def\basicNepi[#1]#2{\Vcase{\NEPI}{#2}{#100}}%

\newcommand{\Nepi}{\@ifnextchar[{\basicNepi}{\basicNepi[50]}}%

\def\basicnepI[#1]#2{\vcasE{\NEPI}{#2}{#100}}%

\newcommand{\nepI}{\@ifnextchar[{\basicnepI}{\basicnepI[50]}}%

\def\basicnbimo[#1]{\vcase{\NBIMO}{#100}}%

\newcommand{\nbimo}{\@ifnextchar[{\basicnbimo}{\basicnbimo[50]}}%

\def\basicNbimo[#1]#2{\Vcase{\NBIMO}{#2}{#100}}%

\newcommand{\Nbimo}{\@ifnextchar[{\basicNbimo}{\basicNbimo[50]}}%

\def\basicnbimO[#1]#2{\vcasE{\NBIMO}{#2}{#100}}%

\newcommand{\nbimO}{\@ifnextchar[{\basicnbimO}{\basicnbimO[50]}}%

\def\basicniso[#1]{\Visocase{\NAR}{\cong}{}{#100}}%

\newcommand{\niso}{\@ifnextchar[{\basicniso}{\basicniso[50]}}%

\def\basicNiso[#1]#2{\Visocase{\NAR}{#2}{\cong}{#100}}%

\newcommand{\Niso}{\@ifnextchar[{\basicNiso}{\basicNiso[50]}}%

\def\basicnisO[#1]#2{\Visocase{\NAR}{\cong}{#2}{#100}}%

\newcommand{\nisO}{\@ifnextchar[{\basicnisO}{\basicnisO[50]}}%

\def\basicnbiar[#1]{\vbicase{\NBIAR}{#100}}%

\newcommand{\nbiar}{\@ifnextchar[{\basicnbiar}{\basicnbiar[50]}}%

\def\basicNbiar[#1]#2#3{\Vbicase{\NBIAR}{#2}{#3}{#100}}%

\newcommand{\Nbiar}{\@ifnextchar[{\basicNbiar}{\basicNbiar[50]}}%

\def\basicnbidist[#1]{\vbicase{\NBIDIST}{#100}}%

\newcommand{\nbidist}{\@ifnextchar[{\basicnbidist}{\basicnbidist[50]}}%

\def\basicNbidist[#1]#2#3%
{\Vbicase{\NBIDIST}{#2\hspace*{2pt}}{\hspace*{2pt}#3}{#100}}%

\newcommand{\Nbidist}{\@ifnextchar[{\basicNbidist}{\basicNbidist[50]}}%

\def\basicnadjar[#1]{\vbicase{\NADJAR}{#100}}%

\newcommand{\nadjar}{\@ifnextchar[{\basicnadjar}{\basicnadjar[50]}}%

\def\basicNadjar[#1]#2#3{\Vbicase{\NADJAR}{#2}{#3}{#100}}%

\newcommand{\Nadjar}{\@ifnextchar[{\basicNadjar}{\basicNadjar[50]}}%

\def\basicnadjdist[#1]{\vbicase{\NADJDIST}{#100}}%

\newcommand{\nadjdist}{\@ifnextchar[{\basicnadjdist}{\basicnadjdist[50]}}%

\def\basicNadjdist[#1]#2#3%
{\Vbicase{\NADJDIST}{#2\hspace*{2pt}}{\hspace*{2pt}#3}{#100}}%

\newcommand{\Nadjdist}{\@ifnextchar[{\basicNadjdist}{\basicNadjdist[50]}}%

\newcommand{\fdcase}[4]{\testdiagrammode\begin{picture}(0,0)%
\put(0,0){#1{#4}}%
\truex{200}\truey{600}\truez{600}%
\put(-\value{x},-\value{x}){\makebox(0,\value{z})[r]{${#2}$}}%
\put(\value{x},-\value{y}){\makebox(0,\value{z})[l]{${#3}$}}%
\end{picture}}%

\newcommand{\fdbicase}[4]{\testdiagrammode\begin{picture}(0,0)%
\put(0,0){#1{#4}}%
\truex{900}\truey{150}%
\put(-\value{x},\value{y}){${#2}$}%
\truex{300}\truey{1050}%
\put(\value{x},-\value{y}){${#3}$}%
\end{picture}}%

\newcommand{\NEAR}[1]{%
\Y=#1%
\divide\Y by 2%
\begin{picture}(0,0)%
\put(-\Y,-\Y){\line(1,1){#1}}%
\put(\Y,\Y){\nehead}%
\end{picture}}%

\newcommand{\NEDIST}[1]{%
\Y=#1%
\divide\Y by 2%
\begin{picture}(0,0)%
\put(-\Y,-\Y){\line(1,1){#1}}%
\put(\Y,\Y){\nehead}%
\truex{400}%
\put(0,0){\circle{\value{x}}}%
\end{picture}}%

\newcommand{\NEDOTAR}[1]%
{\truex{100}\truey{212}%
\Y=#1%
\divide\Y by 2%
\NUMBEROFDOTS=#1%
\divide\NUMBEROFDOTS by \value{y}%
\advance\NUMBEROFDOTS by 1%
\begin{picture}(0,0)%
\multiput(-\Y,-\Y)(\value{y},\value{y}){\NUMBEROFDOTS}%
{\circle*{\value{x}}}%
\put(\Y,\Y){\nehead}%
\end{picture}}%

\newcommand{\NEMONO}[1]{%
\Y=#1%
\divide \Y by 2%
\Truetail%
\bimolength=#1%
\advance\bimolength by -\Truemonotail%
\monolength=\bimolength%
\advance\monolength by -\Y%
\begin{picture}(0,0)%
\put(-\monolength,-\monolength){\line(1,1){\bimolength}}%
\put(-\monolength,-\monolength){\nehead}%
\put(\Y,\Y){\nehead}%
\end{picture}}%

\newcommand{\NEEPI}[1]{%
\Y=#1%
\divide\Y by 2%
\Truehead%
\bimolength=#1%
\advance\bimolength by -\Trueepihead%
\epilength=\bimolength%
\advance\epilength by -\Y%
\begin{picture}(0,0)%
\put(-\Y,-\Y){\line(1,1){\bimolength}}%
\put(\epilength,\epilength){\nehead}%
\put(\Y,\Y){\nehead}%
\end{picture}}%

\newcommand{\NEBIMO}[1]{%
\Y=#1%
\divide\Y by 2%
\Truetail\Truehead%
\bimolength=#1%
\advance\bimolength by -\Truemonotail%
\monolength=\bimolength%
\advance\monolength by -\Y%
\advance\bimolength by -\Trueepihead%
\epilength=\bimolength%
\advance\epilength by -\monolength%
\begin{picture}(0,0)%
\put(-\monolength,-\monolength){\line(1,1){\bimolength}}%
\put(-\monolength,-\monolength){\nehead}%
\put(\epilength,\epilength){\nehead}%
\put(\Y,\Y){\nehead}%
\end{picture}}%

\newcommand{\NEBIAR}[1]{%
\Y=#1%
\divide\Y by 2%
\begin{picture}(0,0)%
\put(-\Y,-\Y){\begin{picture}(0,0)%
\truex{247}%
\put(-\value{x},\value{x}){\line(1,1){#1}}%
\put(\value{x},-\value{x}){\line(1,1){#1}}%
\monolength=#1%
\advance\monolength by -\value{x}%
\epilength=#1%
\advance\epilength by \value{x}%
\put(\monolength,\epilength){\nehead}%
\put(\epilength,\monolength){\nehead}%
\end{picture}}\end{picture}}%

\newcommand{\NEBIDIST}[1]{%
\Y=#1%
\divide\Y by 2%
\truey{400}%
\begin{picture}(0,0)%
\put(-\Y,-\Y){\begin{picture}(0,0)%
\truex{247}%
\monolength=#1%
\advance\monolength by -\value{x}%
\epilength=#1%
\advance\epilength by \value{x}%
\put(\value{x},-\value{x}){\line(1,1){#1}}%
\put(\epilength,\monolength){\nehead}%
\end{picture}}%
\put(-\Y,-\Y){\begin{picture}(0,0)%
\truex{247}%
\monolength=#1%
\advance\monolength by \value{x}%
\epilength=#1%
\advance\epilength by -\value{x}%
\put(-\value{x},\value{x}){\line(1,1){#1}}%
\put(\epilength,\monolength){\nehead}%
\end{picture}}%
\put(-\value{x},\value{x}){\circle{\value{y}}}%
\put(\value{x},-\value{x}){\circle{\value{y}}}%
\end{picture}}%

\newcommand{\NEEQL}[1]{%
\Y=#1%
\divide\Y by 2%
\begin{picture}(0,0)%
\put(-\Y,-\Y){\begin{picture}(0,0)%
\truex{70}%
\put(-\value{x},\value{x}){\line(1,1){#1}}%
\put(\value{x},-\value{x}){\line(1,1){#1}}%
\end{picture}}\end{picture}}%

\newcommand{\NEADJAR}[1]{%
\Y=#1%
\divide\Y by 2%
\begin{picture}(0,0)%
\put(-\Y,-\Y){\begin{picture}(0,0)%
\truex{247}%
\monolength=#1%
\advance\monolength by -\value{x}%
\epilength=#1%
\advance\epilength by \value{x}%
\put(\value{x},-\value{x}){\line(1,1){#1}}%
\put(\epilength,\monolength){\nehead}%
\end{picture}}%
\put(\Y,\Y){\begin{picture}(0,0)%
\truex{247}%
\monolength=#1%
\advance\monolength by -\value{x}%
\epilength=#1%
\advance\epilength by \value{x}%
\put(-\value{x},\value{x}){\line(-1,-1){#1}}%
\put(-\epilength,-\monolength){\swhead}%
\end{picture}}\end{picture}}%

\newcommand{\NEADJDIST}[1]{%
\Y=#1%
\divide\Y by 2%
\truey{400}%
\begin{picture}(0,0)%
\put(-\Y,-\Y){\begin{picture}(0,0)%
\truex{247}%
\monolength=#1%
\advance\monolength by -\value{x}%
\epilength=#1%
\advance\epilength by \value{x}%
\put(\value{x},-\value{x}){\line(1,1){#1}}%
\put(\epilength,\monolength){\nehead}%
\end{picture}}%
\put(\Y,\Y){\begin{picture}(0,0)%
\truex{247}%
\monolength=#1%
\advance\monolength by -\value{x}%
\epilength=#1%
\advance\epilength by \value{x}%
\put(-\value{x},\value{x}){\line(-1,-1){#1}}%
\put(-\epilength,-\monolength){\swhead}%
\end{picture}}%
\put(-\value{x},\value{x}){\circle{\value{y}}}%
\put(\value{x},-\value{x}){\circle{\value{y}}}%
\end{picture}}%

\def\basicnear[#1]{\fdcase{\NEAR}{}{}{#100}}%

\newcommand{\near}{\@ifnextchar[{\basicnear}{\basicnear[59]}}%

\def\basicNear[#1]#2{\fdcase{\NEAR}{#2}{}{#100}}%

\newcommand{\Near}{\@ifnextchar[{\basicNear}{\basicNear[59]}}%

\def\basicneaR[#1]#2{\fdcase{\NEAR}{}{#2}{#100}}%

\newcommand{\neaR}{\@ifnextchar[{\basicneaR}{\basicneaR[59]}}%

\def\basicnedist[#1]{\fdcase{\NEDIST}{}{}{#100}}%

\newcommand{\nedist}{\@ifnextchar[{\basicnedist}{\basicnedist[59]}}%

\def\basicNedist[#1]#2{\fdcase{\NEDIST}{#2}{}{#100}}%

\newcommand{\Nedist}{\@ifnextchar[{\basicNedist}{\basicNedist[59]}}%

\def\basicnedisT[#1]#2{\fdcase{\NEDIST}{}{#2}{#100}}%

\newcommand{\nedisT}{\@ifnextchar[{\basicnedisT}{\basicnedisT[59]}}%

\def\basicnedotar[#1]{\fdcase{\NEDOTAR}{}{}{#100}}%

\newcommand{\nedotar}{\@ifnextchar[{\basicnedotar}{\basicnedotar[59]}}%

\def\basicNedotar[#1]#2{\fdcase{\NEDOTAR}{#2}{}{#100}}%

\newcommand{\Nedotar}{\@ifnextchar[{\basicNedotar}{\basicNedotar[59]}}%

\def\basicnedotaR[#1]#2{\fdcase{\NEDOTAR}{}{#2}{#100}}%

\newcommand{\nedotaR}{\@ifnextchar[{\basicnedotaR}{\basicnedotaR[59]}}%

\def\basicnemono[#1]{\fdcase{\NEMONO}{}{}{#100}}%

\newcommand{\nemono}{\@ifnextchar[{\basicnemono}{\basicnemono[59]}}%

\def\basicNemono[#1]#2{\fdcase{\NEMONO}{#2}{}{#100}}%

\newcommand{\Nemono}{\@ifnextchar[{\basicNemono}{\basicNemono[59]}}%

\def\basicnemonO[#1]#2{\fdcase{\NEMONO}{}{#2}{#100}}%

\newcommand{\nemonO}{\@ifnextchar[{\basicnemonO}{\basicnemonO[59]}}%

\def\basicneepi[#1]{\fdcase{\NEEPI}{}{}{#100}}%

\newcommand{\neepi}{\@ifnextchar[{\basicneepi}{\basicneepi[59]}}%

\def\basicNeepi[#1]#2{\fdcase{\NEEPI}{#2}{}{#100}}%

\newcommand{\Neepi}{\@ifnextchar[{\basicNeepi}{\basicNeepi[59]}}%

\def\basicneepI[#1]#2{\fdcase{\NEEPI}{}{#2}{#100}}%

\newcommand{\neepI}{\@ifnextchar[{\basicneepI}{\basicneepI[59]}}%

\def\basicnebimo[#1]{\fdcase{\NEBIMO}{}{}{#100}}%

\newcommand{\nebimo}{\@ifnextchar[{\basicnebimo}{\basicnebimo[59]}}%

\def\basicNebimo[#1]#2{\fdcase{\NEBIMO}{#2}{}{#100}}%

\newcommand{\Nebimo}{\@ifnextchar[{\basicNebimo}{\basicNebimo[59]}}%

\def\basicnebimO[#1]#2{\fdcase{\NEBIMO}{}{#2}{#100}}%

\newcommand{\nebimO}{\@ifnextchar[{\basicnebimO}{\basicnebimO[59]}}%

\def\basicneiso[#1]{\fdcase{\NEAR}{\hspace{-2pt}\cong}{}{#100}}%

\newcommand{\neiso}{\@ifnextchar[{\basicneiso}{\basicneiso[59]}}%

\def\basicNeiso[#1]#2{\fdcase{\NEAR}{#2}{\cong}{#100}}%

\newcommand{\Neiso}{\@ifnextchar[{\basicNeiso}{\basicNeiso[59]}}%

\def\basicneisO[#1]#2{\fdcase{\NEAR}{\hspace{-2pt}\cong}{#2}{#100}}%

\newcommand{\neisO}{\@ifnextchar[{\basicneisO}{\basicneisO[59]}}%

\def\basicneeql[#1]{\fdcase{\NEEQL}{}{}{#100}}%

\newcommand{\neeql}{\@ifnextchar[{\basicneeql}{\basicneeql[59]}}%

\def\basicNeeql[#1]#2{\fdcase{\NEEQL}{#2}{}{#100}}%

\newcommand{\Neeql}{\@ifnextchar[{\basicNeeql}{\basicNeeql[59]}}%

\def\basicneeqL[#1]#2{\fdcase{\NEEQL}{}{#2}{#100}}%

\newcommand{\neeqL}{\@ifnextchar[{\basicneeqL}{\basicneeqL[59]}}%

\def\basicnebiar[#1]{\fdbicase{\NEBIAR}{}{}{#100}}%

\newcommand{\nebiar}{\@ifnextchar[{\basicnebiar}{\basicnebiar[59]}}%

\def\basicNebiar[#1]#2#3{\fdbicase{\NEBIAR}{#2}{#3}{#100}}%

\newcommand{\Nebiar}{\@ifnextchar[{\basicNebiar}{\basicNebiar[59]}}%

\def\basicneadjar[#1]{\fdbicase{\NEADJAR}{}{}{#100}}%

\newcommand{\neadjar}{\@ifnextchar[{\basicneadjar}{\basicneadjar[59]}}%

\def\basicNeadjar[#1]#2#3{\fdbicase{\NEADJAR}{#2}{#3}{#100}}%

\newcommand{\Neadjar}{\@ifnextchar[{\basicNeadjar}{\basicNeadjar[59]}}%

\def\basicnebidist[#1]{\fdbicase{\NEBIDIST}{}{}{#100}}%

\newcommand{\nebidist}{\@ifnextchar[{\basicnebidist}{\basicnebidist[59]}}%

\def\basicNebidist[#1]#2#3{\fdbicase{\NEBIDIST}{#2}{#3}{#100}}%

\newcommand{\Nebidist}{\@ifnextchar[{\basicNebidist}{\basicNebidist[59]}}%

\def\basicneadjdist[#1]{\fdbicase{\NEADJDIST}{}{}{#100}}%

\newcommand{\neadjdist}{\@ifnextchar[{\basicneadjdist}{\basicneadjdist[59]}}%

\def\basicNeadjdist[#1]#2#3{\fdbicase{\NEADJDIST}{#2}{#3}{#100}}%

\newcommand{\Neadjdist}{\@ifnextchar[{\basicNeadjdist}{\basicNeadjdist[59]}}%

\newcommand{\SWAR}[1]{%
\Y=#1%
\divide\Y by 2%
\begin{picture}(0,0)%
\put(\Y,\Y){\line(-1,-1){#1}}%
\put(-\Y,-\Y){\swhead}%
\end{picture}}%

\newcommand{\SWDIST}[1]{%
\Y=#1%
\divide\Y by 2%
\begin{picture}(0,0)%
\put(\Y,\Y){\line(-1,-1){#1}}%
\put(-\Y,-\Y){\swhead}%
\truex{400}%
\put(0,0){\circle{\value{x}}}%
\end{picture}}%

\newcommand{\SWDOTAR}[1]%
{\truex{100}\truey{212}%
\Y=#1%
\divide\Y by 2%
\NUMBEROFDOTS=#1%
\divide\NUMBEROFDOTS by \value{y}%
\advance\NUMBEROFDOTS by 1%
\begin{picture}(0,0)%
\multiput(\Y,\Y)(-\value{y},-\value{y}){\NUMBEROFDOTS}%
{\circle*{\value{x}}}%
\put(-\Y,-\Y){\swhead}%
\end{picture}}%

\newcommand{\SWMONO}[1]{%
\Y=#1%
\divide \Y by 2%
\Truetail%
\bimolength=#1%
\advance\bimolength by -\Truemonotail%
\monolength=\bimolength%
\advance\monolength by -\Y%
\begin{picture}(0,0)%
\put(\monolength,\monolength){\line(-1,-1){\bimolength}}%
\put(\monolength,\monolength){\swhead}%
\put(-\Y,-\Y){\swhead}%
\end{picture}}%

\newcommand{\SWEPI}[1]{%
\Y=#1%
\divide\Y by 2%
\Truehead%
\bimolength=#1%
\advance\bimolength by -\Trueepihead%
\epilength=\bimolength%
\advance\epilength by -\Y%
\begin{picture}(0,0)%
\put(\Y,\Y){\line(-1,-1){\bimolength}}%
\put(-\epilength,-\epilength){\swhead}%
\put(-\Y,-\Y){\swhead}%
\end{picture}}%

\newcommand{\SWBIMO}[1]{%
\Y=#1%
\divide\Y by 2%
\Truetail\Truehead%
\bimolength=#1%
\advance\bimolength by -\Truemonotail%
\monolength=\bimolength%
\advance\monolength by -\Y%
\advance\bimolength by -\Trueepihead%
\epilength=\bimolength%
\advance\epilength by -\monolength%
\begin{picture}(0,0)%
\put(\monolength,\monolength){\line(-1,-1){\bimolength}}%
\put(\monolength,\monolength){\swhead}%
\put(-\epilength,-\epilength){\swhead}%
\put(-\Y,-\Y){\swhead}%
\end{picture}}%

\newcommand{\SWBIAR}[1]{%
\Y=#1%
\divide\Y by 2%
\begin{picture}(0,0)%
\put(\Y,\Y){\begin{picture}(0,0)%
\truex{247}%
\put(\value{x},-\value{x}){\line(-1,-1){#1}}%
\put(-\value{x},\value{x}){\line(-1,-1){#1}}%
\monolength=#1%
\advance\monolength by -\value{x}%
\epilength=#1%
\advance\epilength by \value{x}%
\put(-\monolength,-\epilength){\swhead}%
\put(-\epilength,-\monolength){\swhead}%
\end{picture}}\end{picture}}%

\newcommand{\SWBIDIST}[1]{%
\Y=#1%
\divide\Y by 2%
\truey{400}%
\begin{picture}(0,0)%
\put(\Y,\Y){\begin{picture}(0,0)%
\truex{247}%
\monolength=#1%
\advance\monolength by -\value{x}%
\epilength=#1%
\advance\epilength by \value{x}%
\put(-\value{x},\value{x}){\line(-1,-1){#1}}%
\put(-\epilength,-\monolength){\swhead}%
\end{picture}}%
\put(\Y,\Y){\begin{picture}(0,0)%
\truex{247}%
\monolength=#1%
\advance\monolength by \value{x}%
\epilength=#1%
\advance\epilength by -\value{x}%
\put(\value{x},-\value{x}){\line(-1,-1){#1}}%
\put(-\epilength,-\monolength){\swhead}%
\end{picture}}%
\put(\value{x},-\value{x}){\circle{\value{y}}}%
\put(-\value{x},\value{x}){\circle{\value{y}}}%
\end{picture}}%

\newcommand{\SWADJAR}[1]{%
\Y=#1%
\divide\Y by 2%
\begin{picture}(0,0)%
\put(\Y,\Y){\begin{picture}(0,0)%
\truex{247}%
\monolength=#1%
\advance\monolength by -\value{x}%
\epilength=#1%
\advance\epilength by \value{x}%
\put(\value{x},-\value{x}){\line(-1,-1){#1}}%
\put(-\monolength,-\epilength){\swhead}%
\end{picture}}%
\put(-\Y,-\Y){\begin{picture}(0,0)%
\truex{247}%
\monolength=#1%
\advance\monolength by -\value{x}%
\epilength=#1%
\advance\epilength by \value{x}%
\put(-\value{x},\value{x}){\line(1,1){#1}}%
\put(\monolength,\epilength){\nehead}%
\end{picture}}\end{picture}}%

\newcommand{\SWADJDIST}[1]{%
\Y=#1%
\divide\Y by 2%
\truey{400}%
\begin{picture}(0,0)%
\put(\Y,\Y){\begin{picture}(0,0)%
\truex{247}%
\monolength=#1%
\advance\monolength by -\value{x}%
\epilength=#1%
\advance\epilength by \value{x}%
\put(\value{x},-\value{x}){\line(-1,-1){#1}}%
\put(-\monolength,-\epilength){\swhead}%
\end{picture}}%
\put(-\Y,-\Y){\begin{picture}(0,0)%
\truex{247}%
\monolength=#1%
\advance\monolength by -\value{x}%
\epilength=#1%
\advance\epilength by \value{x}%
\put(-\value{x},\value{x}){\line(1,1){#1}}%
\put(\monolength,\epilength){\nehead}%
\end{picture}}%
\put(-\value{x},\value{x}){\circle{\value{y}}}%
\put(\value{x},-\value{x}){\circle{\value{y}}}%
\end{picture}}%

\def\basicswar[#1]{\fdcase{\SWAR}{}{}{#100}}%

\newcommand{\swar}{\@ifnextchar[{\basicswar}{\basicswar[59]}}%

\def\basicSwar[#1]#2{\fdcase{\SWAR}{#2}{}{#100}}%

\newcommand{\Swar}{\@ifnextchar[{\basicSwar}{\basicSwar[59]}}%

\def\basicswaR[#1]#2{\fdcase{\SWAR}{}{#2}{#100}}%

\newcommand{\swaR}{\@ifnextchar[{\basicswaR}{\basicswaR[59]}}%

\def\basicswdist[#1]{\fdcase{\SWDIST}{}{}{#100}}%

\newcommand{\swdist}{\@ifnextchar[{\basicswdist}{\basicswdist[59]}}%

\def\basicSwdist[#1]#2{\fdcase{\SWDIST}{#2}{}{#100}}%

\newcommand{\Swdist}{\@ifnextchar[{\basicSwdist}{\basicSwdist[59]}}%

\def\basicswdisT[#1]#2{\fdcase{\SWDIST}{}{#2}{#100}}%

\newcommand{\swdisT}{\@ifnextchar[{\basicswdisT}{\basicswdisT[59]}}%

\def\basicswdotar[#1]{\fdcase{\SWDOTAR}{}{}{#100}}%

\newcommand{\swdotar}{\@ifnextchar[{\basicswdotar}{\basicswdotar[59]}}%

\def\basicSwdotar[#1]#2{\fdcase{\SWDOTAR}{#2}{}{#100}}%

\newcommand{\Swdotar}{\@ifnextchar[{\basicSwdotar}{\basicSwdotar[59]}}%

\def\basicswdotaR[#1]#2{\fdcase{\SWDOTAR}{}{#2}{#100}}%

\newcommand{\swdotaR}{\@ifnextchar[{\basicswdotaR}{\basicswdotaR[59]}}%

\def\basicswmono[#1]{\fdcase{\SWMONO}{}{}{#100}}%

\newcommand{\swmono}{\@ifnextchar[{\basicswmono}{\basicswmono[59]}}%

\def\basicSwmono[#1]#2{\fdcase{\SWMONO}{#2}{}{#100}}%

\newcommand{\Swmono}{\@ifnextchar[{\basicSwmono}{\basicSwmono[59]}}%

\def\basicswmonO[#1]#2{\fdcase{\SWMONO}{}{#2}{#100}}%

\newcommand{\swmonO}{\@ifnextchar[{\basicswmonO}{\basicswmonO[59]}}%

\def\basicswepi[#1]{\fdcase{\SWEPI}{}{}{#100}}%

\newcommand{\swepi}{\@ifnextchar[{\basicswepi}{\basicswepi[59]}}%

\def\basicSwepi[#1]#2{\fdcase{\SWEPI}{#2}{}{#100}}%

\newcommand{\Swepi}{\@ifnextchar[{\basicSwepi}{\basicSwepi[59]}}%

\def\basicswepI[#1]#2{\fdcase{\SWEPI}{}{#2}{#100}}%

\newcommand{\swepI}{\@ifnextchar[{\basicswepI}{\basicswepI[59]}}%

\def\basicswbimo[#1]{\fdcase{\SWBIMO}{}{}{#100}}%

\newcommand{\swbimo}{\@ifnextchar[{\basicswbimo}{\basicswbimo[59]}}%

\def\basicSwbimo[#1]#2{\fdcase{\SWBIMO}{#2}{}{#100}}%

\newcommand{\Swbimo}{\@ifnextchar[{\basicSwbimo}{\basicSwbimo[59]}}%

\def\basicswbimO[#1]#2{\fdcase{\SWBIMO}{}{#2}{#100}}%

\newcommand{\swbimO}{\@ifnextchar[{\basicswbimO}{\basicswbimO[59]}}%

\def\basicswiso[#1]{\fdcase{\SWAR}{\hspace{-2pt}\cong}{}{#100}}%

\newcommand{\swiso}{\@ifnextchar[{\basicswiso}{\basicswiso[59]}}%

\def\basicSwiso[#1]#2{\fdcase{\SWAR}{#2}{\cong}{#100}}%

\newcommand{\Swiso}{\@ifnextchar[{\basicSwiso}{\basicSwiso[59]}}%

\def\basicswisO[#1]#2{\fdcase{\SWAR}{\hspace{-2pt}\cong}{#2}{#100}}%

\newcommand{\swisO}{\@ifnextchar[{\basicswisO}{\basicswisO[59]}}%

\def\basicswbiar[#1]{\fdbicase{\SWBIAR}{}{}{#100}}%

\newcommand{\swbiar}{\@ifnextchar[{\basicswbiar}{\basicswbiar[59]}}%

\def\basicSwbiar[#1]#2#3{\fdbicase{\SWBIAR}{#2}{#3}{#100}}%

\newcommand{\Swbiar}{\@ifnextchar[{\basicSwbiar}{\basicSwbiar[59]}}%

\def\basicswadjar[#1]{\fdbicase{\SWADJAR}{}{}{#100}}%

\newcommand{\swadjar}{\@ifnextchar[{\basicswadjar}{\basicswadjar[59]}}%

\def\basicSwadjar[#1]#2#3{\fdbicase{\SWADJAR}{#2}{#3}{#100}}%

\newcommand{\Swadjar}{\@ifnextchar[{\basicSwadjar}{\basicSwadjar[59]}}%

\def\basicswbidist[#1]{\fdbicase{\SWBIDIST}{}{}{#100}}%

\newcommand{\swbidist}{\@ifnextchar[{\basicswbidist}{\basicswbidist[59]}}%

\def\basicSwbidist[#1]#2#3{\fdbicase{\SWBIDIST}{#2}{#3}{#100}}%

\newcommand{\Swbidist}{\@ifnextchar[{\basicSwbidist}{\basicSwbidist[59]}}%

\def\basicswadjdist[#1]{\fdbicase{\SWADJDIST}{}{}{#100}}%

\newcommand{\swadjdist}{\@ifnextchar[{\basicswadjdist}{\basicswadjdist[59]}}%

\def\basicSwadjdist[#1]#2#3{\fdbicase{\SWADJDIST}{#2}{#3}{#100}}%

\newcommand{\Swadjdist}{\@ifnextchar[{\basicSwadjdist}{\basicSwadjdist[59]}}%

\newcommand{\sdcase}[4]{\testdiagrammode\begin{picture}(0,0)%
\put(0,0){#1{#4}}%
\truex{100}\truez{600}%
\put(\value{x},\value{x}){\makebox(0,\value{z})[l]{${#2}$}}%
\truex{300}\truey{800}%
\put(-\value{x},-\value{y}){\makebox(0,\value{z})[r]{${#3}$}}%
\end{picture}}%

\newcommand{\sdbicase}[4]{\testdiagrammode\begin{picture}(0,0)%
\put(0,0){#1{#4}}%
\truex{350}\truey{600}\truez{950}%
\put(\value{x},\value{x}){\makebox(0,\value{y})[l]{${#2}$}}%
\truex{450}\truey{600}\truez{1050}%
\put(-\value{x},-\value{z}){\makebox(0,\value{y})[r]{${#3}$}}%
\end{picture}}%

\newcommand{\SEAR}[1]{%
\Y=#1%
\divide\Y by 2%
\begin{picture}(0,0)%
\put(-\Y,\Y){\line(1,-1){#1}}%
\put(\Y,-\Y){\sehead}%
\end{picture}}%

\newcommand{\SEDIST}[1]{%
\Y=#1%
\divide\Y by 2%
\begin{picture}(0,0)%
\put(-\Y,\Y){\line(1,-1){#1}}%
\put(\Y,-\Y){\sehead}%
\truex{400}%
\put(0,0){\circle{\value{x}}}%
\end{picture}}%

\newcommand{\SEDOTAR}[1]%
{\truex{100}\truey{212}%
\Y=#1%
\divide\Y by 2%
\NUMBEROFDOTS=#1%
\divide\NUMBEROFDOTS by \value{y}%
\advance\NUMBEROFDOTS by 1%
\begin{picture}(0,0)%
\multiput(-\Y,\Y)(\value{y},-\value{y}){\NUMBEROFDOTS}%
{\circle*{\value{x}}}%
\put(\Y,-\Y){\sehead}%
\end{picture}}%

\newcommand{\SEMONO}[1]{%
\Y=#1%
\divide \Y by 2%
\Truetail%
\bimolength=#1%
\advance\bimolength by -\Truemonotail%
\monolength=\bimolength%
\advance\monolength by -\Y%
\begin{picture}(0,0)%
\put(-\monolength,\monolength){\line(1,-1){\bimolength}}%
\put(-\monolength,\monolength){\sehead}%
\put(\Y,-\Y){\sehead}%
\end{picture}}%

\newcommand{\SEEPI}[1]{%
\Y=#1%
\divide\Y by 2%
\Truehead%
\bimolength=#1%
\advance\bimolength by -\Trueepihead%
\epilength=\bimolength%
\advance\epilength by -\Y%
\begin{picture}(0,0)%
\put(-\Y,\Y){\line(1,-1){\bimolength}}%
\put(\epilength,-\epilength){\sehead}%
\put(\Y,-\Y){\sehead}%
\end{picture}}%

\newcommand{\SEBIMO}[1]{%
\Y=#1%
\divide\Y by 2%
\Truetail\Truehead%
\bimolength=#1%
\advance\bimolength by -\Truemonotail%
\monolength=\bimolength%
\advance\monolength by -\Y%
\advance\bimolength by -\Trueepihead%
\epilength=\bimolength%
\advance\epilength by -\monolength%
\begin{picture}(0,0)%
\put(-\monolength,\monolength){\line(1,-1){\bimolength}}%
\put(-\monolength,\monolength){\sehead}%
\put(\epilength,-\epilength){\sehead}%
\put(\Y,-\Y){\sehead}%
\end{picture}}%

\newcommand{\SEBIAR}[1]{%
\Y=#1%
\divide\Y by 2%
\begin{picture}(0,0)%
\put(-\Y,\Y){\begin{picture}(0,0)%
\truex{247}%
\put(-\value{x},-\value{x}){\line(1,-1){#1}}%
\put(\value{x},\value{x}){\line(1,-1){#1}}%
\monolength=#1%
\advance\monolength by -\value{x}%
\epilength=#1%
\advance\epilength by \value{x}%
\put(\monolength,-\epilength){\sehead}%
\put(\epilength,-\monolength){\sehead}%
\end{picture}}\end{picture}}%

\newcommand{\SEBIDIST}[1]{%
\Y=#1%
\divide\Y by 2%
\truey{400}%
\begin{picture}(0,0)%
\put(-\Y,\Y){\begin{picture}(0,0)%
\truex{247}%
\monolength=#1%
\advance\monolength by -\value{x}%
\epilength=#1%
\advance\epilength by \value{x}%
\put(\value{x},\value{x}){\line(1,-1){#1}}%
\put(\epilength,-\monolength){\sehead}%
\end{picture}}%
\put(-\Y,\Y){\begin{picture}(0,0)%
\truex{247}%
\monolength=#1%
\advance\monolength by \value{x}%
\epilength=#1%
\advance\epilength by -\value{x}%
\put(-\value{x},-\value{x}){\line(1,-1){#1}}%
\put(\epilength,-\monolength){\sehead}%
\end{picture}}%
\put(-\value{x},-\value{x}){\circle{\value{y}}}%
\put(\value{x},\value{x}){\circle{\value{y}}}%
\end{picture}}%

\newcommand{\SEEQL}[1]{%
\Y=#1%
\divide\Y by 2%
\begin{picture}(0,0)%
\put(-\Y,\Y){\begin{picture}(0,0)%
\truex{70}%
\put(-\value{x},-\value{x}){\line(1,-1){#1}}%
\put(\value{x},\value{x}){\line(1,-1){#1}}%
\end{picture}}\end{picture}}%

\newcommand{\SEADJAR}[1]{%
\Y=#1%
\divide\Y by 2%
\begin{picture}(0,0)%
\put(-\Y,\Y){\begin{picture}(0,0)%
\truex{247}%
\monolength=#1%
\advance\monolength by -\value{x}%
\epilength=#1%
\advance\epilength by \value{x}%
\put(-\value{x},-\value{x}){\line(1,-1){#1}}%
\put(\monolength,-\epilength){\sehead}%
\end{picture}}%
\put(\Y,-\Y){\begin{picture}(0,0)%
\truex{247}%
\monolength=#1%
\advance\monolength by -\value{x}%
\epilength=#1%
\advance\epilength by \value{x}%
\put(\value{x},\value{x}){\line(-1,1){#1}}%
\put(-\monolength,\epilength){\nwhead}%
\end{picture}}\end{picture}}%

\newcommand{\SEADJDIST}[1]{%
\Y=#1%
\divide\Y by 2%
\truey{400}%
\begin{picture}(0,0)%
\put(-\Y,\Y){\begin{picture}(0,0)%
\truex{247}%
\monolength=#1%
\advance\monolength by -\value{x}%
\epilength=#1%
\advance\epilength by \value{x}%
\put(-\value{x},-\value{x}){\line(1,-1){#1}}%
\put(\monolength,-\epilength){\sehead}%
\end{picture}}%
\put(\Y,-\Y){\begin{picture}(0,0)%
\truex{247}%
\monolength=#1%
\advance\monolength by -\value{x}%
\epilength=#1%
\advance\epilength by \value{x}%
\put(\value{x},\value{x}){\line(-1,1){#1}}%
\put(-\monolength,\epilength){\nwhead}%
\end{picture}}%
\put(-\value{x},-\value{x}){\circle{\value{y}}}%
\put(\value{x},\value{x}){\circle{\value{y}}}%
\end{picture}}%

\def\basicsear[#1]{\sdcase{\SEAR}{}{}{#100}}%

\newcommand{\sear}{\@ifnextchar[{\basicsear}{\basicsear[59]}}%

\def\basicSear[#1]#2{\sdcase{\SEAR}{#2}{}{#100}}%

\newcommand{\Sear}{\@ifnextchar[{\basicSear}{\basicSear[59]}}%

\def\basicseaR[#1]#2{\sdcase{\SEAR}{}{#2}{#100}}%

\newcommand{\seaR}{\@ifnextchar[{\basicseaR}{\basicseaR[59]}}%

\def\basicsedist[#1]{\sdcase{\SEDIST}{}{}{#100}}%

\newcommand{\sedist}{\@ifnextchar[{\basicsedist}{\basicsedist[59]}}%

\def\basicSedist[#1]#2{\sdcase{\SEDIST}{#2}{}{#100}}%

\newcommand{\Sedist}{\@ifnextchar[{\basicSedist}{\basicSedist[59]}}%

\def\basicsedisT[#1]#2{\sdcase{\SEDIST}{}{#2}{#100}}%

\newcommand{\sedisT}{\@ifnextchar[{\basicsedisT}{\basicsedisT[59]}}%

\def\basicsedotar[#1]{\sdcase{\SEDOTAR}{}{}{#100}}%

\newcommand{\sedotar}{\@ifnextchar[{\basicsedotar}{\basicsedotar[59]}}%

\def\basicSedotar[#1]#2{\sdcase{\SEDOTAR}{#2}{}{#100}}%

\newcommand{\Sedotar}{\@ifnextchar[{\basicSedotar}{\basicSedotar[59]}}%

\def\basicsedotaR[#1]#2{\sdcase{\SEDOTAR}{}{#2}{#100}}%

\newcommand{\sedotaR}{\@ifnextchar[{\basicsedotaR}{\basicsedotaR[59]}}%

\def\basicsemono[#1]{\sdcase{\SEMONO}{}{}{#100}}%

\newcommand{\semono}{\@ifnextchar[{\basicsemono}{\basicsemono[59]}}%

\def\basicSemono[#1]#2{\sdcase{\SEMONO}{#2}{}{#100}}%

\newcommand{\Semono}{\@ifnextchar[{\basicSemono}{\basicSemono[59]}}%

\def\basicsemonO[#1]#2{\sdcase{\SEMONO}{}{#2}{#100}}%

\newcommand{\semonO}{\@ifnextchar[{\basicsemonO}{\basicsemonO[59]}}%

\def\basicseepi[#1]{\sdcase{\SEEPI}{}{}{#100}}%

\newcommand{\seepi}{\@ifnextchar[{\basicseepi}{\basicseepi[59]}}%

\def\basicSeepi[#1]#2{\sdcase{\SEEPI}{#2}{}{#100}}%

\newcommand{\Seepi}{\@ifnextchar[{\basicSeepi}{\basicSeepi[59]}}%

\def\basicseepI[#1]#2{\sdcase{\SEEPI}{}{#2}{#100}}%

\newcommand{\seepI}{\@ifnextchar[{\basicseepI}{\basicseepI[59]}}%

\def\basicsebimo[#1]{\sdcase{\SEBIMO}{}{}{#100}}%

\newcommand{\sebimo}{\@ifnextchar[{\basicsebimo}{\basicsebimo[59]}}%

\def\basicSebimo[#1]#2{\sdcase{\SEBIMO}{#2}{}{#100}}%

\newcommand{\Sebimo}{\@ifnextchar[{\basicSebimo}{\basicSebimo[59]}}%

\def\basicsebimO[#1]#2{\sdcase{\SEBIMO}{}{#2}{#100}}%

\newcommand{\sebimO}{\@ifnextchar[{\basicsebimO}{\basicsebimO[59]}}%

\def\basicseiso[#1]{\sdcase{\SEAR}{\hspace{-2pt}\cong}{}{#100}}%

\newcommand{\seiso}{\@ifnextchar[{\basicseiso}{\basicseiso[59]}}%

\def\basicSeiso[#1]#2{\sdcase{\SEAR}{#2}{\cong}{#100}}%

\newcommand{\Seiso}{\@ifnextchar[{\basicSeiso}{\basicSeiso[59]}}%

\def\basicseisO[#1]#2{\sdcase{\SEAR}{\hspace{-2pt}\cong}{#2}{#100}}%

\newcommand{\seisO}{\@ifnextchar[{\basicseisO}{\basicseisO[59]}}%

\def\basicseeql[#1]{\sdcase{\SEEQL}{}{}{#100}}%

\newcommand{\seeql}{\@ifnextchar[{\basicseeql}{\basicseeql[59]}}%

\def\basicSeeql[#1]#2{\sdcase{\SEEQL}{#2}{}{#100}}%

\newcommand{\Seeql}{\@ifnextchar[{\basicSeeql}{\basicSeeql[59]}}%

\def\basicseeqL[#1]#2{\sdcase{\SEEQL}{}{#2}{#100}}%

\newcommand{\seeqL}{\@ifnextchar[{\basicseeqL}{\basicseeqL[59]}}%

\def\basicsebiar[#1]{\sdbicase{\SEBIAR}{}{}{#100}}%

\newcommand{\sebiar}{\@ifnextchar[{\basicsebiar}{\basicsebiar[59]}}%

\def\basicSebiar[#1]#2#3{\sdbicase{\SEBIAR}{#2}{#3}{#100}}%

\newcommand{\Sebiar}{\@ifnextchar[{\basicSebiar}{\basicSebiar[59]}}%

\def\basicseadjar[#1]{\sdbicase{\SEADJAR}{}{}{#100}}%

\newcommand{\seadjar}{\@ifnextchar[{\basicseadjar}{\basicseadjar[59]}}%

\def\basicSeadjar[#1]#2#3{\sdbicase{\SEADJAR}{#2}{#3}{#100}}%

\newcommand{\Seadjar}{\@ifnextchar[{\basicSeadjar}{\basicSeadjar[59]}}%

\def\basicsebidist[#1]{\sdbicase{\SEBIDIST}{}{}{#100}}%

\newcommand{\sebidist}{\@ifnextchar[{\basicsebidist}{\basicsebidist[59]}}%

\def\basicSebidist[#1]#2#3{\sdbicase{\SEBIDIST}{#2}{#3}{#100}}%

\newcommand{\Sebidist}{\@ifnextchar[{\basicSebidist}{\basicSebidist[59]}}%

\def\basicseadjdist[#1]{\sdbicase{\SEADJDIST}{}{}{#100}}%

\newcommand{\seadjdist}{\@ifnextchar[{\basicseadjdist}{\basicseadjdist[59]}}%

\def\basicSeadjdist[#1]#2#3{\sdbicase{\SEADJDIST}{#2}{#3}{#100}}%

\newcommand{\Seadjdist}{\@ifnextchar[{\basicSeadjdist}{\basicSeadjdist[59]}}%

\newcommand{\NWAR}[1]{%
\Y=#1%
\divide\Y by 2%
\begin{picture}(0,0)%
\put(\Y,-\Y){\line(-1,1){#1}}%
\put(-\Y,\Y){\nwhead}%
\end{picture}}%

\newcommand{\NWDIST}[1]{%
\Y=#1%
\divide\Y by 2%
\begin{picture}(0,0)%
\put(\Y,-\Y){\line(-1,1){#1}}%
\put(-\Y,\Y){\nwhead}%
\truex{400}%
\put(0,0){\circle{\value{x}}}%
\end{picture}}%

\newcommand{\NWDOTAR}[1]%
{\truex{100}\truey{212}%
\Y=#1%
\divide\Y by 2%
\NUMBEROFDOTS=#1%
\divide\NUMBEROFDOTS by \value{y}%
\advance\NUMBEROFDOTS by 1%
\begin{picture}(0,0)%
\multiput(\Y,-\Y)(-\value{y},\value{y}){\NUMBEROFDOTS}%
{\circle*{\value{x}}}%
\put(-\Y,\Y){\nwhead}%
\end{picture}}%

\newcommand{\NWMONO}[1]{%
\Y=#1%
\divide \Y by 2%
\Truetail%
\bimolength=#1%
\advance\bimolength by -\Truemonotail%
\monolength=\bimolength%
\advance\monolength by -\Y%
\begin{picture}(0,0)%
\put(\monolength,-\monolength){\line(-1,1){\bimolength}}%
\put(\monolength,-\monolength){\nwhead}%
\put(-\Y,\Y){\nwhead}%
\end{picture}}%

\newcommand{\NWEPI}[1]{%
\Y=#1%
\divide\Y by 2%
\Truehead%
\bimolength=#1%
\advance\bimolength by -\Trueepihead%
\epilength=\bimolength%
\advance\epilength by -\Y%
\begin{picture}(0,0)%
\put(\Y,-\Y){\line(-1,1){\bimolength}}%
\put(-\epilength,\epilength){\nwhead}%
\put(-\Y,\Y){\nwhead}%
\end{picture}}%

\newcommand{\NWBIMO}[1]{%
\Y=#1%
\divide\Y by 2%
\Truetail\Truehead%
\bimolength=#1%
\advance\bimolength by -\Truemonotail%
\monolength=\bimolength%
\advance\monolength by -\Y%
\advance\bimolength by -\Trueepihead%
\epilength=\bimolength%
\advance\epilength by -\monolength%
\begin{picture}(0,0)%
\put(\monolength,-\monolength){\line(-1,1){\bimolength}}%
\put(\monolength,-\monolength){\nwhead}%
\put(-\epilength,\epilength){\nwhead}%
\put(-\Y,\Y){\nwhead}%
\end{picture}}%

\newcommand{\NWBIAR}[1]{%
\Y=#1%
\divide\Y by 2%
\begin{picture}(0,0)%
\put(\Y,-\Y){\begin{picture}(0,0)%
\truex{247}%
\put(-\value{x},-\value{x}){\line(-1,1){#1}}%
\put(\value{x},\value{x}){\line(-1,1){#1}}%
\monolength=#1%
\advance\monolength by -\value{x}%
\epilength=#1%
\advance\epilength by \value{x}%
\put(-\monolength,\epilength){\nwhead}%
\put(-\epilength,\monolength){\nwhead}%
\end{picture}}\end{picture}}%

\newcommand{\NWBIDIST}[1]{%
\Y=#1%
\divide\Y by 2%
\truey{400}%
\begin{picture}(0,0)%
\put(\Y,-\Y){\begin{picture}(0,0)%
\truex{247}%
\monolength=#1%
\advance\monolength by -\value{x}%
\epilength=#1%
\advance\epilength by \value{x}%
\put(-\value{x},-\value{x}){\line(-1,1){#1}}%
\put(-\epilength,\monolength){\nwhead}%
\end{picture}}%
\put(\Y,-\Y){\begin{picture}(0,0)%
\truex{247}%
\monolength=#1%
\advance\monolength by \value{x}%
\epilength=#1%
\advance\epilength by -\value{x}%
\put(\value{x},\value{x}){\line(-1,1){#1}}%
\put(-\epilength,\monolength){\nwhead}%
\end{picture}}%
\put(-\value{x},-\value{x}){\circle{\value{y}}}%
\put(\value{x},\value{x}){\circle{\value{y}}}%
\end{picture}}%

\newcommand{\NWADJAR}[1]{%
\Y=#1%
\divide\Y by 2%
\begin{picture}(0,0)%
\put(\Y,-\Y){\begin{picture}(0,0)%
\truex{247}%
\monolength=#1%
\advance\monolength by -\value{x}%
\epilength=#1%
\advance\epilength by \value{x}%
\put(-\value{x},-\value{x}){\line(-1,1){#1}}%
\put(-\epilength,\monolength){\nwhead}%
\end{picture}}%
\put(-\Y,\Y){\begin{picture}(0,0)%
\truex{247}%
\monolength=#1%
\advance\monolength by -\value{x}%
\epilength=#1%
\advance\epilength by \value{x}%
\put(\value{x},\value{x}){\line(1,-1){#1}}%
\put(\epilength,-\monolength){\sehead}%
\end{picture}}\end{picture}}%

\newcommand{\NWADJDIST}[1]{%
\Y=#1%
\divide\Y by 2%
\truey{400}%
\begin{picture}(0,0)%
\put(\Y,-\Y){\begin{picture}(0,0)%
\truex{247}%
\monolength=#1%
\advance\monolength by -\value{x}%
\epilength=#1%
\advance\epilength by \value{x}%
\put(-\value{x},-\value{x}){\line(-1,1){#1}}%
\put(-\epilength,\monolength){\nwhead}%
\end{picture}}%
\put(-\Y,\Y){\begin{picture}(0,0)%
\truex{247}%
\monolength=#1%
\advance\monolength by -\value{x}%
\epilength=#1%
\advance\epilength by \value{x}%
\put(\value{x},\value{x}){\line(1,-1){#1}}%
\put(\epilength,-\monolength){\sehead}%
\end{picture}}%
\put(-\value{x},-\value{x}){\circle{\value{y}}}%
\put(\value{x},\value{x}){\circle{\value{y}}}%
\end{picture}}%

\def\basicnwar[#1]{\sdcase{\NWAR}{}{}{#100}}%

\newcommand{\nwar}{\@ifnextchar[{\basicnwar}{\basicnwar[59]}}%

\def\basicNwar[#1]#2{\sdcase{\NWAR}{#2}{}{#100}}%

\newcommand{\Nwar}{\@ifnextchar[{\basicNwar}{\basicNwar[59]}}%

\def\basicnwaR[#1]#2{\sdcase{\NWAR}{}{#2}{#100}}%

\newcommand{\nwaR}{\@ifnextchar[{\basicnwaR}{\basicnwaR[59]}}%

\def\basicnwdist[#1]{\sdcase{\NWDIST}{}{}{#100}}%

\newcommand{\nwdist}{\@ifnextchar[{\basicnwdist}{\basicnwdist[59]}}%

\def\basicNwdist[#1]#2{\sdcase{\NWDIST}{#2}{}{#100}}%

\newcommand{\Nwdist}{\@ifnextchar[{\basicNwdist}{\basicNwdist[59]}}%

\def\basicnwdisT[#1]#2{\sdcase{\NWDIST}{}{#2}{#100}}%

\newcommand{\nwdisT}{\@ifnextchar[{\basicnwdisT}{\basicnwdisT[59]}}%

\def\basicnwdotar[#1]{\sdcase{\NWDOTAR}{}{}{#100}}%

\newcommand{\nwdotar}{\@ifnextchar[{\basicnwdotar}{\basicnwdotar[59]}}%

\def\basicNwdotar[#1]#2{\sdcase{\NWDOTAR}{#2}{}{#100}}%

\newcommand{\Nwdotar}{\@ifnextchar[{\basicNwdotar}{\basicNwdotar[59]}}%

\def\basicnwdotaR[#1]#2{\sdcase{\NWDOTAR}{}{#2}{#100}}%

\newcommand{\nwdotaR}{\@ifnextchar[{\basicnwdotaR}{\basicnwdotaR[59]}}%

\def\basicnwmono[#1]{\sdcase{\NWMONO}{}{}{#100}}%

\newcommand{\nwmono}{\@ifnextchar[{\basicnwmono}{\basicnwmono[59]}}%

\def\basicNwmono[#1]#2{\sdcase{\NWMONO}{#2}{}{#100}}%

\newcommand{\Nwmono}{\@ifnextchar[{\basicNwmono}{\basicNwmono[59]}}%

\def\basicnwmonO[#1]#2{\sdcase{\NWMONO}{}{#2}{#100}}%

\newcommand{\nwmonO}{\@ifnextchar[{\basicnwmonO}{\basicnwmonO[59]}}%

\def\basicnwepi[#1]{\sdcase{\NWEPI}{}{}{#100}}%

\newcommand{\nwepi}{\@ifnextchar[{\basicnwepi}{\basicnwepi[59]}}%

\def\basicNwepi[#1]#2{\sdcase{\NWEPI}{#2}{}{#100}}%

\newcommand{\Nwepi}{\@ifnextchar[{\basicNwepi}{\basicNwepi[59]}}%

\def\basicnwepI[#1]#2{\sdcase{\NWEPI}{}{#2}{#100}}%

\newcommand{\nwepI}{\@ifnextchar[{\basicnwepI}{\basicnwepI[59]}}%

\def\basicnwbimo[#1]{\sdcase{\NWBIMO}{}{}{#100}}%

\newcommand{\nwbimo}{\@ifnextchar[{\basicnwbimo}{\basicnwbimo[59]}}%

\def\basicNwbimo[#1]#2{\sdcase{\NWBIMO}{#2}{}{#100}}%

\newcommand{\Nwbimo}{\@ifnextchar[{\basicNwbimo}{\basicNwbimo[59]}}%

\def\basicnwbimO[#1]#2{\sdcase{\NWBIMO}{}{#2}{#100}}%

\newcommand{\nwbimO}{\@ifnextchar[{\basicnwbimO}{\basicnwbimO[59]}}%

\def\basicnwiso[#1]{\sdcase{\NWAR}{\hspace{-2pt}\cong}{}{#100}}%

\newcommand{\nwiso}{\@ifnextchar[{\basicnwiso}{\basicnwiso[59]}}%

\def\basicNwiso[#1]#2{\sdcase{\NWAR}{#2}{\cong}{#100}}%

\newcommand{\Nwiso}{\@ifnextchar[{\basicNwiso}{\basicNwiso[59]}}%

\def\basicnwisO[#1]#2{\sdcase{\NWAR}{\hspace{-2pt}\cong}{#2}{#100}}%

\newcommand{\nwisO}{\@ifnextchar[{\basicnwisO}{\basicnwisO[59]}}%

\def\basicnwbiar[#1]{\sdbicase{\NWBIAR}{}{}{#100}}%

\newcommand{\nwbiar}{\@ifnextchar[{\basicnwbiar}{\basicnwbiar[59]}}%

\def\basicNwbiar[#1]#2#3{\sdbicase{\NWBIAR}{#2}{#3}{#100}}%

\newcommand{\Nwbiar}{\@ifnextchar[{\basicNwbiar}{\basicNwbiar[59]}}%

\def\basicnwadjar[#1]{\sdbicase{\NWADJAR}{}{}{#100}}%

\newcommand{\nwadjar}{\@ifnextchar[{\basicnwadjar}{\basicnwadjar[59]}}%

\def\basicNwadjar[#1]#2#3{\sdbicase{\NWADJAR}{#2}{#3}{#100}}%

\newcommand{\Nwadjar}{\@ifnextchar[{\basicNwadjar}{\basicNwadjar[59]}}%

\def\basicnwbidist[#1]{\sdbicase{\NWBIDIST}{}{}{#100}}%

\newcommand{\nwbidist}{\@ifnextchar[{\basicnwbidist}{\basicnwbidist[59]}}%

\def\basicNwbidist[#1]#2#3{\sdbicase{\NWBIDIST}{#2}{#3}{#100}}%

\newcommand{\Nwbidist}{\@ifnextchar[{\basicNwbidist}{\basicNwbidist[59]}}%

\def\basicnwadjdist[#1]{\sdbicase{\NWADJDIST}{}{}{#100}}%

\newcommand{\nwadjdist}{\@ifnextchar[{\basicnwadjdist}{\basicnwadjdist[59]}}%

\def\basicNwadjdist[#1]#2#3{\sdbicase{\NWADJDIST}{#2}{#3}{#100}}%

\newcommand{\Nwadjdist}{\@ifnextchar[{\basicNwadjdist}{\basicNwadjdist[59]}}%

\newcommand{\ENEAR}[3]{\testdiagrammode%
\Y=#3%
\divide\Y by 2%
\Z=\Y%
\divide\Z by 2%
\begin{picture}(0,0)%
\put(-\Y,-\Z){\line(2,1){#3}}%
\put(\Y,\Z){\enehead}%
\truex{200}\truey{800}\truez{600}%
\put(-\value{x},\value{x}){\makebox(0,\value{z})[r]{${#1}$}}%
\put(\value{x},-\value{y}){\makebox(0,\value{z})[l]{${#2}$}}%
\end{picture}}%

\newcommand{\ENEDIST}[3]{\testdiagrammode%
\Y=#3%
\divide\Y by 2%
\Z=\Y%
\divide\Z by 2%
\begin{picture}(0,0)%
\put(-\Y,-\Z){\line(2,1){#3}}%
\put(\Y,\Z){\enehead}%
\truex{400}%
\put(0,0){\circle{\value{x}}}%
\truex{200}\truey{800}\truez{600}%
\put(-\value{x},\value{x}){\makebox(0,\value{z})[r]{${#1}$}}%
\put(\value{x},-\value{y}){\makebox(0,\value{z})[l]{${#2}$}}%
\end{picture}}%

\newcommand{\ENEDOTAR}[3]{\testdiagrammode%
\truex{100}\truey{268}\truez{134}%
\Y=#3%
\divide\Y by 2%
\Z=\Y%
\divide\Z by 2%
\NUMBEROFDOTS=#3%
\divide\NUMBEROFDOTS by \value{y}%
\advance\NUMBEROFDOTS by 1%
\begin{picture}(0,0)%
\multiput(-\Y,-\Z)(\value{y},\value{z}){\NUMBEROFDOTS}%
{\circle*{\value{x}}}%
\put(\Y,\Z){\enehead}%
\truex{200}\truey{800}\truez{600}%
\put(-\value{x},\value{x}){\makebox(0,\value{z})[r]{${#1}$}}%
\put(\value{x},-\value{y}){\makebox(0,\value{z})[l]{${#2}$}}%
\end{picture}}%

\newcommand{\ENEMONO}[3]{\testdiagrammode%
\Y=#3%
\divide\Y by 2%
\Z=\Y%
\divide\Z by 2%
\TrueTail%
\bimolength=#3%
\advance\bimolength by -\TrueMonoTail%
\monolength=\bimolength%
\advance\monolength by -\Y%
\secondmonolength=\monolength%
\divide\secondmonolength by 2%
\begin{picture}(0,0)%
\put(-\monolength,-\secondmonolength){\line(2,1){\bimolength}}%
\put(-\monolength,-\secondmonolength){\enehead}%
\put(\Y,\Z){\enehead}%
\truex{200}\truey{800}\truez{600}%
\put(-\value{x},\value{x}){\makebox(0,\value{z})[r]{${#1}$}}%
\put(\value{x},-\value{y}){\makebox(0,\value{z})[l]{${#2}$}}%
\end{picture}}%

\newcommand{\ENEEPI}[3]{\testdiagrammode%
\Y=#3%
\divide\Y by 2%
\Z=\Y%
\divide\Z by 2%
\TrueHead%
\bimolength=#3%
\advance\bimolength by -\TrueEpiHead%
\epilength=\bimolength%
\advance\epilength by -\Y%
\secondepilength=\epilength%
\divide\secondepilength by 2%
\begin{picture}(0,0)%
\put(-\Y,-\Z){\line(2,1){\bimolength}}%
\put(\epilength,\secondepilength){\enehead}%
\put(\Y,\Z){\enehead}%
\truex{200}\truey{800}\truez{600}%
\put(-\value{x},\value{x}){\makebox(0,\value{z})[r]{${#1}$}}%
\put(\value{x},-\value{y}){\makebox(0,\value{z})[l]{${#2}$}}%
\end{picture}}%

\newcommand{\ENEBIMO}[3]{\testdiagrammode%
\Y=#3%
\divide\Y by 2%
\Z=\Y%
\divide\Z by 2%
\TrueTail\TrueHead%
\bimolength=#3%
\advance\bimolength by -\TrueMonoTail%
\monolength=\bimolength%
\advance\monolength by -\Y%
\advance\bimolength by -\TrueEpiHead%
\epilength=\bimolength%
\advance\epilength by -\monolength%
\secondmonolength=\monolength%
\divide\secondmonolength by 2%
\secondepilength=\epilength%
\divide\secondepilength by 2%
\begin{picture}(0,0)%
\put(-\monolength,-\secondmonolength){\line(2,1){\bimolength}}%
\put(-\monolength,-\secondmonolength){\enehead}%
\put(\epilength,\secondepilength){\enehead}%
\put(\Y,\Z){\enehead}%
\truex{200}\truey{800}\truez{600}%
\put(-\value{x},\value{x}){\makebox(0,\value{z})[r]{${#1}$}}%
\put(\value{x},-\value{y}){\makebox(0,\value{z})[l]{${#2}$}}%
\end{picture}}%

\newcommand{\ENEEQL}[3]{\testdiagrammode%
\Y=#3%
\divide\Y by 2%
\Z=\Y%
\divide\Z by 2%
\begin{picture}(0,0)%
\put(-\Y,-\Z){\begin{picture}(0,0)%
\truex{44}\truey{89}%
\put(-\value{x},\value{y}){\line(2,1){#3}}%
\put(\value{x},-\value{y}){\line(2,1){#3}}%
\end{picture}}%
\truex{200}\truey{800}\truez{600}%
\put(-\value{x},\value{x}){\makebox(0,\value{z})[r]{${#1}$}}%
\put(\value{x},-\value{y}){\makebox(0,\value{z})[l]{${#2}$}}%
\end{picture}}%

\newcommand{\ENEBIAR}[3]{\testdiagrammode%
\Y=#3%
\divide\Y by 2%
\Z=\Y%
\divide\Z by 2%
\begin{picture}(0,0)%
\put(-\Y,-\Z){\begin{picture}(0,0)%
\truex{156}\truey{313}%
\put(-\value{x},\value{y}){\line(2,1){#3}}%
\put(\value{x},-\value{y}){\line(2,1){#3}}%
\monolength=#3%
\advance\monolength by -\value{x}%
\epilength=#3%
\advance\epilength by \value{x}%
\secondmonolength=\Y%
\advance\secondmonolength by -\value{y}%
\secondepilength=\Y%
\advance\secondepilength by \value{y}%
\put(\monolength,\secondepilength){\enehead}%
\put(\epilength,\secondmonolength){\enehead}%
\end{picture}}
\truex{300}\truey{1000}\truez{600}%
\put(-\value{x},\value{x}){\makebox(0,\value{z})[r]{${#1}$}}%
\put(\value{x},-\value{y}){\makebox(0,\value{z})[l]{${#2}$}}%
\end{picture}}%

\newcommand{\ENEBIDIST}[3]{\testdiagrammode%
\Y=#3%
\divide\Y by 2%
\Z=\Y%
\divide\Z by 2%
\begin{picture}(0,0)%
\truex{156}\truey{313}\truez{400}%
\put(-\Y,-\Z){\begin{picture}(0,0)%
\put(-\value{x},\value{y}){\line(2,1){#3}}%
\put(\value{x},-\value{y}){\line(2,1){#3}}%
\monolength=#3%
\advance\monolength by -\value{x}%
\epilength=#3%
\advance\epilength by \value{x}%
\secondmonolength=\Y%
\advance\secondmonolength by -\value{y}%
\secondepilength=\Y%
\advance\secondepilength by \value{y}%
\put(\monolength,\secondepilength){\enehead}%
\put(\epilength,\secondmonolength){\enehead}%
\end{picture}}
\put(-\value{x},\value{y}){\circle{\value{z}}}%
\put(\value{x},-\value{y}){\circle{\value{z}}}%
\truex{300}\truey{1000}\truez{600}%
\put(-\value{x},\value{x}){\makebox(0,\value{z})[r]{${#1}$}}%
\put(\value{x},-\value{y}){\makebox(0,\value{z})[l]{${#2}$}}%
\end{picture}}%

\newcommand{\ENEADJAR}[3]{\testdiagrammode%
\Y=#3%
\divide\Y by 2%
\Z=\Y%
\divide\Z by 2%
\begin{picture}(0,0)%
\put(-\Y,-\Z){\begin{picture}(0,0)%
\truex{156}\truey{313}%
\monolength=#3%
\advance\monolength by -\value{x}%
\epilength=#3%
\advance\epilength by \value{x}%
\secondmonolength=\Y%
\advance\secondmonolength by -\value{y}%
\secondepilength=\Y%
\advance\secondepilength by \value{y}%
\put(\value{x},-\value{y}){\line(2,1){#3}}%
\put(\epilength,\secondmonolength){\enehead}%
\put(\monolength,\secondepilength){\line(-2,-1){#3}}%
\put(-\value{x},\value{y}){\wswhead}%
\end{picture}}
\truex{300}\truey{1000}\truez{600}%
\put(-\value{x},\value{x}){\makebox(0,\value{z})[r]{${#1}$}}%
\put(\value{x},-\value{y}){\makebox(0,\value{z})[l]{${#2}$}}%
\end{picture}}%

\newcommand{\ENEADJDIST}[3]{\testdiagrammode%
\Y=#3%
\divide\Y by 2%
\Z=\Y%
\divide\Z by 2%
\begin{picture}(0,0)%
\truex{156}\truey{313}\truez{400}%
\put(-\Y,-\Z){\begin{picture}(0,0)%
\monolength=#3%
\advance\monolength by -\value{x}%
\epilength=#3%
\advance\epilength by \value{x}%
\secondmonolength=\Y%
\advance\secondmonolength by -\value{y}%
\secondepilength=\Y%
\advance\secondepilength by \value{y}%
\put(\value{x},-\value{y}){\line(2,1){#3}}%
\put(\epilength,\secondmonolength){\enehead}%
\put(\monolength,\secondepilength){\line(-2,-1){#3}}%
\put(-\value{x},\value{y}){\wswhead}%
\end{picture}}
\put(-\value{x},\value{y}){\circle{\value{z}}}%
\put(\value{x},-\value{y}){\circle{\value{z}}}%
\truex{300}\truey{1000}\truez{600}%
\put(-\value{x},\value{x}){\makebox(0,\value{z})[r]{${#1}$}}%
\put(\value{x},-\value{y}){\makebox(0,\value{z})[l]{${#2}$}}%
\end{picture}}%

\def\basicenear[#1]{\ENEAR{}{}{#100}}%

\newcommand{\enear}{\@ifnextchar[{\basicenear}{\basicenear[133]}}%

\def\basicEnear[#1]#2{\ENEAR{#2}{}{#100}}%

\newcommand{\Enear}{\@ifnextchar[{\basicEnear}{\basicEnear[133]}}%

\def\basiceneaR[#1]#2{\ENEAR{}{#2}{#100}}%

\newcommand{\eneaR}{\@ifnextchar[{\basiceneaR}{\basiceneaR[133]}}%

\def\basicenedist[#1]{\ENEDIST{}{}{#100}}%

\newcommand{\enedist}{\@ifnextchar[{\basicenedist}{\basicenedist[133]}}%

\def\basicEnedist[#1]#2{\ENEDIST{#2}{}{#100}}%

\newcommand{\Enedist}{\@ifnextchar[{\basicEnedist}{\basicEnedist[133]}}%

\def\basicenedisT[#1]#2{\ENEDIST{}{#2}{#100}}%

\newcommand{\enedisT}{\@ifnextchar[{\basicenedisT}{\basicenedisT[133]}}%

\def\basicenedotar[#1]{\ENEDOTAR{}{}{#100}}%

\newcommand{\enedotar}{\@ifnextchar[{\basicenedotar}{\basicenedotar[133]}}%

\def\basicEnedotar[#1]#2{\ENEDOTAR{#2}{}{#100}}%

\newcommand{\Enedotar}{\@ifnextchar[{\basicEnedotar}{\basicEnedotar[133]}}%

\def\basicenedotaR[#1]#2{\ENEDOTAR{}{#2}{#100}}%

\newcommand{\enedotaR}{\@ifnextchar[{\basicenedotaR}{\basicenedotaR[133]}}%

\def\basicenemono[#1]{\ENEMONO{}{}{#100}}%

\newcommand{\enemono}{\@ifnextchar[{\basicenemono}{\basicenemono[133]}}%

\def\basicEnemono[#1]#2{\ENEMONO{#2}{}{#100}}%

\newcommand{\Enemono}{\@ifnextchar[{\basicEnemono}{\basicEnemono[133]}}%

\def\basicenemonO[#1]#2{\ENEMONO{}{#2}{#100}}%

\newcommand{\enemonO}{\@ifnextchar[{\basicenemonO}{\basicenemonO[133]}}%

\def\basiceneepi[#1]{\ENEEPI{}{}{#100}}%

\newcommand{\eneepi}{\@ifnextchar[{\basiceneepi}{\basiceneepi[133]}}%

\def\basicEneepi[#1]#2{\ENEEPI{#2}{}{#100}}%

\newcommand{\Eneepi}{\@ifnextchar[{\basicEneepi}{\basicEneepi[133]}}%

\def\basiceneepI[#1]#2{\ENEEPI{}{#2}{#100}}%

\newcommand{\eneepI}{\@ifnextchar[{\basiceneepI}{\basiceneepI[133]}}%

\def\basicenebimo[#1]{\ENEBIMO{}{}{#100}}%

\newcommand{\enebimo}{\@ifnextchar[{\basicenebimo}{\basicenebimo[133]}}%

\def\basicEnebimo[#1]#2{\ENEBIMO{#2}{}{#100}}%

\newcommand{\Enebimo}{\@ifnextchar[{\basicEnebimo}{\basicEnebimo[133]}}%

\def\basicenebimO[#1]#2{\ENEBIMO{}{#2}{#100}}%

\newcommand{\enebimO}{\@ifnextchar[{\basicenebimO}{\basicenebimO[133]}}%

\def\basiceneiso[#1]{\ENEAR{\cong}{}{#100}}%

\newcommand{\eneiso}{\@ifnextchar[{\basiceneiso}{\basiceneiso[133]}}%

\def\basicEneiso[#1]#2{\ENEAR{#2}{\cong}{#100}}%

\newcommand{\Eneiso}{\@ifnextchar[{\basicEneiso}{\basicEneiso[133]}}%

\def\basiceneisO[#1]#2{\ENEAR{\cong}{#2}{#100}}%

\newcommand{\eneisO}{\@ifnextchar[{\basiceneisO}{\basiceneisO[133]}}%

\def\basiceneeql[#1]{\ENEEQL{}{}{#100}}%

\newcommand{\eneeql}{\@ifnextchar[{\basiceneeql}{\basiceneeql[133]}}%

\def\basicEneeql[#1]#2{\ENEEQL{#2}{}{#100}}%

\newcommand{\Eneeql}{\@ifnextchar[{\basicEneeql}{\basicEneeql[133]}}%

\def\basiceneeqL[#1]#2{\ENEEQL{}{#2}{#100}}%

\newcommand{\eneeqL}{\@ifnextchar[{\basiceneeqL}{\basiceneeqL[133]}}%

\def\basicenebiar[#1]{\ENEBIAR{}{}{#100}}%

\newcommand{\enebiar}{\@ifnextchar[{\basicenebiar}{\basicenebiar[133]}}%

\def\basicEnebiar[#1]#2#3{\ENEBIAR{#2}{#3}{#100}}%

\newcommand{\Enebiar}{\@ifnextchar[{\basicEnebiar}{\basicEnebiar[133]}}%

\def\basicenebidist[#1]{\ENEBIDIST{}{}{#100}}%

\newcommand{\enebidist}{\@ifnextchar[{\basicenebidist}{\basicenebidist[133]}}%

\def\basicEnebidist[#1]#2#3{\ENEBIDIST{#2}{#3}{#100}}%

\newcommand{\Enebidist}{\@ifnextchar[{\basicEnebidist}{\basicEnebidist[133]}}%

\def\basiceneadjar[#1]{\ENEADJAR{}{}{#100}}%

\newcommand{\eneadjar}{\@ifnextchar[{\basiceneadjar}{\basiceneadjar[133]}}%

\def\basicEneadjar[#1]#2#3{\ENEADJAR{#2}{#3}{#100}}%

\newcommand{\Eneadjar}{\@ifnextchar[{\basicEneadjar}{\basicEneadjar[133]}}%

\def\basiceneadjdist[#1]{\ENEADJDIST{}{}{#100}}%

\newcommand{\eneadjdist}{\@ifnextchar[{\basiceneadjdist}{\basiceneadjdist[133]}}%

\def\basicEneadjdist[#1]#2#3{\ENEADJDIST{#2}{#3}{#100}}%

\newcommand{\Eneadjdist}{\@ifnextchar[{\basicEneadjdist}{\basicEneadjdist[133]}}%

\newcommand{\ESEAR}[3]{\testdiagrammode%
\Y=#3%
\divide\Y by 2%
\Z=\Y%
\divide\Z by 2%
\begin{picture}(0,0)%
\put(-\Y,\Z){\line(2,-1){#3}}%
\put(\Y,-\Z){\esehead}%
\truex{200}\truey{800}\truez{600}%
\put(\value{x},\value{x}){\makebox(0,\value{z})[l]{${#1}$}}%
\put(-\value{x},-\value{y}){\makebox(0,\value{z})[r]{${#2}$}}%
\end{picture}}%

\newcommand{\ESEDIST}[3]{\testdiagrammode%
\Y=#3%
\divide\Y by 2%
\Z=\Y%
\divide\Z by 2%
\begin{picture}(0,0)%
\put(-\Y,\Z){\line(2,-1){#3}}%
\put(\Y,-\Z){\esehead}%
\truex{400}%
\put(0,0){\circle{\value{x}}}%
\truex{200}\truey{800}\truez{600}%
\put(\value{x},\value{x}){\makebox(0,\value{z})[l]{${#1}$}}%
\put(-\value{x},-\value{y}){\makebox(0,\value{z})[r]{${#2}$}}%
\end{picture}}%

\newcommand{\ESEDOTAR}[3]{\testdiagrammode%
\truex{100}\truey{268}\truez{134}%
\Y=#3%
\divide\Y by 2%
\Z=\Y%
\divide\Z by 2%
\NUMBEROFDOTS=#3%
\divide\NUMBEROFDOTS by \value{y}%
\advance\NUMBEROFDOTS by 1%
\begin{picture}(0,0)%
\multiput(-\Y,\Z)(\value{y},-\value{z}){\NUMBEROFDOTS}%
{\circle*{\value{x}}}%
\put(\Y,-\Z){\esehead}%
\truex{200}\truey{800}\truez{600}%
\put(\value{x},\value{x}){\makebox(0,\value{z})[l]{${#1}$}}%
\put(-\value{x},-\value{y}){\makebox(0,\value{z})[r]{${#2}$}}%
\end{picture}}%

\newcommand{\ESEMONO}[3]{\testdiagrammode%
\Y=#3%
\divide\Y by 2%
\Z=\Y%
\divide\Z by 2%
\TrueTail%
\bimolength=#3%
\advance\bimolength by -\TrueMonoTail%
\monolength=\bimolength%
\advance\monolength by -\Y%
\secondmonolength=\monolength%
\divide\secondmonolength by 2%
\begin{picture}(0,0)%
\put(-\monolength,\secondmonolength){\line(2,-1){\bimolength}}%
\put(-\monolength,\secondmonolength){\esehead}%
\put(\Y,-\Z){\esehead}%
\truex{200}\truey{800}\truez{600}%
\put(\value{x},\value{x}){\makebox(0,\value{z})[l]{${#1}$}}%
\put(-\value{x},-\value{y}){\makebox(0,\value{z})[r]{${#2}$}}%
\end{picture}}%

\newcommand{\ESEEPI}[3]{\testdiagrammode%
\Y=#3%
\divide\Y by 2%
\Z=\Y%
\divide\Z by 2%
\TrueHead%
\bimolength=#3%
\advance\bimolength by -\TrueEpiHead%
\epilength=\bimolength%
\advance\epilength by -\Y%
\secondepilength=\epilength%
\divide\secondepilength by 2%
\begin{picture}(0,0)%
\put(-\Y,\Z){\line(2,-1){\bimolength}}%
\put(\epilength,-\secondepilength){\esehead}%
\put(\Y,-\Z){\esehead}%
\truex{200}\truey{800}\truez{600}%
\put(\value{x},\value{x}){\makebox(0,\value{z})[l]{${#1}$}}%
\put(-\value{x},-\value{y}){\makebox(0,\value{z})[r]{${#2}$}}%
\end{picture}}%

\newcommand{\ESEBIMO}[3]{\testdiagrammode%
\Y=#3%
\divide\Y by 2%
\Z=\Y%
\divide\Z by 2%
\TrueTail\TrueHead%
\bimolength=#3%
\advance\bimolength by -\TrueMonoTail%
\monolength=\bimolength%
\advance\monolength by -\Y%
\advance\bimolength by -\TrueEpiHead%
\epilength=\bimolength%
\advance\epilength by -\monolength%
\secondmonolength=\monolength%
\divide\secondmonolength by 2%
\secondepilength=\epilength%
\divide\secondepilength by 2%
\begin{picture}(0,0)%
\put(-\monolength,\secondmonolength){\line(2,-1){\bimolength}}%
\put(-\monolength,\secondmonolength){\esehead}%
\put(\epilength,-\secondepilength){\esehead}%
\put(\Y,-\Z){\esehead}%
\truex{200}\truey{800}\truez{600}%
\put(\value{x},\value{x}){\makebox(0,\value{z})[l]{${#1}$}}%
\put(-\value{x},-\value{y}){\makebox(0,\value{z})[r]{${#2}$}}%
\end{picture}}%

\newcommand{\ESEEQL}[3]{\testdiagrammode%
\Y=#3%
\divide\Y by 2%
\Z=\Y%
\divide\Z by 2%
\begin{picture}(0,0)%
\put(-\Y,\Z){\begin{picture}(0,0)%
\truex{44}\truey{89}%
\put(-\value{x},-\value{y}){\line(2,-1){#3}}%
\put(\value{x},\value{y}){\line(2,-1){#3}}%
\end{picture}}%
\truex{200}\truey{800}\truez{600}%
\put(\value{x},\value{x}){\makebox(0,\value{z})[l]{${#1}$}}%
\put(-\value{x},-\value{y}){\makebox(0,\value{z})[r]{${#2}$}}%
\end{picture}}%

\newcommand{\ESEBIAR}[3]{%
\Y=#3%
\divide\Y by 2%
\Z=\Y%
\divide\Z by 2%
\begin{picture}(0,0)%
\put(-\Y,\Z){\begin{picture}(0,0)%
\truex{156}\truey{313}%
\put(-\value{x},-\value{y}){\line(2,-1){#3}}%
\put(\value{x},\value{y}){\line(2,-1){#3}}%
\monolength=#3%
\advance\monolength by -\value{x}%
\epilength=#3%
\advance\epilength by \value{x}%
\secondmonolength=\Y%
\advance\secondmonolength by -\value{y}%
\secondepilength=\Y%
\advance\secondepilength by \value{y}%
\put(\monolength,-\secondepilength){\esehead}%
\put(\epilength,-\secondmonolength){\esehead}%
\end{picture}}
\truex{400}\truey{1000}\truez{600}%
\put(\value{x},\value{x}){\makebox(0,\value{z})[l]{${#1}$}}%
\put(-\value{x},-\value{y}){\makebox(0,\value{z})[r]{${#2}$}}%
\end{picture}}%

\newcommand{\ESEBIDIST}[3]{\testdiagrammode%
\Y=#3%
\divide\Y by 2%
\Z=\Y%
\divide\Z by 2%
\begin{picture}(0,0)%
\truex{156}\truey{313}\truez{400}%
\put(-\Y,\Z){\begin{picture}(0,0)%
\put(-\value{x},-\value{y}){\line(2,-1){#3}}%
\put(\value{x},\value{y}){\line(2,-1){#3}}%
\monolength=#3%
\advance\monolength by -\value{x}%
\epilength=#3%
\advance\epilength by \value{x}%
\secondmonolength=\Y%
\advance\secondmonolength by -\value{y}%
\secondepilength=\Y%
\advance\secondepilength by \value{y}%
\put(\monolength,-\secondepilength){\esehead}%
\put(\epilength,-\secondmonolength){\esehead}%
\end{picture}}
\put(\value{x},\value{y}){\circle{\value{z}}}%
\put(-\value{x},-\value{y}){\circle{\value{z}}}%
\truex{400}\truey{1000}\truez{600}%
\put(\value{x},\value{x}){\makebox(0,\value{z})[l]{${#1}$}}%
\put(-\value{x},-\value{y}){\makebox(0,\value{z})[r]{${#2}$}}%
\end{picture}}%

\newcommand{\ESEADJAR}[3]{\testdiagrammode%
\Y=#3%
\divide\Y by 2%
\Z=\Y%
\divide\Z by 2%
\begin{picture}(0,0)%
\put(-\Y,\Z){\begin{picture}(0,0)%
\truex{156}\truey{313}%
\monolength=#3%
\advance\monolength by -\value{x}%
\epilength=#3%
\advance\epilength by \value{x}%
\secondmonolength=\Y%
\advance\secondmonolength by -\value{y}%
\secondepilength=\Y%
\advance\secondepilength by \value{y}%
\put(-\value{x},-\value{y}){\line(2,-1){#3}}%
\put(\monolength,-\secondepilength){\esehead}%
\put(\epilength,-\secondmonolength){\line(-2,1){#3}}%
\put(\value{x},\value{y}){\wnwhead}%
\end{picture}}
\truex{400}\truey{1000}\truez{600}%
\put(\value{x},\value{x}){\makebox(0,\value{z})[l]{${#1}$}}%
\put(-\value{x},-\value{y}){\makebox(0,\value{z})[r]{${#2}$}}%
\end{picture}}%

\newcommand{\ESEADJDIST}[3]{\testdiagrammode%
\Y=#3%
\divide\Y by 2%
\Z=\Y%
\divide\Z by 2%
\begin{picture}(0,0)%
\truex{156}\truey{313}\truez{400}%
\put(-\Y,\Z){\begin{picture}(0,0)%
\monolength=#3%
\advance\monolength by -\value{x}%
\epilength=#3%
\advance\epilength by \value{x}%
\secondmonolength=\Y%
\advance\secondmonolength by -\value{y}%
\secondepilength=\Y%
\advance\secondepilength by \value{y}%
\put(-\value{x},-\value{y}){\line(2,-1){#3}}%
\put(\monolength,-\secondepilength){\esehead}%
\put(\epilength,-\secondmonolength){\line(-2,1){#3}}%
\put(\value{x},\value{y}){\wnwhead}%
\end{picture}}
\put(\value{x},\value{y}){\circle{\value{z}}}%
\put(-\value{x},-\value{y}){\circle{\value{z}}}%
\truex{400}\truey{1000}\truez{600}%
\put(\value{x},\value{x}){\makebox(0,\value{z})[l]{${#1}$}}%
\put(-\value{x},-\value{y}){\makebox(0,\value{z})[r]{${#2}$}}%
\end{picture}}%

\def\basicesear[#1]{\ESEAR{}{}{#100}}%

\newcommand{\esear}{\@ifnextchar[{\basicesear}{\basicesear[133]}}%

\def\basicEsear[#1]#2{\ESEAR{#2}{}{#100}}%

\newcommand{\Esear}{\@ifnextchar[{\basicEsear}{\basicEsear[133]}}%

\def\basiceseaR[#1]#2{\ESEAR{}{#2}{#100}}%

\newcommand{\eseaR}{\@ifnextchar[{\basiceseaR}{\basiceseaR[133]}}%

\def\basicesedist[#1]{\ESEDIST{}{}{#100}}%

\newcommand{\esedist}{\@ifnextchar[{\basicesedist}{\basicesedist[133]}}%

\def\basicEsedist[#1]#2{\ESEDIST{#2}{}{#100}}%

\newcommand{\Esedist}{\@ifnextchar[{\basicEsedist}{\basicEsedist[133]}}%

\def\basicesedisT[#1]#2{\ESEDIST{}{#2}{#100}}%

\newcommand{\esedisT}{\@ifnextchar[{\basicesedisT}{\basicesedisT[133]}}%

\def\basicesedotar[#1]{\ESEDOTAR{}{}{#100}}%

\newcommand{\esedotar}{\@ifnextchar[{\basicesedotar}{\basicesedotar[133]}}%

\def\basicEsedotar[#1]#2{\ESEDOTAR{#2}{}{#100}}%

\newcommand{\Esedotar}{\@ifnextchar[{\basicEsedotar}{\basicEsedotar[133]}}%

\def\basicesedotaR[#1]#2{\ESEDOTAR{}{#2}{#100}}%

\newcommand{\esedotaR}{\@ifnextchar[{\basicesedotaR}{\basicesedotaR[133]}}%

\def\basicesemono[#1]{\ESEMONO{}{}{#100}}%

\newcommand{\esemono}{\@ifnextchar[{\basicesemono}{\basicesemono[133]}}%

\def\basicEsemono[#1]#2{\ESEMONO{#2}{}{#100}}%

\newcommand{\Esemono}{\@ifnextchar[{\basicEsemono}{\basicEsemono[133]}}%

\def\basicesemonO[#1]#2{\ESEMONO{}{#2}{#100}}%

\newcommand{\esemonO}{\@ifnextchar[{\basicesemonO}{\basicesemonO[133]}}%

\def\basiceseepi[#1]{\ESEEPI{}{}{#100}}%

\newcommand{\eseepi}{\@ifnextchar[{\basiceseepi}{\basiceseepi[133]}}%

\def\basicEseepi[#1]#2{\ESEEPI{#2}{}{#100}}%

\newcommand{\Eseepi}{\@ifnextchar[{\basicEseepi}{\basicEseepi[133]}}%

\def\basiceseepI[#1]#2{\ESEEPI{}{#2}{#100}}%

\newcommand{\eseepI}{\@ifnextchar[{\basiceseepI}{\basiceseepI[133]}}%

\def\basicesebimo[#1]{\ESEBIMO{}{}{#100}}%

\newcommand{\esebimo}{\@ifnextchar[{\basicesebimo}{\basicesebimo[133]}}%

\def\basicEsebimo[#1]#2{\ESEBIMO{#2}{}{#100}}%

\newcommand{\Esebimo}{\@ifnextchar[{\basicEsebimo}{\basicEsebimo[133]}}%

\def\basicesebimO[#1]#2{\ESEBIMO{}{#2}{#100}}%

\newcommand{\esebimO}{\@ifnextchar[{\basicesebimO}{\basicesebimO[133]}}%

\def\basiceseiso[#1]{\ESEAR{\cong}{}{#100}}%

\newcommand{\eseiso}{\@ifnextchar[{\basiceseiso}{\basiceseiso[133]}}%

\def\basicEseiso[#1]#2{\ESEAR{#2}{\cong}{#100}}%

\newcommand{\Eseiso}{\@ifnextchar[{\basicEseiso}{\basicEseiso[133]}}%

\def\basiceseisO[#1]#2{\ESEAR{\cong}{#2}{#100}}%

\newcommand{\eseisO}{\@ifnextchar[{\basiceseisO}{\basiceseisO[133]}}%

\def\basiceseeql[#1]{\ESEEQL{}{}{#100}}%

\newcommand{\eseeql}{\@ifnextchar[{\basiceseeql}{\basiceseeql[133]}}%

\def\basicEseeql[#1]#2{\ESEEQL{#2}{}{#100}}%

\newcommand{\Eseeql}{\@ifnextchar[{\basicEseeql}{\basicEseeql[133]}}%

\def\basiceseeqL[#1]#2{\ESEEQL{}{#2}{#100}}%

\newcommand{\eseeqL}{\@ifnextchar[{\basiceseeqL}{\basiceseeqL[133]}}%

\def\basicesebiar[#1]{\ESEBIAR{}{}{#100}}%

\newcommand{\esebiar}{\@ifnextchar[{\basicesebiar}{\basicesebiar[133]}}%

\def\basicEsebiar[#1]#2#3{\ESEBIAR{#2}{#3}{#100}}%

\newcommand{\Esebiar}{\@ifnextchar[{\basicEsebiar}{\basicEsebiar[133]}}%

\def\basicesebidist[#1]{\ESEBIDIST{}{}{#100}}%

\newcommand{\esebidist}{\@ifnextchar[{\basicesebidist}{\basicesebidist[133]}}%

\def\basicEsebidist[#1]#2#3{\ESEBIDIST{#2}{#3}{#100}}%

\newcommand{\Esebidist}{\@ifnextchar[{\basicEsebidist}{\basicEsebidist[133]}}%

\def\basiceseadjar[#1]{\ESEADJAR{}{}{#100}}%

\newcommand{\eseadjar}{\@ifnextchar[{\basiceseadjar}{\basiceseadjar[133]}}%

\def\basicEseadjar[#1]#2#3{\ESEADJAR{#2}{#3}{#100}}%

\newcommand{\Eseadjar}{\@ifnextchar[{\basicEseadjar}{\basicEseadjar[133]}}%

\def\basiceseadjdist[#1]{\ESEADJDIST{}{}{#100}}%

\newcommand{\eseadjdist}{\@ifnextchar[{\basiceseadjdist}{\basiceseadjdist[133]}}%

\def\basicEseadjdist[#1]#2#3{\ESEADJDIST{#2}{#3}{#100}}%

\newcommand{\Eseadjdist}{\@ifnextchar[{\basicEseadjdist}{\basicEseadjdist[133]}}%

\newcommand{\WSWAR}[3]{\testdiagrammode%
\Y=#3%
\divide\Y by 2%
\Z=\Y%
\divide\Z by 2%
\begin{picture}(0,0)%
\put(\Y,\Z){\line(-2,-1){#3}}%
\put(-\Y,-\Z){\wswhead}%
\truex{200}\truey{800}\truez{600}%
\put(-\value{x},\value{x}){\makebox(0,\value{z})[r]{${#1}$}}%
\put(\value{x},-\value{y}){\makebox(0,\value{z})[l]{${#2}$}}%
\end{picture}}%

\newcommand{\WSWDIST}[3]{\testdiagrammode%
\Y=#3%
\divide\Y by 2%
\Z=\Y%
\divide\Z by 2%
\begin{picture}(0,0)%
\put(\Y,\Z){\line(-2,-1){#3}}%
\put(-\Y,-\Z){\wswhead}%
\truex{400}%
\put(0,0){\circle{\value{x}}}%
\truex{200}\truey{800}\truez{600}%
\put(-\value{x},\value{x}){\makebox(0,\value{z})[r]{${#1}$}}%
\put(\value{x},-\value{y}){\makebox(0,\value{z})[l]{${#2}$}}%
\end{picture}}%

\newcommand{\WSWDOTAR}[3]{\testdiagrammode%
\truex{100}\truey{268}\truez{134}%
\Y=#3%
\divide\Y by 2%
\Z=\Y%
\divide\Z by 2%
\NUMBEROFDOTS=#3%
\divide\NUMBEROFDOTS by \value{y}%
\advance\NUMBEROFDOTS by 1%
\begin{picture}(0,0)%
\multiput(\Y,\Z)(-\value{y},-\value{z}){\NUMBEROFDOTS}%
{\circle*{\value{x}}}%
\put(-\Y,-\Z){\wswhead}%
\truex{200}\truey{800}\truez{600}%
\put(-\value{x},\value{x}){\makebox(0,\value{z})[r]{${#1}$}}%
\put(\value{x},-\value{y}){\makebox(0,\value{z})[l]{${#2}$}}%
\end{picture}}%

\newcommand{\WSWMONO}[3]{\testdiagrammode%
\Y=#3%
\divide\Y by 2%
\Z=\Y%
\divide\Z by 2%
\TrueTail%
\bimolength=#3%
\advance\bimolength by -\TrueMonoTail%
\monolength=\bimolength%
\advance\monolength by -\Y%
\secondmonolength=\monolength%
\divide\secondmonolength by 2%
\begin{picture}(0,0)%
\put(\monolength,\secondmonolength){\line(-2,-1){\bimolength}}%
\put(\monolength,\secondmonolength){\wswhead}%
\put(-\Y,-\Z){\wswhead}%
\truex{200}\truey{800}\truez{600}%
\put(-\value{x},\value{x}){\makebox(0,\value{z})[r]{${#1}$}}%
\put(\value{x},-\value{y}){\makebox(0,\value{z})[l]{${#2}$}}%
\end{picture}}%

\newcommand{\WSWEPI}[3]{\testdiagrammode%
\Y=#3%
\divide\Y by 2%
\Z=\Y%
\divide\Z by 2%
\TrueHead%
\bimolength=#3%
\advance\bimolength by -\TrueEpiHead%
\epilength=\bimolength%
\advance\epilength by -\Y%
\secondepilength=\epilength%
\divide\secondepilength by 2%
\begin{picture}(0,0)%
\put(\Y,\Z){\line(-2,-1){\bimolength}}%
\put(-\epilength,-\secondepilength){\wswhead}%
\put(-\Y,-\Z){\wswhead}%
\truex{200}\truey{800}\truez{600}%
\put(-\value{x},\value{x}){\makebox(0,\value{z})[r]{${#1}$}}%
\put(\value{x},-\value{y}){\makebox(0,\value{z})[l]{${#2}$}}%
\end{picture}}%

\newcommand{\WSWBIMO}[3]{\testdiagrammode%
\Y=#3%
\divide\Y by 2%
\Z=\Y%
\divide\Z by 2%
\TrueTail\TrueHead%
\bimolength=#3%
\advance\bimolength by -\TrueMonoTail%
\monolength=\bimolength%
\advance\monolength by -\Y%
\advance\bimolength by -\TrueEpiHead%
\epilength=\bimolength%
\advance\epilength by -\monolength%
\secondmonolength=\monolength%
\divide\secondmonolength by 2%
\secondepilength=\epilength%
\divide\secondepilength by 2%
\begin{picture}(0,0)%
\put(\monolength,\secondmonolength){\line(-2,-1){\bimolength}}%
\put(\monolength,\secondmonolength){\wswhead}%
\put(-\epilength,-\secondepilength){\wswhead}%
\put(-\Y,-\Z){\wswhead}%
\truex{200}\truey{800}\truez{600}%
\put(-\value{x},\value{x}){\makebox(0,\value{z})[r]{${#1}$}}%
\put(\value{x},-\value{y}){\makebox(0,\value{z})[l]{${#2}$}}%
\end{picture}}%

\newcommand{\WSWBIAR}[3]{\testdiagrammode%
\Y=#3%
\divide\Y by 2%
\Z=\Y%
\divide\Z by 2%
\begin{picture}(0,0)%
\put(\Y,\Z){\begin{picture}(0,0)%
\truex{156}\truey{313}%
\put(-\value{x},\value{y}){\line(-2,-1){#3}}%
\put(\value{x},-\value{y}){\line(-2,-1){#3}}%
\monolength=#3%
\advance\monolength by -\value{x}%
\epilength=#3%
\advance\epilength by \value{x}%
\secondmonolength=\Y%
\advance\secondmonolength by -\value{y}%
\secondepilength=\Y%
\advance\secondepilength by \value{y}%
\put(-\monolength,-\secondepilength){\wswhead}%
\put(-\epilength,-\secondmonolength){\wswhead}%
\end{picture}}
\truex{300}\truey{1000}\truez{600}%
\put(-\value{x},\value{x}){\makebox(0,\value{z})[r]{${#1}$}}%
\put(\value{x},-\value{y}){\makebox(0,\value{z})[l]{${#2}$}}%
\end{picture}}%

\newcommand{\WSWBIDIST}[3]{\testdiagrammode%
\Y=#3%
\divide\Y by 2%
\Z=\Y%
\divide\Z by 2%
\begin{picture}(0,0)%
\truex{156}\truey{313}\truez{400}%
\put(\Y,\Z){\begin{picture}(0,0)%
\put(-\value{x},\value{y}){\line(-2,-1){#3}}%
\put(\value{x},-\value{y}){\line(-2,-1){#3}}%
\monolength=#3%
\advance\monolength by -\value{x}%
\epilength=#3%
\advance\epilength by \value{x}%
\secondmonolength=\Y%
\advance\secondmonolength by -\value{y}%
\secondepilength=\Y%
\advance\secondepilength by \value{y}%
\put(-\monolength,-\secondepilength){\wswhead}%
\put(-\epilength,-\secondmonolength){\wswhead}%
\end{picture}}
\put(-\value{x},\value{y}){\circle{\value{z}}}%
\put(\value{x},-\value{y}){\circle{\value{z}}}%
\truex{300}\truey{1000}\truez{600}%
\put(-\value{x},\value{x}){\makebox(0,\value{z})[r]{${#1}$}}%
\put(\value{x},-\value{y}){\makebox(0,\value{z})[l]{${#2}$}}%
\end{picture}}%

\newcommand{\WSWADJAR}[3]{\testdiagrammode%
\Y=#3%
\divide\Y by 2%
\Z=\Y%
\divide\Z by 2%
\begin{picture}(0,0)%
\put(\Y,\Z){\begin{picture}(0,0)%
\truex{156}\truey{313}%
\monolength=#3%
\advance\monolength by -\value{x}%
\epilength=#3%
\advance\epilength by \value{x}%
\secondmonolength=\Y%
\advance\secondmonolength by -\value{y}%
\secondepilength=\Y%
\advance\secondepilength by \value{y}%
\put(\value{x},-\value{y}){\line(-2,-1){#3}}%
\put(-\monolength,-\secondepilength){\wswhead}%
\put(-\epilength,-\secondmonolength){\line(2,1){#3}}%
\put(-\value{x},\value{y}){\enehead}%
\end{picture}}
\truex{300}\truey{1000}\truez{600}%
\put(-\value{x},\value{x}){\makebox(0,\value{z})[r]{${#1}$}}%
\put(\value{x},-\value{y}){\makebox(0,\value{z})[l]{${#2}$}}%
\end{picture}}%

\newcommand{\WSWADJDIST}[3]{\testdiagrammode%
\Y=#3%
\divide\Y by 2%
\Z=\Y%
\divide\Z by 2%
\begin{picture}(0,0)%
\truex{156}\truey{313}\truez{400}%
\put(\Y,\Z){\begin{picture}(0,0)%
\monolength=#3%
\advance\monolength by -\value{x}%
\epilength=#3%
\advance\epilength by \value{x}%
\secondmonolength=\Y%
\advance\secondmonolength by -\value{y}%
\secondepilength=\Y%
\advance\secondepilength by \value{y}%
\put(\value{x},-\value{y}){\line(-2,-1){#3}}%
\put(-\monolength,-\secondepilength){\wswhead}%
\put(-\epilength,-\secondmonolength){\line(2,1){#3}}%
\put(-\value{x},\value{y}){\enehead}%
\end{picture}}
\put(-\value{x},\value{y}){\circle{\value{z}}}%
\put(\value{x},-\value{y}){\circle{\value{z}}}%
\truex{300}\truey{1000}\truez{600}%
\put(-\value{x},\value{x}){\makebox(0,\value{z})[r]{${#1}$}}%
\put(\value{x},-\value{y}){\makebox(0,\value{z})[l]{${#2}$}}%
\end{picture}}%

\def\basicwswar[#1]{\WSWAR{}{}{#100}}%

\newcommand{\wswar}{\@ifnextchar[{\basicwswar}{\basicwswar[133]}}%

\def\basicWswar[#1]#2{\WSWAR{#2}{}{#100}}%

\newcommand{\Wswar}{\@ifnextchar[{\basicWswar}{\basicWswar[133]}}%

\def\basicwswaR[#1]#2{\WSWAR{}{#2}{#100}}%

\newcommand{\wswaR}{\@ifnextchar[{\basicwswaR}{\basicwswaR[133]}}%

\def\basicwswdist[#1]{\WSWDIST{}{}{#100}}%

\newcommand{\wswdist}{\@ifnextchar[{\basicwswdist}{\basicwswdist[133]}}%

\def\basicWswdist[#1]#2{\WSWDIST{#2}{}{#100}}%

\newcommand{\Wswdist}{\@ifnextchar[{\basicWswdist}{\basicWswdist[133]}}%

\def\basicwswdisT[#1]#2{\WSWDIST{}{#2}{#100}}%

\newcommand{\wswdisT}{\@ifnextchar[{\basicwswdisT}{\basicwswdisT[133]}}%

\def\basicwswdotar[#1]{\WSWDOTAR{}{}{#100}}%

\newcommand{\wswdotar}{\@ifnextchar[{\basicwswdotar}{\basicwswdotar[133]}}%

\def\basicWswdotar[#1]#2{\WSWDOTAR{#2}{}{#100}}%

\newcommand{\Wswdotar}{\@ifnextchar[{\basicWswdotar}{\basicWswdotar[133]}}%

\def\basicwswdotaR[#1]#2{\WSWDOTAR{}{#2}{#100}}%

\newcommand{\wswdotaR}{\@ifnextchar[{\basicwswdotaR}{\basicwswdotaR[133]}}%

\def\basicwswmono[#1]{\WSWMONO{}{}{#100}}%

\newcommand{\wswmono}{\@ifnextchar[{\basicwswmono}{\basicwswmono[133]}}%

\def\basicWswmono[#1]#2{\WSWMONO{#2}{}{#100}}%

\newcommand{\Wswmono}{\@ifnextchar[{\basicWswmono}{\basicWswmono[133]}}%

\def\basicwswmonO[#1]#2{\WSWMONO{}{#2}{#100}}%

\newcommand{\wswmonO}{\@ifnextchar[{\basicwswmonO}{\basicwswmonO[133]}}%

\def\basicwswepi[#1]{\WSWEPI{}{}{#100}}%

\newcommand{\wswepi}{\@ifnextchar[{\basicwswepi}{\basicwswepi[133]}}%

\def\basicWswepi[#1]#2{\WSWEPI{#2}{}{#100}}%

\newcommand{\Wswepi}{\@ifnextchar[{\basicWswepi}{\basicWswepi[133]}}%

\def\basicwswepI[#1]#2{\WSWEPI{}{#2}{#100}}%

\newcommand{\wswepI}{\@ifnextchar[{\basicwswepI}{\basicwswepI[133]}}%

\def\basicwswbimo[#1]{\WSWBIMO{}{}{#100}}%

\newcommand{\wswbimo}{\@ifnextchar[{\basicwswbimo}{\basicwswbimo[133]}}%

\def\basicWswbimo[#1]#2{\WSWBIMO{#2}{}{#100}}%

\newcommand{\Wswbimo}{\@ifnextchar[{\basicWswbimo}{\basicWswbimo[133]}}%

\def\basicwswbimO[#1]#2{\WSWBIMO{}{#2}{#100}}%

\newcommand{\wswbimO}{\@ifnextchar[{\basicwswbimO}{\basicwswbimO[133]}}%

\def\basicwswiso[#1]{\WSWAR{\cong}{}{#100}}%

\newcommand{\wswiso}{\@ifnextchar[{\basicwswiso}{\basicwswiso[133]}}%

\def\basicWswiso[#1]#2{\WSWAR{#2}{\cong}{#100}}%

\newcommand{\Wswiso}{\@ifnextchar[{\basicWswiso}{\basicWswiso[133]}}%

\def\basicwswisO[#1]#2{\WSWAR{\cong}{#2}{#100}}%

\newcommand{\wswisO}{\@ifnextchar[{\basicwswisO}{\basicwswisO[133]}}%

\def\basicwswbiar[#1]{\WSWBIAR{}{}{#100}}%

\newcommand{\wswbiar}{\@ifnextchar[{\basicwswbiar}{\basicwswbiar[133]}}%

\def\basicWswbiar[#1]#2#3{\WSWBIAR{#2}{#3}{#100}}%

\newcommand{\Wswbiar}{\@ifnextchar[{\basicWswbiar}{\basicWswbiar[133]}}%

\def\basicwswbidist[#1]{\WSWBIDIST{}{}{#100}}%

\newcommand{\wswbidist}{\@ifnextchar[{\basicwswbidist}{\basicwswbidist[133]}}%

\def\basicWswbidist[#1]#2#3{\WSWBIDIST{#2}{#3}{#100}}%

\newcommand{\Wswbidist}{\@ifnextchar[{\basicWswbidist}{\basicWswbidist[133]}}%

\def\basicwswadjar[#1]{\WSWADJAR{}{}{#100}}%

\newcommand{\wswadjar}{\@ifnextchar[{\basicwswadjar}{\basicwswadjar[133]}}%

\def\basicWswadjar[#1]#2#3{\WSWADJAR{#2}{#3}{#100}}%

\newcommand{\Wswadjar}{\@ifnextchar[{\basicWswadjar}{\basicWswadjar[133]}}%

\def\basicwswadjdist[#1]{\WSWADJDIST{}{}{#100}}%

\newcommand{\wswadjdist}{\@ifnextchar[{\basicwswadjdist}{\basicwswadjdist[133]}}%

\def\basicWswadjdist[#1]#2#3{\WSWADJDIST{#2}{#3}{#100}}%

\newcommand{\Wswadjdist}{\@ifnextchar[{\basicWswadjdist}{\basicWswadjdist[133]}}%

\newcommand{\WNWAR}[3]{\testdiagrammode%
\Y=#3%
\divide\Y by 2%
\Z=\Y%
\divide\Z by 2%
\begin{picture}(0,0)%
\put(\Y,-\Z){\line(-2,1){#3}}%
\put(-\Y,\Z){\wnwhead}%
\truex{200}\truey{800}\truez{600}%
\put(\value{x},\value{x}){\makebox(0,\value{z})[l]{${#1}$}}%
\put(-\value{x},-\value{y}){\makebox(0,\value{z})[r]{${#2}$}}%
\end{picture}}%

\newcommand{\WNWDIST}[3]{\testdiagrammode%
\Y=#3%
\divide\Y by 2%
\Z=\Y%
\divide\Z by 2%
\begin{picture}(0,0)%
\put(\Y,-\Z){\line(-2,1){#3}}%
\put(-\Y,\Z){\wnwhead}%
\truex{400}%
\put(0,0){\circle{\value{x}}}%
\truex{200}\truey{800}\truez{600}%
\put(\value{x},\value{x}){\makebox(0,\value{z})[l]{${#1}$}}%
\put(-\value{x},-\value{y}){\makebox(0,\value{z})[r]{${#2}$}}%
\end{picture}}%

\newcommand{\WNWDOTAR}[3]{\testdiagrammode%
\truex{100}\truey{268}\truez{134}%
\Y=#3%
\divide\Y by 2%
\Z=\Y%
\divide\Z by 2%
\NUMBEROFDOTS=#3%
\divide\NUMBEROFDOTS by \value{y}%
\advance\NUMBEROFDOTS by 1%
\begin{picture}(0,0)%
\multiput(\Y,-\Z)(-\value{y},\value{z}){\NUMBEROFDOTS}%
{\circle*{\value{x}}}%
\put(-\Y,\Z){\wnwhead}%
\truex{200}\truey{800}\truez{600}%
\put(\value{x},\value{x}){\makebox(0,\value{z})[l]{${#1}$}}%
\put(-\value{x},-\value{y}){\makebox(0,\value{z})[r]{${#2}$}}%
\end{picture}}%

\newcommand{\WNWMONO}[3]{\testdiagrammode%
\Y=#3%
\divide\Y by 2%
\Z=\Y%
\divide\Z by 2%
\TrueTail%
\bimolength=#3%
\advance\bimolength by -\TrueMonoTail%
\monolength=\bimolength%
\advance\monolength by -\Y%
\secondmonolength=\monolength%
\divide\secondmonolength by 2%
\begin{picture}(0,0)%
\put(\monolength,-\secondmonolength){\line(-2,1){\bimolength}}%
\put(\monolength,-\secondmonolength){\wnwhead}%
\put(-\Y,\Z){\wnwhead}%
\truex{200}\truey{800}\truez{600}%
\put(\value{x},\value{x}){\makebox(0,\value{z})[l]{${#1}$}}%
\put(-\value{x},-\value{y}){\makebox(0,\value{z})[r]{${#2}$}}%
\end{picture}}%

\newcommand{\WNWEPI}[3]{\testdiagrammode%
\Y=#3%
\divide\Y by 2%
\Z=\Y%
\divide\Z by 2%
\TrueHead%
\bimolength=#3%
\advance\bimolength by -\TrueEpiHead%
\epilength=\bimolength%
\advance\epilength by -\Y%
\secondepilength=\epilength%
\divide\secondepilength by 2%
\begin{picture}(0,0)%
\put(\Y,-\Z){\line(-2,1){\bimolength}}%
\put(-\epilength,\secondepilength){\wnwhead}%
\put(-\Y,\Z){\wnwhead}%
\truex{200}\truey{800}\truez{600}%
\put(\value{x},\value{x}){\makebox(0,\value{z})[l]{${#1}$}}%
\put(-\value{x},-\value{y}){\makebox(0,\value{z})[r]{${#2}$}}%
\end{picture}}%

\newcommand{\WNWBIMO}[3]{\testdiagrammode%
\Y=#3%
\divide\Y by 2%
\Z=\Y%
\divide\Z by 2%
\TrueTail\TrueHead%
\bimolength=#3%
\advance\bimolength by -\TrueMonoTail%
\monolength=\bimolength%
\advance\monolength by -\Y%
\advance\bimolength by -\TrueEpiHead%
\epilength=\bimolength%
\advance\epilength by -\monolength%
\secondmonolength=\monolength%
\divide\secondmonolength by 2%
\secondepilength=\epilength%
\divide\secondepilength by 2%
\begin{picture}(0,0)%
\put(\monolength,-\secondmonolength){\line(-2,1){\bimolength}}%
\put(\monolength,-\secondmonolength){\wnwhead}%
\put(-\epilength,\secondepilength){\wnwhead}%
\put(-\Y,\Z){\wnwhead}%
\truex{200}\truey{800}\truez{600}%
\put(\value{x},\value{x}){\makebox(0,\value{z})[l]{${#1}$}}%
\put(-\value{x},-\value{y}){\makebox(0,\value{z})[r]{${#2}$}}%
\end{picture}}%

\newcommand{\WNWBIAR}[3]{\testdiagrammode%
\Y=#3%
\divide\Y by 2%
\Z=\Y%
\divide\Z by 2%
\begin{picture}(0,0)%
\put(\Y,-\Z){\begin{picture}(0,0)%
\truex{156}\truey{313}%
\put(-\value{x},-\value{y}){\line(-2,1){#3}}%
\put(\value{x},\value{y}){\line(-2,1){#3}}%
\monolength=#3%
\advance\monolength by -\value{x}%
\epilength=#3%
\advance\epilength by \value{x}%
\secondmonolength=\Y%
\advance\secondmonolength by -\value{y}%
\secondepilength=\Y%
\advance\secondepilength by \value{y}%
\put(-\monolength,\secondepilength){\wnwhead}%
\put(-\epilength,\secondmonolength){\wnwhead}%
\end{picture}}
\truex{400}\truey{1000}\truez{600}%
\put(\value{x},\value{x}){\makebox(0,\value{z})[l]{${#1}$}}%
\put(-\value{x},-\value{y}){\makebox(0,\value{z})[r]{${#2}$}}%
\end{picture}}%

\newcommand{\WNWBIDIST}[3]{\testdiagrammode%
\Y=#3%
\divide\Y by 2%
\Z=\Y%
\divide\Z by 2%
\begin{picture}(0,0)%
\truex{156}\truey{313}\truez{400}%
\put(\Y,-\Z){\begin{picture}(0,0)%
\put(-\value{x},-\value{y}){\line(-2,1){#3}}%
\put(\value{x},\value{y}){\line(-2,1){#3}}%
\monolength=#3%
\advance\monolength by -\value{x}%
\epilength=#3%
\advance\epilength by \value{x}%
\secondmonolength=\Y%
\advance\secondmonolength by -\value{y}%
\secondepilength=\Y%
\advance\secondepilength by \value{y}%
\put(-\monolength,\secondepilength){\wnwhead}%
\put(-\epilength,\secondmonolength){\wnwhead}%
\end{picture}}
\put(\value{x},\value{y}){\circle{\value{z}}}%
\put(-\value{x},-\value{y}){\circle{\value{z}}}%
\truex{400}\truey{1000}\truez{600}%
\put(\value{x},\value{x}){\makebox(0,\value{z})[l]{${#1}$}}%
\put(-\value{x},-\value{y}){\makebox(0,\value{z})[r]{${#2}$}}%
\end{picture}}%

\newcommand{\WNWADJAR}[3]{\testdiagrammode%
\Y=#3%
\divide\Y by 2%
\Z=\Y%
\divide\Z by 2%
\begin{picture}(0,0)%
\put(\Y,-\Z){\begin{picture}(0,0)%
\truex{156}\truey{313}%
\monolength=#3%
\advance\monolength by -\value{x}%
\epilength=#3%
\advance\epilength by \value{x}%
\secondmonolength=\Y%
\advance\secondmonolength by -\value{y}%
\secondepilength=\Y%
\advance\secondepilength by \value{y}%
\put(-\value{x},-\value{y}){\line(-2,1){#3}}%
\put(-\epilength,\secondmonolength){\wnwhead}%
\put(-\monolength,\secondepilength){\line(2,-1){#3}}%
\put(\value{x},\value{y}){\esehead}%
\end{picture}}
\truex{400}\truey{1000}\truez{600}%
\put(\value{x},\value{x}){\makebox(0,\value{z})[l]{${#1}$}}%
\put(-\value{x},-\value{y}){\makebox(0,\value{z})[r]{${#2}$}}%
\end{picture}}%

\newcommand{\WNWADJDIST}[3]{\testdiagrammode%
\Y=#3%
\divide\Y by 2%
\Z=\Y%
\divide\Z by 2%
\begin{picture}(0,0)%
\truex{156}\truey{313}\truez{400}%
\put(\Y,-\Z){\begin{picture}(0,0)%
\monolength=#3%
\advance\monolength by -\value{x}%
\epilength=#3%
\advance\epilength by \value{x}%
\secondmonolength=\Y%
\advance\secondmonolength by -\value{y}%
\secondepilength=\Y%
\advance\secondepilength by \value{y}%
\put(-\value{x},-\value{y}){\line(-2,1){#3}}%
\put(-\epilength,\secondmonolength){\wnwhead}%
\put(-\monolength,\secondepilength){\line(2,-1){#3}}%
\put(\value{x},\value{y}){\esehead}%
\end{picture}}
\put(\value{x},\value{y}){\circle{\value{z}}}%
\put(-\value{x},-\value{y}){\circle{\value{z}}}%
\truex{400}\truey{1000}\truez{600}%
\put(\value{x},\value{x}){\makebox(0,\value{z})[l]{${#1}$}}%
\put(-\value{x},-\value{y}){\makebox(0,\value{z})[r]{${#2}$}}%
\end{picture}}%

\def\basicwnwar[#1]{\WNWAR{}{}{#100}}%

\newcommand{\wnwar}{\@ifnextchar[{\basicwnwar}{\basicwnwar[133]}}%

\def\basicWnwar[#1]#2{\WNWAR{#2}{}{#100}}%

\newcommand{\Wnwar}{\@ifnextchar[{\basicWnwar}{\basicWnwar[133]}}%

\def\basicwnwaR[#1]#2{\WNWAR{}{#2}{#100}}%

\newcommand{\wnwaR}{\@ifnextchar[{\basicwnwaR}{\basicwnwaR[133]}}%

\def\basicwnwdist[#1]{\WNWDIST{}{}{#100}}%

\newcommand{\wnwdist}{\@ifnextchar[{\basicwnwdist}{\basicwnwdist[133]}}%

\def\basicWnwdist[#1]#2{\WNWDIST{#2}{}{#100}}%

\newcommand{\Wnwdist}{\@ifnextchar[{\basicWnwdist}{\basicWnwdist[133]}}%

\def\basicwnwdisT[#1]#2{\WNWDIST{}{#2}{#100}}%

\newcommand{\wnwdisT}{\@ifnextchar[{\basicwnwdisT}{\basicwnwdisT[133]}}%

\def\basicwnwdotar[#1]{\WNWDOTAR{}{}{#100}}%

\newcommand{\wnwdotar}{\@ifnextchar[{\basicwnwdotar}{\basicwnwdotar[133]}}%

\def\basicWnwdotar[#1]#2{\WNWDOTAR{#2}{}{#100}}%

\newcommand{\Wnwdotar}{\@ifnextchar[{\basicWnwdotar}{\basicWnwdotar[133]}}%

\def\basicwnwdotaR[#1]#2{\WNWDOTAR{}{#2}{#100}}%

\newcommand{\wnwdotaR}{\@ifnextchar[{\basicwnwdotaR}{\basicwnwdotaR[133]}}%

\def\basicwnwmono[#1]{\WNWMONO{}{}{#100}}%

\newcommand{\wnwmono}{\@ifnextchar[{\basicwnwmono}{\basicwnwmono[133]}}%

\def\basicWnwmono[#1]#2{\WNWMONO{#2}{}{#100}}%

\newcommand{\Wnwmono}{\@ifnextchar[{\basicWnwmono}{\basicWnwmono[133]}}%

\def\basicwnwmonO[#1]#2{\WNWMONO{}{#2}{#100}}%

\newcommand{\wnwmonO}{\@ifnextchar[{\basicwnwmonO}{\basicwnwmonO[133]}}%

\def\basicwnwepi[#1]{\WNWEPI{}{}{#100}}%

\newcommand{\wnwepi}{\@ifnextchar[{\basicwnwepi}{\basicwnwepi[133]}}%

\def\basicWnwepi[#1]#2{\WNWEPI{#2}{}{#100}}%

\newcommand{\Wnwepi}{\@ifnextchar[{\basicWnwepi}{\basicWnwepi[133]}}%

\def\basicwnwepI[#1]#2{\WNWEPI{}{#2}{#100}}%

\newcommand{\wnwepI}{\@ifnextchar[{\basicwnwepI}{\basicwnwepI[133]}}%

\def\basicwnwbimo[#1]{\WNWBIMO{}{}{#100}}%

\newcommand{\wnwbimo}{\@ifnextchar[{\basicwnwbimo}{\basicwnwbimo[133]}}%

\def\basicWnwbimo[#1]#2{\WNWBIMO{#2}{}{#100}}%

\newcommand{\Wnwbimo}{\@ifnextchar[{\basicWnwbimo}{\basicWnwbimo[133]}}%

\def\basicwnwbimO[#1]#2{\WNWBIMO{}{#2}{#100}}%

\newcommand{\wnwbimO}{\@ifnextchar[{\basicwnwbimO}{\basicwnwbimO[133]}}%

\def\basicwnwiso[#1]{\WNWAR{\cong}{}{#100}}%

\newcommand{\wnwiso}{\@ifnextchar[{\basicwnwiso}{\basicwnwiso[133]}}%

\def\basicWnwiso[#1]#2{\WNWAR{#2}{\cong}{#100}}%

\newcommand{\Wnwiso}{\@ifnextchar[{\basicWnwiso}{\basicWnwiso[133]}}%

\def\basicwnwisO[#1]#2{\WNWAR{\cong}{#2}{#100}}%

\newcommand{\wnwisO}{\@ifnextchar[{\basicwnwisO}{\basicwnwisO[133]}}%

\def\basicwnwbiar[#1]{\WNWBIAR{}{}{#100}}%

\newcommand{\wnwbiar}{\@ifnextchar[{\basicwnwbiar}{\basicwnwbiar[133]}}%

\def\basicWnwbiar[#1]#2#3{\WNWBIAR{#2}{#3}{#100}}%

\newcommand{\Wnwbiar}{\@ifnextchar[{\basicWnwbiar}{\basicWnwbiar[133]}}%

\def\basicwnwbidist[#1]{\WNWBIDIST{}{}{#100}}%

\newcommand{\wnwbidist}{\@ifnextchar[{\basicwnwbidist}{\basicwnwbidist[133]}}%

\def\basicWnwbidist[#1]#2#3{\WNWBIDIST{#2}{#3}{#100}}%

\newcommand{\Wnwbidist}{\@ifnextchar[{\basicWnwbidist}{\basicWnwbidist[133]}}%

\def\basicwnwadjar[#1]{\WNWADJAR{}{}{#100}}%

\newcommand{\wnwadjar}{\@ifnextchar[{\basicwnwadjar}{\basicwnwadjar[133]}}%

\def\basicWnwadjar[#1]#2#3{\WNWADJAR{#2}{#3}{#100}}%

\newcommand{\Wnwadjar}{\@ifnextchar[{\basicWnwadjar}{\basicWnwadjar[133]}}%

\def\basicwnwadjdist[#1]{\WNWADJDIST{}{}{#100}}%

\newcommand{\wnwadjdist}{\@ifnextchar[{\basicwnwadjdist}{\basicwnwadjdist[133]}}%

\def\basicWnwadjdist[#1]#2#3{\WNWADJDIST{#2}{#3}{#100}}%

\newcommand{\Wnwadjdist}{\@ifnextchar[{\basicWnwadjdist}{\basicWnwadjdist[133]}}%

\newcommand{\NNEAR}[3]{\testdiagrammode%
\Z=#3%
\divide\Z by 2%
\begin{picture}(0,0)%
\put(-\Z,-#3){\line(1,2){#3}}%
\put(\Z,#3){\nnehead}%
\truex{200}\truey{800}\truez{600}%
\put(-\value{x},\value{x}){\makebox(0,\value{z})[r]{${#1}$}}%
\put(\value{x},-\value{y}){\makebox(0,\value{z})[l]{${#2}$}}%
\end{picture}}%

\newcommand{\NNEDIST}[3]{\testdiagrammode%
\Z=#3%
\divide\Z by 2%
\begin{picture}(0,0)%
\put(-\Z,-#3){\line(1,2){#3}}%
\put(\Z,#3){\nnehead}%
\truex{400}%
\put(0,0){\circle{\value{x}}}%
\truex{200}\truey{800}\truez{600}%
\put(-\value{x},\value{x}){\makebox(0,\value{z})[r]{${#1}$}}%
\put(\value{x},-\value{y}){\makebox(0,\value{z})[l]{${#2}$}}%
\end{picture}}%

\newcommand{\NNEDOTAR}[3]{\testdiagrammode%
\truex{100}\truey{268}\truez{134}%
\Z=#3%
\divide\Z by 2%
\NUMBEROFDOTS=#3%
\divide\NUMBEROFDOTS by \value{z}%
\advance\NUMBEROFDOTS by 1%
\begin{picture}(0,0)%
\multiput(-\Z,-#3)(\value{z},\value{y}){\NUMBEROFDOTS}%
{\circle*{\value{x}}}%
\put(\Z,#3){\nnehead}%
\truex{200}\truey{800}\truez{600}%
\put(-\value{x},\value{x}){\makebox(0,\value{z})[r]{${#1}$}}%
\put(\value{x},-\value{y}){\makebox(0,\value{z})[l]{${#2}$}}%
\end{picture}}%

\newcommand{\NNEMONO}[3]{\testdiagrammode%
\Z=#3%
\divide\Z by 2%
\truetaiL%
\bimolength=#3%
\advance\bimolength by -\truemonotaiL%
\monolength=\bimolength%
\advance\monolength by -\Z%
\secondmonolength=\monolength%
\multiply\secondmonolength by 2%
\begin{picture}(0,0)%
\put(-\monolength,-\secondmonolength){\line(1,2){\bimolength}}%
\put(-\monolength,-\secondmonolength){\nnehead}%
\put(\Z,#3){\nnehead}%
\truex{200}\truey{800}\truez{600}%
\put(-\value{x},\value{x}){\makebox(0,\value{z})[r]{${#1}$}}%
\put(\value{x},-\value{y}){\makebox(0,\value{z})[l]{${#2}$}}%
\end{picture}}%

\newcommand{\NNEEPI}[3]{\testdiagrammode%
\Z=#3%
\divide\Z by 2%
\trueheaD%
\bimolength=#3%
\advance\bimolength by -\trueepiheaD%
\epilength=\bimolength%
\advance\epilength by -\Z%
\secondepilength=\epilength%
\multiply\secondepilength by 2%
\begin{picture}(0,0)%
\put(-\Z,-#3){\line(1,2){\bimolength}}%
\put(\epilength,\secondepilength){\nnehead}%
\put(\Z,#3){\nnehead}%
\truex{200}\truey{800}\truez{600}%
\put(-\value{x},\value{x}){\makebox(0,\value{z})[r]{${#1}$}}%
\put(\value{x},-\value{y}){\makebox(0,\value{z})[l]{${#2}$}}%
\end{picture}}%

\newcommand{\NNEBIMO}[3]{\testdiagrammode%
\Z=#3%
\divide\Z by 2%
\truetaiL\trueheaD%
\bimolength=#3%
\advance\bimolength by -\truemonotaiL%
\monolength=\bimolength%
\advance\monolength by -\Z%
\advance\bimolength by -\trueepiheaD%
\epilength=\bimolength%
\advance\epilength by -\monolength%
\secondmonolength=\monolength%
\multiply\secondmonolength by 2%
\secondepilength=\epilength%
\multiply\secondepilength by 2%
\begin{picture}(0,0)%
\put(-\monolength,-\secondmonolength){\line(1,2){\bimolength}}%
\put(-\monolength,-\secondmonolength){\nnehead}%
\put(\epilength,\secondepilength){\nnehead}%
\put(\Z,#3){\nnehead}%
\truex{200}\truey{800}\truez{600}%
\put(-\value{x},\value{x}){\makebox(0,\value{z})[r]{${#1}$}}%
\put(\value{x},-\value{y}){\makebox(0,\value{z})[l]{${#2}$}}%
\end{picture}}%

\newcommand{\NNEEQL}[3]{\testdiagrammode%
\Z=#3%
\divide\Z by 2%
\begin{picture}(0,0)%
\put(-\Z,-#3){\begin{picture}(0,0)%
\truex{44}\truey{89}%
\put(-\value{y},\value{x}){\line(1,2){#3}}%
\put(\value{y},-\value{x}){\line(1,2){#3}}%
\end{picture}}%
\truex{200}\truey{800}\truez{600}%
\put(-\value{x},\value{x}){\makebox(0,\value{z})[r]{${#1}$}}%
\put(\value{x},-\value{y}){\makebox(0,\value{z})[l]{${#2}$}}%
\end{picture}}%

\newcommand{\NNEBIAR}[3]{\testdiagrammode%
\Y=#3%
\divide\Y by 2%
\Z=#3%
\multiply \Z by 2%
\begin{picture}(0,0)%
\put(-\Y,-#3){\begin{picture}(0,0)%
\truex{313}\truey{156}%
\put(-\value{x},\value{y}){\line(1,2){#3}}%
\put(\value{x},-\value{y}){\line(1,2){#3}}%
\monolength=#3%
\advance\monolength by -\value{x}%
\epilength=#3%
\advance\epilength by \value{x}%
\secondmonolength=\Z%
\advance\secondmonolength by -\value{y}%
\secondepilength=\Z%
\advance\secondepilength by \value{y}%
\put(\monolength,\secondepilength){\nnehead}%
\put(\epilength,\secondmonolength){\nnehead}%
\end{picture}}
\truex{300}\truey{1000}\truez{600}%
\put(-\value{x},\value{x}){\makebox(0,\value{z})[r]{${#1}$}}%
\put(\value{x},-\value{y}){\makebox(0,\value{z})[l]{${#2}$}}%
\end{picture}}%

\newcommand{\NNEBIDIST}[3]{\testdiagrammode%
\Y=#3%
\divide\Y by 2%
\Z=#3%
\multiply \Z by 2%
\begin{picture}(0,0)%
\truex{313}\truey{156}\truez{400}%
\put(-\Y,-#3){\begin{picture}(0,0)%
\put(-\value{x},\value{y}){\line(1,2){#3}}%
\put(\value{x},-\value{y}){\line(1,2){#3}}%
\monolength=#3%
\advance\monolength by -\value{x}%
\epilength=#3%
\advance\epilength by \value{x}%
\secondmonolength=\Z%
\advance\secondmonolength by -\value{y}%
\secondepilength=\Z%
\advance\secondepilength by \value{y}%
\put(\monolength,\secondepilength){\nnehead}%
\put(\epilength,\secondmonolength){\nnehead}%
\end{picture}}
\put(-\value{x},\value{y}){\circle{\value{z}}}%
\put(\value{x},-\value{y}){\circle{\value{z}}}%
\truex{300}\truey{1000}\truez{600}%
\put(-\value{x},\value{x}){\makebox(0,\value{z})[r]{${#1}$}}%
\put(\value{x},-\value{y}){\makebox(0,\value{z})[l]{${#2}$}}%
\end{picture}}%

\newcommand{\NNEADJAR}[3]{\testdiagrammode%
\Y=#3%
\divide\Y by 2%
\Z=#3%
\multiply \Z by 2%
\begin{picture}(0,0)%
\put(-\Y,-#3){\begin{picture}(0,0)%
\truex{313}\truey{156}%
\monolength=#3%
\advance\monolength by -\value{x}%
\epilength=#3%
\advance\epilength by \value{x}%
\secondmonolength=\Z%
\advance\secondmonolength by -\value{y}%
\secondepilength=\Z%
\advance\secondepilength by \value{y}%
\put(\value{x},-\value{y}){\line(1,2){#3}}%
\put(\epilength,\secondmonolength){\nnehead}%
\put(\monolength,\secondepilength){\line(-1,-2){#3}}%
\put(-\value{x},\value{y}){\sswhead}%
\end{picture}}
\truex{300}\truey{1000}\truez{600}%
\put(-\value{x},\value{x}){\makebox(0,\value{z})[r]{${#1}$}}%
\put(\value{x},-\value{y}){\makebox(0,\value{z})[l]{${#2}$}}%
\end{picture}}%

\newcommand{\NNEADJDIST}[3]{\testdiagrammode%
\Y=#3%
\divide\Y by 2%
\Z=#3%
\multiply \Z by 2%
\begin{picture}(0,0)%
\truex{313}\truey{156}\truez{400}%
\put(-\Y,-#3){\begin{picture}(0,0)%
\monolength=#3%
\advance\monolength by -\value{x}%
\epilength=#3%
\advance\epilength by \value{x}%
\secondmonolength=\Z%
\advance\secondmonolength by -\value{y}%
\secondepilength=\Z%
\advance\secondepilength by \value{y}%
\put(\value{x},-\value{y}){\line(1,2){#3}}%
\put(\epilength,\secondmonolength){\nnehead}%
\put(\monolength,\secondepilength){\line(-1,-2){#3}}%
\put(-\value{x},\value{y}){\sswhead}%
\end{picture}}
\put(-\value{x},\value{y}){\circle{\value{z}}}%
\put(\value{x},-\value{y}){\circle{\value{z}}}%
\truex{300}\truey{1000}\truez{600}%
\put(-\value{x},\value{x}){\makebox(0,\value{z})[r]{${#1}$}}%
\put(\value{x},-\value{y}){\makebox(0,\value{z})[l]{${#2}$}}%
\end{picture}}%

\def\basicnnear[#1]{\NNEAR{}{}{#100}}%

\newcommand{\nnear}{\@ifnextchar[{\basicnnear}{\basicnnear[67]}}%

\def\basicNnear[#1]#2{\NNEAR{#2}{}{#100}}%

\newcommand{\Nnear}{\@ifnextchar[{\basicNnear}{\basicNnear[67]}}%

\def\basicnneaR[#1]#2{\NNEAR{}{#2}{#100}}%

\newcommand{\nneaR}{\@ifnextchar[{\basicnneaR}{\basicnneaR[67]}}%

\def\basicnnedist[#1]{\NNEDIST{}{}{#100}}%

\newcommand{\nnedist}{\@ifnextchar[{\basicnnedist}{\basicnnedist[67]}}%

\def\basicNnedist[#1]#2{\NNEDIST{#2}{}{#100}}%

\newcommand{\Nnedist}{\@ifnextchar[{\basicNnedist}{\basicNnedist[67]}}%

\def\basicnnedisT[#1]#2{\NNEDIST{}{#2}{#100}}%

\newcommand{\nnedisT}{\@ifnextchar[{\basicnnedisT}{\basicnnedisT[67]}}%

\def\basicnnedotar[#1]{\NNEDOTAR{}{}{#100}}%

\newcommand{\nnedotar}{\@ifnextchar[{\basicnnedotar}{\basicnnedotar[67]}}%

\def\basicNnedotar[#1]#2{\NNEDOTAR{#2}{}{#100}}%

\newcommand{\Nnedotar}{\@ifnextchar[{\basicNnedotar}{\basicNnedotar[67]}}%

\def\basicnnedotaR[#1]#2{\NNEDOTAR{}{#2}{#100}}%

\newcommand{\nnedotaR}{\@ifnextchar[{\basicnnedotaR}{\basicnnedotaR[67]}}%

\def\basicnnemono[#1]{\NNEMONO{}{}{#100}}%

\newcommand{\nnemono}{\@ifnextchar[{\basicnnemono}{\basicnnemono[67]}}%

\def\basicNnemono[#1]#2{\NNEMONO{#2}{}{#100}}%

\newcommand{\Nnemono}{\@ifnextchar[{\basicNnemono}{\basicNnemono[67]}}%

\def\basicnnemonO[#1]#2{\NNEMONO{}{#2}{#100}}%

\newcommand{\nnemonO}{\@ifnextchar[{\basicnnemonO}{\basicnnemonO[67]}}%

\def\basicnneepi[#1]{\NNEEPI{}{}{#100}}%

\newcommand{\nneepi}{\@ifnextchar[{\basicnneepi}{\basicnneepi[67]}}%

\def\basicNneepi[#1]#2{\NNEEPI{#2}{}{#100}}%

\newcommand{\Nneepi}{\@ifnextchar[{\basicNneepi}{\basicNneepi[67]}}%

\def\basicnneepI[#1]#2{\NNEEPI{}{#2}{#100}}%

\newcommand{\nneepI}{\@ifnextchar[{\basicnneepI}{\basicnneepI[67]}}%

\def\basicnnebimo[#1]{\NNEBIMO{}{}{#100}}%

\newcommand{\nnebimo}{\@ifnextchar[{\basicnnebimo}{\basicnnebimo[67]}}%

\def\basicNnebimo[#1]#2{\NNEBIMO{#2}{}{#100}}%

\newcommand{\Nnebimo}{\@ifnextchar[{\basicNnebimo}{\basicNnebimo[67]}}%

\def\basicnnebimO[#1]#2{\NNEBIMO{}{#2}{#100}}%

\newcommand{\nnebimO}{\@ifnextchar[{\basicnnebimO}{\basicnnebimO[67]}}%

\def\basicnneiso[#1]{\NNEAR{\cong}{}{#100}}%

\newcommand{\nneiso}{\@ifnextchar[{\basicnneiso}{\basicnneiso[67]}}%

\def\basicNneiso[#1]#2{\NNEAR{#2}{\cong}{#100}}%

\newcommand{\Nneiso}{\@ifnextchar[{\basicNneiso}{\basicNneiso[67]}}%

\def\basicnneisO[#1]#2{\NNEAR{\cong}{#2}{#100}}%

\newcommand{\nneisO}{\@ifnextchar[{\basicnneisO}{\basicnneisO[67]}}%

\def\basicnneeql[#1]{\NNEEQL{}{}{#100}}%

\newcommand{\nneeql}{\@ifnextchar[{\basicnneeql}{\basicnneeql[67]}}%

\def\basicNneeql[#1]#2{\NNEEQL{#2}{}{#100}}%

\newcommand{\Nneeql}{\@ifnextchar[{\basicNneeql}{\basicNneeql[67]}}%

\def\basicnneeqL[#1]#2{\NNEEQL{}{#2}{#100}}%

\newcommand{\nneeqL}{\@ifnextchar[{\basicnneeqL}{\basicnneeqL[67]}}%

\def\basicnnebiar[#1]{\NNEBIAR{}{}{#100}}%

\newcommand{\nnebiar}{\@ifnextchar[{\basicnnebiar}{\basicnnebiar[67]}}%

\def\basicNnebiar[#1]#2#3{\NNEBIAR{#2}{#3}{#100}}%

\newcommand{\Nnebiar}{\@ifnextchar[{\basicNnebiar}{\basicNnebiar[67]}}%

\def\basicnnebidist[#1]{\NNEBIDIST{}{}{#100}}%

\newcommand{\nnebidist}{\@ifnextchar[{\basicnnebidist}{\basicnnebidist[67]}}%

\def\basicNnebidist[#1]#2#3{\NNEBIDIST{#2}{#3}{#100}}%

\newcommand{\Nnebidist}{\@ifnextchar[{\basicNnebidist}{\basicNnebidist[67]}}%

\def\basicnneadjar[#1]{\NNEADJAR{}{}{#100}}%

\newcommand{\nneadjar}{\@ifnextchar[{\basicnneadjar}{\basicnneadjar[67]}}%

\def\basicNneadjar[#1]#2#3{\NNEADJAR{#2}{#3}{#100}}%

\newcommand{\Nneadjar}{\@ifnextchar[{\basicNneadjar}{\basicNneadjar[67]}}%

\def\basicnneadjdist[#1]{\NNEADJDIST{}{}{#100}}%

\newcommand{\nneadjdist}{\@ifnextchar[{\basicnneadjdist}{\basicnneadjdist[67]}}%

\def\basicNneadjdist[#1]#2#3{\NNEADJDIST{#2}{#3}{#100}}%

\newcommand{\Nneadjdist}{\@ifnextchar[{\basicNneadjdist}{\basicNneadjdist[67]}}%

\newcommand{\SSEAR}[3]{\testdiagrammode%
\Z=#3%
\divide\Z by 2%
\begin{picture}(0,0)%
\put(-\Z,#3){\line(1,-2){#3}}%
\put(\Z,-#3){\ssehead}%
\truex{200}\truey{800}\truez{600}%
\put(\value{x},\value{x}){\makebox(0,\value{z})[l]{${#1}$}}%
\put(-\value{x},-\value{y}){\makebox(0,\value{z})[r]{${#2}$}}%
\end{picture}}%

\newcommand{\SSEDIST}[3]{\testdiagrammode%
\Z=#3%
\divide\Z by 2%
\begin{picture}(0,0)%
\put(-\Z,#3){\line(1,-2){#3}}%
\put(\Z,-#3){\ssehead}%
\truex{400}%
\put(0,0){\circle{\value{x}}}%
\truex{200}\truey{800}\truez{600}%
\put(\value{x},\value{x}){\makebox(0,\value{z})[l]{${#1}$}}%
\put(-\value{x},-\value{y}){\makebox(0,\value{z})[r]{${#2}$}}%
\end{picture}}%

\newcommand{\SSEDOTAR}[3]{\testdiagrammode%
\truex{100}\truey{268}\truez{134}%
\Z=#3%
\divide\Z by 2%
\NUMBEROFDOTS=#3%
\divide\NUMBEROFDOTS by \value{z}%
\advance\NUMBEROFDOTS by 1%
\begin{picture}(0,0)%
\multiput(-\Z,#3)(\value{z},-\value{y}){\NUMBEROFDOTS}%
{\circle*{\value{x}}}%
\put(\Z,-#3){\ssehead}%
\truex{200}\truey{800}\truez{600}%
\put(\value{x},\value{x}){\makebox(0,\value{z})[l]{${#1}$}}%
\put(-\value{x},-\value{y}){\makebox(0,\value{z})[r]{${#2}$}}%
\end{picture}}%

\newcommand{\SSEMONO}[3]{\testdiagrammode%
\Z=#3%
\divide\Z by 2%
\truetaiL%
\bimolength=#3%
\advance\bimolength by -\truemonotaiL%
\monolength=\bimolength%
\advance\monolength by -\Z%
\secondmonolength=\monolength%
\multiply\secondmonolength by 2%
\begin{picture}(0,0)%
\put(-\monolength,\secondmonolength){\line(1,-2){\bimolength}}%
\put(-\monolength,\secondmonolength){\ssehead}%
\put(\Z,-#3){\ssehead}%
\truex{200}\truey{800}\truez{600}%
\put(\value{x},\value{x}){\makebox(0,\value{z})[l]{${#1}$}}%
\put(-\value{x},-\value{y}){\makebox(0,\value{z})[r]{${#2}$}}%
\end{picture}}%

\newcommand{\SSEEPI}[3]{\testdiagrammode%
\Z=#3%
\divide\Z by 2%
\trueheaD%
\bimolength=#3%
\advance\bimolength by -\trueepiheaD%
\epilength=\bimolength%
\advance\epilength by -\Z%
\secondepilength=\epilength%
\multiply\secondepilength by 2%
\begin{picture}(0,0)%
\put(-\Z,#3){\line(1,-2){\bimolength}}%
\put(\epilength,-\secondepilength){\ssehead}%
\put(\Z,-#3){\ssehead}%
\truex{200}\truey{800}\truez{600}%
\put(\value{x},\value{x}){\makebox(0,\value{z})[l]{${#1}$}}%
\put(-\value{x},-\value{y}){\makebox(0,\value{z})[r]{${#2}$}}%
\end{picture}}%

\newcommand{\SSEBIMO}[3]{\testdiagrammode%
\Z=#3%
\divide\Z by 2%
\truetaiL\trueheaD%
\bimolength=#3%
\advance\bimolength by -\truemonotaiL%
\monolength=\bimolength%
\advance\monolength by -\Z%
\advance\bimolength by -\trueepiheaD%
\epilength=\bimolength%
\advance\epilength by -\monolength%
\secondmonolength=\monolength%
\multiply\secondmonolength by 2%
\secondepilength=\epilength%
\multiply\secondepilength by 2%
\begin{picture}(0,0)%
\put(-\monolength,\secondmonolength){\line(1,-2){\bimolength}}%
\put(-\monolength,\secondmonolength){\ssehead}%
\put(\epilength,-\secondepilength){\ssehead}%
\put(\Z,-#3){\ssehead}%
\truex{200}\truey{800}\truez{600}%
\put(\value{x},\value{x}){\makebox(0,\value{z})[l]{${#1}$}}%
\put(-\value{x},-\value{y}){\makebox(0,\value{z})[r]{${#2}$}}%
\end{picture}}%

\newcommand{\SSEEQL}[3]{\testdiagrammode%
\Z=#3%
\divide\Z by 2%
\begin{picture}(0,0)%
\put(-\Z,#3){\begin{picture}(0,0)%
\truex{44}\truey{89}%
\put(-\value{y},-\value{x}){\line(1,-2){#3}}%
\put(\value{y},\value{x}){\line(1,-2){#3}}%
\end{picture}}%
\truex{200}\truey{800}\truez{600}%
\put(\value{x},\value{x}){\makebox(0,\value{z})[l]{${#1}$}}%
\put(-\value{x},-\value{y}){\makebox(0,\value{z})[r]{${#2}$}}%
\end{picture}}%

\newcommand{\SSEBIAR}[3]{\testdiagrammode%
\Y=#3%
\divide\Y by 2%
\Z=#3%
\multiply \Z by 2%
\begin{picture}(0,0)%
\put(-\Y,#3){\begin{picture}(0,0)%
\truex{313}\truey{156}%
\put(-\value{x},-\value{y}){\line(1,-2){#3}}%
\put(\value{x},\value{y}){\line(1,-2){#3}}%
\monolength=#3%
\advance\monolength by -\value{x}%
\epilength=#3%
\advance\epilength by \value{x}%
\secondmonolength=\Z%
\advance\secondmonolength by -\value{y}%
\secondepilength=\Z%
\advance\secondepilength by \value{y}%
\put(\monolength,-\secondepilength){\ssehead}%
\put(\epilength,-\secondmonolength){\ssehead}%
\end{picture}}
\truex{400}\truey{1000}\truez{600}%
\put(\value{x},\value{x}){\makebox(0,\value{z})[l]{${#1}$}}%
\put(-\value{x},-\value{y}){\makebox(0,\value{z})[r]{${#2}$}}%
\end{picture}}%

\newcommand{\SSEBIDIST}[3]{\testdiagrammode%
\Y=#3%
\divide\Y by 2%
\Z=#3%
\multiply \Z by 2%
\begin{picture}(0,0)%
\truex{313}\truey{156}\truez{400}%
\put(-\Y,#3){\begin{picture}(0,0)%
\put(-\value{x},-\value{y}){\line(1,-2){#3}}%
\put(\value{x},\value{y}){\line(1,-2){#3}}%
\monolength=#3%
\advance\monolength by -\value{x}%
\epilength=#3%
\advance\epilength by \value{x}%
\secondmonolength=\Z%
\advance\secondmonolength by -\value{y}%
\secondepilength=\Z%
\advance\secondepilength by \value{y}%
\put(\monolength,-\secondepilength){\ssehead}%
\put(\epilength,-\secondmonolength){\ssehead}%
\end{picture}}
\put(-\value{x},-\value{y}){\circle{\value{z}}}%
\put(\value{x},\value{y}){\circle{\value{z}}}%
\truex{500}\truey{1000}\truez{600}%
\put(\value{x},\value{x}){\makebox(0,\value{z})[l]{${#1}$}}%
\put(-\value{x},-\value{y}){\makebox(0,\value{z})[r]{${#2}$}}%
\end{picture}}%

\newcommand{\SSEADJAR}[3]{\testdiagrammode%
\Y=#3%
\divide\Y by 2%
\Z=#3%
\multiply \Z by 2%
\begin{picture}(0,0)%
\put(-\Y,#3){\begin{picture}(0,0)%
\truex{313}\truey{156}%
\monolength=#3%
\advance\monolength by -\value{x}%
\epilength=#3%
\advance\epilength by \value{x}%
\secondmonolength=\Z%
\advance\secondmonolength by -\value{y}%
\secondepilength=\Z%
\advance\secondepilength by \value{y}%
\put(-\value{x},-\value{y}){\line(1,-2){#3}}%
\put(\monolength,-\secondepilength){\ssehead}%
\put(\epilength,-\secondmonolength){\line(-1,2){#3}}%
\put(\value{x},\value{y}){\nnwhead}%
\end{picture}}
\truex{400}\truey{1000}\truez{600}%
\put(\value{x},\value{x}){\makebox(0,\value{z})[l]{${#1}$}}%
\put(-\value{x},-\value{y}){\makebox(0,\value{z})[r]{${#2}$}}%
\end{picture}}%

\newcommand{\SSEADJDIST}[3]{\testdiagrammode%
\Y=#3%
\divide\Y by 2%
\Z=#3%
\multiply \Z by 2%
\begin{picture}(0,0)%
\truex{313}\truey{156}\truez{400}%
\put(-\Y,#3){\begin{picture}(0,0)%
\monolength=#3%
\advance\monolength by -\value{x}%
\epilength=#3%
\advance\epilength by \value{x}%
\secondmonolength=\Z%
\advance\secondmonolength by -\value{y}%
\secondepilength=\Z%
\advance\secondepilength by \value{y}%
\put(-\value{x},-\value{y}){\line(1,-2){#3}}%
\put(\monolength,-\secondepilength){\ssehead}%
\put(\epilength,-\secondmonolength){\line(-1,2){#3}}%
\put(\value{x},\value{y}){\nnwhead}%
\end{picture}}
\put(\value{x},\value{y}){\circle{\value{z}}}%
\put(-\value{x},-\value{y}){\circle{\value{z}}}%
\truex{500}\truey{1000}\truez{600}%
\put(\value{x},\value{x}){\makebox(0,\value{z})[l]{${#1}$}}%
\put(-\value{x},-\value{y}){\makebox(0,\value{z})[r]{${#2}$}}%
\end{picture}}%

\def\basicssear[#1]{\SSEAR{}{}{#100}}%

\newcommand{\ssear}{\@ifnextchar[{\basicssear}{\basicssear[67]}}%

\def\basicSsear[#1]#2{\SSEAR{#2}{}{#100}}%

\newcommand{\Ssear}{\@ifnextchar[{\basicSsear}{\basicSsear[67]}}%

\def\basicsseaR[#1]#2{\SSEAR{}{#2}{#100}}%

\newcommand{\sseaR}{\@ifnextchar[{\basicsseaR}{\basicsseaR[67]}}%

\def\basicssedist[#1]{\SSEDIST{}{}{#100}}%

\newcommand{\ssedist}{\@ifnextchar[{\basicssedist}{\basicssedist[67]}}%

\def\basicSsedist[#1]#2{\SSEDIST{#2}{}{#100}}%

\newcommand{\Ssedist}{\@ifnextchar[{\basicSsedist}{\basicSsedist[67]}}%

\def\basicssedisT[#1]#2{\SSEDIST{}{#2}{#100}}%

\newcommand{\ssedisT}{\@ifnextchar[{\basicssedisT}{\basicssedisT[67]}}%

\def\basicssedotar[#1]{\SSEDOTAR{}{}{#100}}%

\newcommand{\ssedotar}{\@ifnextchar[{\basicssedotar}{\basicssedotar[67]}}%

\def\basicSsedotar[#1]#2{\SSEDOTAR{#2}{}{#100}}%

\newcommand{\Ssedotar}{\@ifnextchar[{\basicSsedotar}{\basicSsedotar[67]}}%

\def\basicssedotaR[#1]#2{\SSEDOTAR{}{#2}{#100}}%

\newcommand{\ssedotaR}{\@ifnextchar[{\basicssedotaR}{\basicssedotaR[67]}}%

\def\basicssemono[#1]{\SSEMONO{}{}{#100}}%

\newcommand{\ssemono}{\@ifnextchar[{\basicssemono}{\basicssemono[67]}}%

\def\basicSsemono[#1]#2{\SSEMONO{#2}{}{#100}}%

\newcommand{\Ssemono}{\@ifnextchar[{\basicSsemono}{\basicSsemono[67]}}%

\def\basicssemonO[#1]#2{\SSEMONO{}{#2}{#100}}%

\newcommand{\ssemonO}{\@ifnextchar[{\basicssemonO}{\basicssemonO[67]}}%

\def\basicsseepi[#1]{\SSEEPI{}{}{#100}}%

\newcommand{\sseepi}{\@ifnextchar[{\basicsseepi}{\basicsseepi[67]}}%

\def\basicSseepi[#1]#2{\SSEEPI{#2}{}{#100}}%

\newcommand{\Sseepi}{\@ifnextchar[{\basicSseepi}{\basicSseepi[67]}}%

\def\basicsseepI[#1]#2{\SSEEPI{}{#2}{#100}}%

\newcommand{\sseepI}{\@ifnextchar[{\basicsseepI}{\basicsseepI[67]}}%

\def\basicssebimo[#1]{\SSEBIMO{}{}{#100}}%

\newcommand{\ssebimo}{\@ifnextchar[{\basicssebimo}{\basicssebimo[67]}}%

\def\basicSsebimo[#1]#2{\SSEBIMO{#2}{}{#100}}%

\newcommand{\Ssebimo}{\@ifnextchar[{\basicSsebimo}{\basicSsebimo[67]}}%

\def\basicssebimO[#1]#2{\SSEBIMO{}{#2}{#100}}%

\newcommand{\ssebimO}{\@ifnextchar[{\basicssebimO}{\basicssebimO[67]}}%

\def\basicsseiso[#1]{\SSEAR{\cong}{}{#100}}%

\newcommand{\sseiso}{\@ifnextchar[{\basicsseiso}{\basicsseiso[67]}}%

\def\basicSseiso[#1]#2{\SSEAR{#2}{\cong}{#100}}%

\newcommand{\Sseiso}{\@ifnextchar[{\basicSseiso}{\basicSseiso[67]}}%

\def\basicsseisO[#1]#2{\SSEAR{\cong}{#2}{#100}}%

\newcommand{\sseisO}{\@ifnextchar[{\basicsseisO}{\basicsseisO[67]}}%

\def\basicsseeql[#1]{\SSEEQL{}{}{#100}}%

\newcommand{\sseeql}{\@ifnextchar[{\basicsseeql}{\basicsseeql[67]}}%

\def\basicSseeql[#1]#2{\SSEEQL{#2}{}{#100}}%

\newcommand{\Sseeql}{\@ifnextchar[{\basicSseeql}{\basicSseeql[67]}}%

\def\basicsseeqL[#1]#2{\SSEEQL{}{#2}{#100}}%

\newcommand{\sseeqL}{\@ifnextchar[{\basicsseeqL}{\basicsseeqL[67]}}%

\def\basicssebiar[#1]{\SSEBIAR{}{}{#100}}%

\newcommand{\ssebiar}{\@ifnextchar[{\basicssebiar}{\basicssebiar[67]}}%

\def\basicSsebiar[#1]#2#3{\SSEBIAR{#2}{#3}{#100}}%

\newcommand{\Ssebiar}{\@ifnextchar[{\basicSsebiar}{\basicSsebiar[67]}}%

\def\basicssebidist[#1]{\SSEBIDIST{}{}{#100}}%

\newcommand{\ssebidist}{\@ifnextchar[{\basicssebidist}{\basicssebidist[67]}}%

\def\basicSsebidist[#1]#2#3{\SSEBIDIST{#2}{#3}{#100}}%

\newcommand{\Ssebidist}{\@ifnextchar[{\basicSsebidist}{\basicSsebidist[67]}}%

\def\basicsseadjar[#1]{\SSEADJAR{}{}{#100}}%

\newcommand{\sseadjar}{\@ifnextchar[{\basicsseadjar}{\basicsseadjar[67]}}%

\def\basicSseadjar[#1]#2#3{\SSEADJAR{#2}{#3}{#100}}%

\newcommand{\Sseadjar}{\@ifnextchar[{\basicSseadjar}{\basicSseadjar[67]}}%

\def\basicsseadjdist[#1]{\SSEADJDIST{}{}{#100}}%

\newcommand{\sseadjdist}{\@ifnextchar[{\basicsseadjdist}{\basicsseadjdist[67]}}%

\def\basicSseadjdist[#1]#2#3{\SSEADJDIST{#2}{#3}{#100}}%

\newcommand{\Sseadjdist}{\@ifnextchar[{\basicSseadjdist}{\basicSseadjdist[67]}}%

\newcommand{\SSWAR}[3]{\testdiagrammode%
\Z=#3%
\divide\Z by 2%
\begin{picture}(0,0)%
\put(\Z,#3){\line(-1,-2){#3}}%
\put(-\Z,-#3){\sswhead}%
\truex{200}\truey{800}\truez{600}%
\put(-\value{x},\value{x}){\makebox(0,\value{z})[r]{${#1}$}}%
\put(\value{x},-\value{y}){\makebox(0,\value{z})[l]{${#2}$}}%
\end{picture}}%

\newcommand{\SSWDIST}[3]{\testdiagrammode%
\Z=#3%
\divide\Z by 2%
\begin{picture}(0,0)%
\put(\Z,#3){\line(-1,-2){#3}}%
\put(-\Z,-#3){\sswhead}%
\truex{400}%
\put(0,0){\circle{\value{x}}}%
\truex{200}\truey{800}\truez{600}%
\put(-\value{x},\value{x}){\makebox(0,\value{z})[r]{${#1}$}}%
\put(\value{x},-\value{y}){\makebox(0,\value{z})[l]{${#2}$}}%
\end{picture}}%

\newcommand{\SSWDOTAR}[3]{\testdiagrammode%
\truex{100}\truey{268}\truez{134}%
\Z=#3%
\divide\Z by 2%
\NUMBEROFDOTS=#3%
\divide\NUMBEROFDOTS by \value{z}%
\advance\NUMBEROFDOTS by 1%
\begin{picture}(0,0)%
\multiput(\Z,#3)(-\value{z},-\value{y}){\NUMBEROFDOTS}%
{\circle*{\value{x}}}%
\put(-\Z,-#3){\sswhead}%
\truex{200}\truey{800}\truez{600}%
\put(-\value{x},\value{x}){\makebox(0,\value{z})[r]{${#1}$}}%
\put(\value{x},-\value{y}){\makebox(0,\value{z})[l]{${#2}$}}%
\end{picture}}%

\newcommand{\SSWMONO}[3]{\testdiagrammode%
\Z=#3%
\divide\Z by 2%
\truetaiL%
\bimolength=#3%
\advance\bimolength by -\truemonotaiL%
\monolength=\bimolength%
\advance\monolength by -\Z%
\secondmonolength=\monolength%
\multiply\secondmonolength by 2%
\begin{picture}(0,0)%
\put(\monolength,\secondmonolength){\line(-1,-2){\bimolength}}%
\put(\monolength,\secondmonolength){\sswhead}%
\put(-\Z,-#3){\sswhead}%
\truex{200}\truey{800}\truez{600}%
\put(-\value{x},\value{x}){\makebox(0,\value{z})[r]{${#1}$}}%
\put(\value{x},-\value{y}){\makebox(0,\value{z})[l]{${#2}$}}%
\end{picture}}%

\newcommand{\SSWEPI}[3]{\testdiagrammode%
\Z=#3%
\divide\Z by 2%
\trueheaD%
\bimolength=#3%
\advance\bimolength by -\trueepiheaD%
\epilength=\bimolength%
\advance\epilength by -\Z%
\secondepilength=\epilength%
\multiply\secondepilength by 2%
\begin{picture}(0,0)%
\put(\Z,#3){\line(-1,-2){\bimolength}}%
\put(-\epilength,-\secondepilength){\sswhead}%
\put(-\Z,-#3){\sswhead}%
\truex{200}\truey{800}\truez{600}%
\put(-\value{x},\value{x}){\makebox(0,\value{z})[r]{${#1}$}}%
\put(\value{x},-\value{y}){\makebox(0,\value{z})[l]{${#2}$}}%
\end{picture}}%

\newcommand{\SSWBIMO}[3]{\testdiagrammode%
\Z=#3%
\divide\Z by 2%
\truetaiL\trueheaD%
\bimolength=#3%
\advance\bimolength by -\truemonotaiL%
\monolength=\bimolength%
\advance\monolength by -\Z%
\advance\bimolength by -\trueepiheaD%
\epilength=\bimolength%
\advance\epilength by -\monolength%
\secondmonolength=\monolength%
\multiply\secondmonolength by 2%
\secondepilength=\epilength%
\multiply\secondepilength by 2%
\begin{picture}(0,0)%
\put(\monolength,\secondmonolength){\line(-1,-2){\bimolength}}%
\put(\monolength,\secondmonolength){\sswhead}%
\put(-\epilength,-\secondepilength){\sswhead}%
\put(-\Z,-#3){\sswhead}%
\truex{200}\truey{800}\truez{600}%
\put(-\value{x},\value{x}){\makebox(0,\value{z})[r]{${#1}$}}%
\put(\value{x},-\value{y}){\makebox(0,\value{z})[l]{${#2}$}}%
\end{picture}}%

\newcommand{\SSWBIAR}[3]{\testdiagrammode%
\Y=#3%
\divide\Y by 2%
\Z=#3%
\multiply \Z by 2%
\begin{picture}(0,0)%
\put(\Y,#3){\begin{picture}(0,0)%
\truex{313}\truey{156}%
\put(-\value{x},\value{y}){\line(-1,-2){#3}}%
\put(\value{x},-\value{y}){\line(-1,-2){#3}}%
\monolength=#3%
\advance\monolength by -\value{x}%
\epilength=#3%
\advance\epilength by \value{x}%
\secondmonolength=\Z%
\advance\secondmonolength by -\value{y}%
\secondepilength=\Z%
\advance\secondepilength by \value{y}%
\put(-\monolength,-\secondepilength){\sswhead}%
\put(-\epilength,-\secondmonolength){\sswhead}%
\end{picture}}
\truex{300}\truey{1000}\truez{600}%
\put(-\value{x},\value{x}){\makebox(0,\value{z})[r]{${#1}$}}%
\put(\value{x},-\value{y}){\makebox(0,\value{z})[l]{${#2}$}}%
\end{picture}}%

\newcommand{\SSWBIDIST}[3]{\testdiagrammode%
\Y=#3%
\divide\Y by 2%
\Z=#3%
\multiply \Z by 2%
\begin{picture}(0,0)%
\truex{313}\truey{156}\truez{400}%
\put(\Y,#3){\begin{picture}(0,0)%
\put(-\value{x},\value{y}){\line(-1,-2){#3}}%
\put(\value{x},-\value{y}){\line(-1,-2){#3}}%
\monolength=#3%
\advance\monolength by -\value{x}%
\epilength=#3%
\advance\epilength by \value{x}%
\secondmonolength=\Z%
\advance\secondmonolength by -\value{y}%
\secondepilength=\Z%
\advance\secondepilength by \value{y}%
\put(-\monolength,-\secondepilength){\sswhead}%
\put(-\epilength,-\secondmonolength){\sswhead}%
\end{picture}}
\put(-\value{x},\value{y}){\circle{\value{z}}}%
\put(\value{x},-\value{y}){\circle{\value{z}}}%
\truex{300}\truey{1000}\truez{600}%
\put(-\value{x},\value{x}){\makebox(0,\value{z})[r]{${#1}$}}%
\put(\value{x},-\value{y}){\makebox(0,\value{z})[l]{${#2}$}}%
\end{picture}}%

\newcommand{\SSWADJAR}[3]{\testdiagrammode%
\Y=#3%
\divide\Y by 2%
\Z=#3%
\multiply \Z by 2%
\begin{picture}(0,0)%
\put(\Y,#3){\begin{picture}(0,0)%
\truex{313}\truey{156}%
\monolength=#3%
\advance\monolength by -\value{x}%
\epilength=#3%
\advance\epilength by \value{x}%
\secondmonolength=\Z%
\advance\secondmonolength by -\value{y}%
\secondepilength=\Z%
\advance\secondepilength by \value{y}%
\put(\value{x},-\value{y}){\line(-1,-2){#3}}%
\put(-\monolength,-\secondepilength){\sswhead}%
\put(-\epilength,-\secondmonolength){\line(1,2){#3}}%
\put(-\value{x},\value{y}){\nnehead}%
\end{picture}}
\truex{300}\truey{1000}\truez{600}%
\put(-\value{x},\value{x}){\makebox(0,\value{z})[r]{${#1}$}}%
\put(\value{x},-\value{y}){\makebox(0,\value{z})[l]{${#2}$}}%
\end{picture}}%

\newcommand{\SSWADJDIST}[3]{\testdiagrammode%
\Y=#3%
\divide\Y by 2%
\Z=#3%
\multiply \Z by 2%
\begin{picture}(0,0)%
\truex{313}\truey{156}\truez{400}%
\put(\Y,#3){\begin{picture}(0,0)%
\monolength=#3%
\advance\monolength by -\value{x}%
\epilength=#3%
\advance\epilength by \value{x}%
\secondmonolength=\Z%
\advance\secondmonolength by -\value{y}%
\secondepilength=\Z%
\advance\secondepilength by \value{y}%
\put(\value{x},-\value{y}){\line(-1,-2){#3}}%
\put(-\monolength,-\secondepilength){\sswhead}%
\put(-\epilength,-\secondmonolength){\line(1,2){#3}}%
\put(-\value{x},\value{y}){\nnehead}%
\end{picture}}
\put(-\value{x},\value{y}){\circle{\value{z}}}%
\put(\value{x},-\value{y}){\circle{\value{z}}}%
\truex{300}\truey{1000}\truez{600}%
\put(-\value{x},\value{x}){\makebox(0,\value{z})[r]{${#1}$}}%
\put(\value{x},-\value{y}){\makebox(0,\value{z})[l]{${#2}$}}%
\end{picture}}%

\def\basicsswar[#1]{\SSWAR{}{}{#100}}%

\newcommand{\sswar}{\@ifnextchar[{\basicsswar}{\basicsswar[67]}}%

\def\basicSswar[#1]#2{\SSWAR{#2}{}{#100}}%

\newcommand{\Sswar}{\@ifnextchar[{\basicSswar}{\basicSswar[67]}}%

\def\basicsswaR[#1]#2{\SSWAR{}{#2}{#100}}%

\newcommand{\sswaR}{\@ifnextchar[{\basicsswaR}{\basicsswaR[67]}}%

\def\basicsswdist[#1]{\SSWDIST{}{}{#100}}%

\newcommand{\sswdist}{\@ifnextchar[{\basicsswdist}{\basicsswdist[67]}}%

\def\basicSswdist[#1]#2{\SSWDIST{#2}{}{#100}}%

\newcommand{\Sswdist}{\@ifnextchar[{\basicSswdist}{\basicSswdist[67]}}%

\def\basicsswdisT[#1]#2{\SSWDIST{}{#2}{#100}}%

\newcommand{\sswdisT}{\@ifnextchar[{\basicsswdisT}{\basicsswdisT[67]}}%

\def\basicsswdotar[#1]{\SSWDOTAR{}{}{#100}}%

\newcommand{\sswdotar}{\@ifnextchar[{\basicsswdotar}{\basicsswdotar[67]}}%

\def\basicSswdotar[#1]#2{\SSWDOTAR{#2}{}{#100}}%

\newcommand{\Sswdotar}{\@ifnextchar[{\basicSswdotar}{\basicSswdotar[67]}}%

\def\basicsswdotaR[#1]#2{\SSWDOTAR{}{#2}{#100}}%

\newcommand{\sswdotaR}{\@ifnextchar[{\basicsswdotaR}{\basicsswdotaR[67]}}%

\def\basicsswmono[#1]{\SSWMONO{}{}{#100}}%

\newcommand{\sswmono}{\@ifnextchar[{\basicsswmono}{\basicsswmono[67]}}%

\def\basicSswmono[#1]#2{\SSWMONO{#2}{}{#100}}%

\newcommand{\Sswmono}{\@ifnextchar[{\basicSswmono}{\basicSswmono[67]}}%

\def\basicsswmonO[#1]#2{\SSWMONO{}{#2}{#100}}%

\newcommand{\sswmonO}{\@ifnextchar[{\basicsswmonO}{\basicsswmonO[67]}}%

\def\basicsswepi[#1]{\SSWEPI{}{}{#100}}%

\newcommand{\sswepi}{\@ifnextchar[{\basicsswepi}{\basicsswepi[67]}}%

\def\basicSswepi[#1]#2{\SSWEPI{#2}{}{#100}}%

\newcommand{\Sswepi}{\@ifnextchar[{\basicSswepi}{\basicSswepi[67]}}%

\def\basicsswepI[#1]#2{\SSWEPI{}{#2}{#100}}%

\newcommand{\sswepI}{\@ifnextchar[{\basicsswepI}{\basicsswepI[67]}}%

\def\basicsswbimo[#1]{\SSWBIMO{}{}{#100}}%

\newcommand{\sswbimo}{\@ifnextchar[{\basicsswbimo}{\basicsswbimo[67]}}%

\def\basicSswbimo[#1]#2{\SSWBIMO{#2}{}{#100}}%

\newcommand{\Sswbimo}{\@ifnextchar[{\basicSswbimo}{\basicSswbimo[67]}}%

\def\basicsswbimO[#1]#2{\SSWBIMO{}{#2}{#100}}%

\newcommand{\sswbimO}{\@ifnextchar[{\basicsswbimO}{\basicsswbimO[67]}}%

\def\basicsswiso[#1]{\SSWAR{\cong}{}{#100}}%

\newcommand{\sswiso}{\@ifnextchar[{\basicsswiso}{\basicsswiso[67]}}%

\def\basicSswiso[#1]#2{\SSWAR{#2}{\cong}{#100}}%

\newcommand{\Sswiso}{\@ifnextchar[{\basicSswiso}{\basicSswiso[67]}}%

\def\basicsswisO[#1]#2{\SSWAR{\cong}{#2}{#100}}%

\newcommand{\sswisO}{\@ifnextchar[{\basicsswisO}{\basicsswisO[67]}}%

\def\basicsswbiar[#1]{\SSWBIAR{}{}{#100}}%

\newcommand{\sswbiar}{\@ifnextchar[{\basicsswbiar}{\basicsswbiar[67]}}%

\def\basicSswbiar[#1]#2#3{\SSWBIAR{#2}{#3}{#100}}%

\newcommand{\Sswbiar}{\@ifnextchar[{\basicSswbiar}{\basicSswbiar[67]}}%

\def\basicsswbidist[#1]{\SSWBIDIST{}{}{#100}}%

\newcommand{\sswbidist}{\@ifnextchar[{\basicsswbidist}{\basicsswbidist[67]}}%

\def\basicSswbidist[#1]#2#3{\SSWBIDIST{#2}{#3}{#100}}%

\newcommand{\Sswbidist}{\@ifnextchar[{\basicSswbidist}{\basicSswbidist[67]}}%

\def\basicsswadjar[#1]{\SSWADJAR{}{}{#100}}%

\newcommand{\sswadjar}{\@ifnextchar[{\basicsswadjar}{\basicsswadjar[67]}}%

\def\basicSswadjar[#1]#2#3{\SSWADJAR{#2}{#3}{#100}}%

\newcommand{\Sswadjar}{\@ifnextchar[{\basicSswadjar}{\basicSswadjar[67]}}%

\def\basicsswadjdist[#1]{\SSWADJDIST{}{}{#100}}%

\newcommand{\sswadjdist}{\@ifnextchar[{\basicsswadjdist}{\basicsswadjdist[67]}}%

\def\basicSswadjdist[#1]#2#3{\SSWADJDIST{#2}{#3}{#100}}%

\newcommand{\Sswadjdist}{\@ifnextchar[{\basicSswadjdist}{\basicSswadjdist[67]}}%

\newcommand{\NNWAR}[3]{\testdiagrammode%
\Z=#3%
\divide\Z by 2%
\begin{picture}(0,0)%
\put(\Z,-#3){\line(-1,2){#3}}%
\put(-\Z,#3){\nnwhead}%
\truex{200}\truey{800}\truez{600}%
\put(\value{x},\value{x}){\makebox(0,\value{z})[l]{${#1}$}}%
\put(-\value{x},-\value{y}){\makebox(0,\value{z})[r]{${#2}$}}%
\end{picture}}%

\newcommand{\NNWDIST}[3]{\testdiagrammode%
\Z=#3%
\divide\Z by 2%
\begin{picture}(0,0)%
\put(\Z,-#3){\line(-1,2){#3}}%
\put(-\Z,#3){\nnwhead}%
\truex{400}%
\put(0,0){\circle{\value{x}}}%
\truex{200}\truey{800}\truez{600}%
\put(\value{x},\value{x}){\makebox(0,\value{z})[l]{${#1}$}}%
\put(-\value{x},-\value{y}){\makebox(0,\value{z})[r]{${#2}$}}%
\end{picture}}%

\newcommand{\NNWDOTAR}[3]{\testdiagrammode%
\truex{100}\truey{268}\truez{134}%
\Z=#3%
\divide\Z by 2%
\NUMBEROFDOTS=#3%
\divide\NUMBEROFDOTS by \value{z}%
\advance\NUMBEROFDOTS by 1%
\begin{picture}(0,0)%
\multiput(\Z,-#3)(-\value{z},\value{y}){\NUMBEROFDOTS}%
{\circle*{\value{x}}}%
\put(-\Z,#3){\nnwhead}%
\truex{200}\truey{800}\truez{600}%
\put(\value{x},\value{x}){\makebox(0,\value{z})[l]{${#1}$}}%
\put(-\value{x},-\value{y}){\makebox(0,\value{z})[r]{${#2}$}}%
\end{picture}}%

\newcommand{\NNWMONO}[3]{\testdiagrammode%
\Z=#3%
\divide\Z by 2%
\truetaiL%
\bimolength=#3%
\advance\bimolength by -\truemonotaiL%
\monolength=\bimolength%
\advance\monolength by -\Z%
\secondmonolength=\monolength%
\multiply\secondmonolength by 2%
\begin{picture}(0,0)%
\put(\monolength,-\secondmonolength){\line(-1,2){\bimolength}}%
\put(\monolength,-\secondmonolength){\nnwhead}%
\put(-\Z,#3){\nnwhead}%
\truex{200}\truey{800}\truez{600}%
\put(\value{x},\value{x}){\makebox(0,\value{z})[l]{${#1}$}}%
\put(-\value{x},-\value{y}){\makebox(0,\value{z})[r]{${#2}$}}%
\end{picture}}%

\newcommand{\NNWEPI}[3]{\testdiagrammode%
\Z=#3%
\divide\Z by 2%
\trueheaD%
\bimolength=#3%
\advance\bimolength by -\trueepiheaD%
\epilength=\bimolength%
\advance\epilength by -\Z%
\secondepilength=\epilength%
\multiply\secondepilength by 2%
\begin{picture}(0,0)%
\put(\Z,-#3){\line(-1,2){\bimolength}}%
\put(-\epilength,\secondepilength){\nnwhead}%
\put(-\Z,#3){\nnwhead}%
\truex{200}\truey{800}\truez{600}%
\put(\value{x},\value{x}){\makebox(0,\value{z})[l]{${#1}$}}%
\put(-\value{x},-\value{y}){\makebox(0,\value{z})[r]{${#2}$}}%
\end{picture}}%

\newcommand{\NNWBIMO}[3]{\testdiagrammode%
\Z=#3%
\divide\Z by 2%
\truetaiL\trueheaD%
\bimolength=#3%
\advance\bimolength by -\truemonotaiL%
\monolength=\bimolength%
\advance\monolength by -\Z%
\advance\bimolength by -\trueepiheaD%
\epilength=\bimolength%
\advance\epilength by -\monolength%
\secondmonolength=\monolength%
\multiply\secondmonolength by 2%
\secondepilength=\epilength%
\multiply\secondepilength by 2%
\begin{picture}(0,0)%
\put(\monolength,-\secondmonolength){\line(-1,2){\bimolength}}%
\put(\monolength,-\secondmonolength){\nnwhead}%
\put(-\epilength,\secondepilength){\nnwhead}%
\put(-\Z,#3){\nnwhead}%
\truex{200}\truey{800}\truez{600}%
\put(\value{x},\value{x}){\makebox(0,\value{z})[l]{${#1}$}}%
\put(-\value{x},-\value{y}){\makebox(0,\value{z})[r]{${#2}$}}%
\end{picture}}%

\newcommand{\NNWBIAR}[3]{\testdiagrammode%
\Y=#3%
\divide\Y by 2%
\Z=#3%
\multiply \Z by 2%
\begin{picture}(0,0)%
\put(\Y,-#3){\begin{picture}(0,0)%
\truex{313}\truey{156}%
\put(-\value{x},-\value{y}){\line(-1,2){#3}}%
\put(\value{x},\value{y}){\line(-1,2){#3}}%
\monolength=#3%
\advance\monolength by -\value{x}%
\epilength=#3%
\advance\epilength by \value{x}%
\secondmonolength=\Z%
\advance\secondmonolength by -\value{y}%
\secondepilength=\Z%
\advance\secondepilength by \value{y}%
\put(-\monolength,\secondepilength){\nnwhead}%
\put(-\epilength,\secondmonolength){\nnwhead}%
\end{picture}}
\truex{400}\truey{1000}\truez{600}%
\put(\value{x},\value{x}){\makebox(0,\value{z})[l]{${#1}$}}%
\put(-\value{x},-\value{y}){\makebox(0,\value{z})[r]{${#2}$}}%
\end{picture}}%

\newcommand{\NNWBIDIST}[3]{\testdiagrammode%
\Y=#3%
\divide\Y by 2%
\Z=#3%
\multiply \Z by 2%
\begin{picture}(0,0)%
\truex{313}\truey{156}\truez{400}%
\put(\Y,-#3){\begin{picture}(0,0)%
\put(-\value{x},-\value{y}){\line(-1,2){#3}}%
\put(\value{x},\value{y}){\line(-1,2){#3}}%
\monolength=#3%
\advance\monolength by -\value{x}%
\epilength=#3%
\advance\epilength by \value{x}%
\secondmonolength=\Z%
\advance\secondmonolength by -\value{y}%
\secondepilength=\Z%
\advance\secondepilength by \value{y}%
\put(-\monolength,\secondepilength){\nnwhead}%
\put(-\epilength,\secondmonolength){\nnwhead}%
\end{picture}}
\put(-\value{x},-\value{y}){\circle{\value{z}}}%
\put(\value{x},\value{y}){\circle{\value{z}}}%
\truex{500}\truey{1000}\truez{600}%
\put(\value{x},\value{x}){\makebox(0,\value{z})[l]{${#1}$}}%
\put(-\value{x},-\value{y}){\makebox(0,\value{z})[r]{${#2}$}}%
\end{picture}}%

\newcommand{\NNWADJAR}[3]{\testdiagrammode%
\Y=#3%
\divide\Y by 2%
\Z=#3%
\multiply \Z by 2%
\begin{picture}(0,0)%
\put(\Y,-#3){\begin{picture}(0,0)%
\truex{313}\truey{156}%
\monolength=#3%
\advance\monolength by -\value{x}%
\epilength=#3%
\advance\epilength by \value{x}%
\secondmonolength=\Z%
\advance\secondmonolength by -\value{y}%
\secondepilength=\Z%
\advance\secondepilength by \value{y}%
\put(-\value{x},-\value{y}){\line(-1,2){#3}}%
\put(-\epilength,\secondmonolength){\nnwhead}%
\put(-\monolength,\secondepilength){\line(1,-2){#3}}%
\put(\value{x},\value{y}){\ssehead}%
\end{picture}}
\truex{400}\truey{1000}\truez{600}%
\put(\value{x},\value{x}){\makebox(0,\value{z})[l]{${#1}$}}%
\put(-\value{x},-\value{y}){\makebox(0,\value{z})[r]{${#2}$}}%
\end{picture}}%

\newcommand{\NNWADJDIST}[3]{\testdiagrammode%
\Y=#3%
\divide\Y by 2%
\Z=#3%
\multiply \Z by 2%
\begin{picture}(0,0)%
\truex{313}\truey{156}\truez{400}%
\put(\Y,-#3){\begin{picture}(0,0)%
\monolength=#3%
\advance\monolength by -\value{x}%
\epilength=#3%
\advance\epilength by \value{x}%
\secondmonolength=\Z%
\advance\secondmonolength by -\value{y}%
\secondepilength=\Z%
\advance\secondepilength by \value{y}%
\put(-\value{x},-\value{y}){\line(-1,2){#3}}%
\put(-\epilength,\secondmonolength){\nnwhead}%
\put(-\monolength,\secondepilength){\line(1,-2){#3}}%
\put(\value{x},\value{y}){\ssehead}%
\end{picture}}
\put(\value{x},\value{y}){\circle{\value{z}}}%
\put(-\value{x},-\value{y}){\circle{\value{z}}}%
\truex{500}\truey{1000}\truez{600}%
\put(\value{x},\value{x}){\makebox(0,\value{z})[l]{${#1}$}}%
\put(-\value{x},-\value{y}){\makebox(0,\value{z})[r]{${#2}$}}%
\end{picture}}%

\def\basicnnwar[#1]{\NNWAR{}{}{#100}}%

\newcommand{\nnwar}{\@ifnextchar[{\basicnnwar}{\basicnnwar[67]}}%

\def\basicNnwar[#1]#2{\NNWAR{#2}{}{#100}}%

\newcommand{\Nnwar}{\@ifnextchar[{\basicNnwar}{\basicNnwar[67]}}%

\def\basicnnwaR[#1]#2{\NNWAR{}{#2}{#100}}%

\newcommand{\nnwaR}{\@ifnextchar[{\basicnnwaR}{\basicnnwaR[67]}}%

\def\basicnnwdist[#1]{\NNWDIST{}{}{#100}}%

\newcommand{\nnwdist}{\@ifnextchar[{\basicnnwdist}{\basicnnwdist[67]}}%

\def\basicNnwdist[#1]#2{\NNWDIST{#2}{}{#100}}%

\newcommand{\Nnwdist}{\@ifnextchar[{\basicNnwdist}{\basicNnwdist[67]}}%

\def\basicnnwdisT[#1]#2{\NNWDIST{}{#2}{#100}}%

\newcommand{\nnwdisT}{\@ifnextchar[{\basicnnwdisT}{\basicnnwdisT[67]}}%

\def\basicnnwdotar[#1]{\NNWDOTAR{}{}{#100}}%

\newcommand{\nnwdotar}{\@ifnextchar[{\basicnnwdotar}{\basicnnwdotar[67]}}%

\def\basicNnwdotar[#1]#2{\NNWDOTAR{#2}{}{#100}}%

\newcommand{\Nnwdotar}{\@ifnextchar[{\basicNnwdotar}{\basicNnwdotar[67]}}%

\def\basicnnwdotaR[#1]#2{\NNWDOTAR{}{#2}{#100}}%

\newcommand{\nnwdotaR}{\@ifnextchar[{\basicnnwdotaR}{\basicnnwdotaR[67]}}%

\def\basicnnwmono[#1]{\NNWMONO{}{}{#100}}%

\newcommand{\nnwmono}{\@ifnextchar[{\basicnnwmono}{\basicnnwmono[67]}}%

\def\basicNnwmono[#1]#2{\NNWMONO{#2}{}{#100}}%

\newcommand{\Nnwmono}{\@ifnextchar[{\basicNnwmono}{\basicNnwmono[67]}}%

\def\basicnnwmonO[#1]#2{\NNWMONO{}{#2}{#100}}%

\newcommand{\nnwmonO}{\@ifnextchar[{\basicnnwmonO}{\basicnnwmonO[67]}}%

\def\basicnnwepi[#1]{\NNWEPI{}{}{#100}}%

\newcommand{\nnwepi}{\@ifnextchar[{\basicnnwepi}{\basicnnwepi[67]}}%

\def\basicNnwepi[#1]#2{\NNWEPI{#2}{}{#100}}%

\newcommand{\Nnwepi}{\@ifnextchar[{\basicNnwepi}{\basicNnwepi[67]}}%

\def\basicnnwepI[#1]#2{\NNWEPI{}{#2}{#100}}%

\newcommand{\nnwepI}{\@ifnextchar[{\basicnnwepI}{\basicnnwepI[67]}}%

\def\basicnnwbimo[#1]{\NNWBIMO{}{}{#100}}%

\newcommand{\nnwbimo}{\@ifnextchar[{\basicnnwbimo}{\basicnnwbimo[67]}}%

\def\basicNnwbimo[#1]#2{\NNWBIMO{#2}{}{#100}}%

\newcommand{\Nnwbimo}{\@ifnextchar[{\basicNnwbimo}{\basicNnwbimo[67]}}%

\def\basicnnwbimO[#1]#2{\NNWBIMO{}{#2}{#100}}%

\newcommand{\nnwbimO}{\@ifnextchar[{\basicnnwbimO}{\basicnnwbimO[67]}}%

\def\basicnnwiso[#1]{\NNWAR{\cong}{}{#100}}%

\newcommand{\nnwiso}{\@ifnextchar[{\basicnnwiso}{\basicnnwiso[67]}}%

\def\basicNnwiso[#1]#2{\NNWAR{#2}{\cong}{#100}}%

\newcommand{\Nnwiso}{\@ifnextchar[{\basicNnwiso}{\basicNnwiso[67]}}%

\def\basicnnwisO[#1]#2{\NNWAR{\cong}{#2}{#100}}%

\newcommand{\nnwisO}{\@ifnextchar[{\basicnnwisO}{\basicnnwisO[67]}}%

\def\basicnnwbiar[#1]{\NNWBIAR{}{}{#100}}%

\newcommand{\nnwbiar}{\@ifnextchar[{\basicnnwbiar}{\basicnnwbiar[67]}}%

\def\basicNnwbiar[#1]#2#3{\NNWBIAR{#2}{#3}{#100}}%

\newcommand{\Nnwbiar}{\@ifnextchar[{\basicNnwbiar}{\basicNnwbiar[67]}}%

\def\basicnnwbidist[#1]{\NNWBIDIST{}{}{#100}}%

\newcommand{\nnwbidist}{\@ifnextchar[{\basicnnwbidist}{\basicnnwbidist[67]}}%

\def\basicNnwbidist[#1]#2#3{\NNWBIDIST{#2}{#3}{#100}}%

\newcommand{\Nnwbidist}{\@ifnextchar[{\basicNnwbidist}{\basicNnwbidist[67]}}%

\def\basicnnwadjar[#1]{\NNWADJAR{}{}{#100}}%

\newcommand{\nnwadjar}{\@ifnextchar[{\basicnnwadjar}{\basicnnwadjar[67]}}%

\def\basicNnwadjar[#1]#2#3{\NNWADJAR{#2}{#3}{#100}}%

\newcommand{\Nnwadjar}{\@ifnextchar[{\basicNnwadjar}{\basicNnwadjar[67]}}%

\def\basicnnwadjdist[#1]{\NNWADJDIST{}{}{#100}}%

\newcommand{\nnwadjdist}{\@ifnextchar[{\basicnnwadjdist}{\basicnnwadjdist[67]}}%

\def\basicNnwadjdist[#1]#2#3{\NNWADJDIST{#2}{#3}{#100}}%

\newcommand{\Nnwadjdist}{\@ifnextchar[{\basicNnwadjdist}{\basicNnwadjdist[67]}}%

\newcommand{\EENEAR}[3]{\testdiagrammode%
\Y=#3%
\divide \Y by 2%
\Z=\Y%
\divide \Z by 3%
\begin{picture}(0,0)%
\put(-\Y,-\Z){\line(3,1){#3}}%
\put(\Y,\Z){\eenehead}%
\truex{200}\truey{800}\truez{600}%
\put(-\value{x},\value{x}){\makebox(0,\value{z})[r]{${#1}$}}%
\put(\value{x},-\value{y}){\makebox(0,\value{z})[l]{${#2}$}}%
\end{picture}}%

\def\basiceenear[#1]{\EENEAR{}{}{#100}}%

\newcommand{\eenear}{\@ifnextchar[{\basiceenear}{\basiceenear[211]}}%

\def\basicEenear[#1]#2{\EENEAR{#2}{}{#100}}%

\newcommand{\Eenear}{\@ifnextchar[{\basicEenear}{\basicEenear[211]}}%

\def\basiceeneaR[#1]#2{\EENEAR{}{#2}{#100}}%

\newcommand{\eeneaR}{\@ifnextchar[{\basiceeneaR}{\basiceeneaR[211]}}%

\newcommand{\EESEAR}[3]{\testdiagrammode%
\Y=#3%
\divide \Y by 2%
\Z=\Y%
\divide \Z by 3%
\begin{picture}(0,0)%
\put(-\Y,\Z){\line(3,-1){#3}}%
\put(\Y,-\Z){\eesehead}%
\truex{200}\truey{800}\truez{600}%
\put(\value{x},\value{x}){\makebox(0,\value{z})[l]{${#1}$}}%
\put(-\value{x},-\value{y}){\makebox(0,\value{z})[r]{${#2}$}}%
\end{picture}}%

\def\basiceesear[#1]{\EESEAR{}{}{#100}}%

\newcommand{\eesear}{\@ifnextchar[{\basiceesear}{\basiceesear[211]}}%

\def\basicEesear[#1]#2{\EESEAR{#2}{}{#100}}%

\newcommand{\Eesear}{\@ifnextchar[{\basicEesear}{\basicEesear[211]}}%

\def\basiceeseaR[#1]#2{\EESEAR{}{#2}{#100}}%

\newcommand{\eeseaR}{\@ifnextchar[{\basiceeseaR}{\basiceeseaR[211]}}%

\newcommand{\WWNWAR}[3]{\testdiagrammode%
\Y=#3%
\divide \Y by 2%
\Z=\Y%
\divide \Z by 3%
\begin{picture}(0,0)%
\put(\Y,-\Z){\line(-3,1){#3}}%
\put(-\Y,\Z){\wwnwhead}%
\truex{200}\truey{800}\truez{600}%
\put(\value{x},\value{x}){\makebox(0,\value{z})[l]{${#1}$}}%
\put(-\value{x},-\value{y}){\makebox(0,\value{z})[r]{${#2}$}}%
\end{picture}}%

\def\basicwwnwar[#1]{\WWNWAR{}{}{#100}}%

\newcommand{\wwnwar}{\@ifnextchar[{\basicwwnwar}{\basicwwnwar[211]}}%

\def\basicWwnwar[#1]#2{\WWNWAR{#2}{}{#100}}%

\newcommand{\Wwnwar}{\@ifnextchar[{\basicWwnwar}{\basicWwnwar[211]}}%

\def\basicwwnwaR[#1]#2{\WWNWAR{}{#2}{#100}}%

\newcommand{\wwnwaR}{\@ifnextchar[{\basicwwnwaR}{\basicwwnwaR[211]}}%

\newcommand{\WWSWAR}[3]{\testdiagrammode%
\Y=#3%
\divide \Y by 2%
\Z=\Y%
\divide \Z by 3%
\begin{picture}(0,0)%
\put(\Y,\Z){\line(-3,-1){#3}}%
\put(-\Y,-\Z){\wwswhead}%
\truex{200}\truey{800}\truez{600}%
\put(-\value{x},\value{x}){\makebox(0,\value{z})[r]{${#1}$}}%
\put(\value{x},-\value{y}){\makebox(0,\value{z})[l]{${#2}$}}%
\end{picture}}%

\def\basicwwswar[#1]{\WWSWAR{}{}{#100}}%

\newcommand{\wwswar}{\@ifnextchar[{\basicwwswar}{\basicwwswar[211]}}%

\def\basicWwswar[#1]#2{\WWSWAR{#2}{}{#100}}%

\newcommand{\Wwswar}{\@ifnextchar[{\basicWwswar}{\basicWwswar[211]}}%

\def\basicwwswaR[#1]#2{\WWSWAR{}{#2}{#100}}%

\newcommand{\wwswaR}{\@ifnextchar[{\basicwwswaR}{\basicwwswaR[211]}}%

\newcommand{\NNNEAR}[3]{\testdiagrammode%
\Y=#3%
\divide \Y by 2%
\Z=\Y%
\multiply \Z by 3%
\begin{picture}(0,0)%
\put(-\Y,-\Z){\line(1,3){#3}}%
\put(\Y,\Z){\nnnehead}%
\truex{100}\truez{600}%
\put(-\value{x},\value{x}){\makebox(0,\value{z})[r]{${#1}$}}%
\put(\value{x},-\value{z}){\makebox(0,\value{z})[l]{${#2}$}}%
\end{picture}}%

\def\basicnnnear[#1]{\NNNEAR{}{}{#100}}%

\newcommand{\nnnear}{\@ifnextchar[{\basicnnnear}{\basicnnnear[71]}}%

\def\basicNnnear[#1]#2{\NNNEAR{#2}{}{#100}}%

\newcommand{\Nnnear}{\@ifnextchar[{\basicNnnear}{\basicNnnear[71]}}%

\def\basicnnneaR[#1]#2{\NNNEAR{}{#2}{#100}}%

\newcommand{\nnneaR}{\@ifnextchar[{\basicnnneaR}{\basicnnneaR[71]}}%

\newcommand{\SSSWAR}[3]{\testdiagrammode%
\Y=#3%
\divide \Y by 2%
\Z=\Y%
\multiply \Z by 3%
\begin{picture}(0,0)%
\put(\Y,\Z){\line(-1,-3){#3}}%
\put(-\Y,-\Z){\ssswhead}%
\truex{100}\truez{600}%
\put(-\value{x},\value{x}){\makebox(0,\value{z})[r]{${#1}$}}%
\put(\value{x},-\value{z}){\makebox(0,\value{z})[l]{${#2}$}}%
\end{picture}}%

\def\basicssswar[#1]{\SSSWAR{}{}{#100}}%

\newcommand{\ssswar}{\@ifnextchar[{\basicssswar}{\basicssswar[71]}}%

\def\basicSsswar[#1]#2{\SSSWAR{#2}{}{#100}}%

\newcommand{\Ssswar}{\@ifnextchar[{\basicSsswar}{\basicSsswar[71]}}%

\def\basicssswaR[#1]#2{\SSSWAR{}{#2}{#100}}%

\newcommand{\ssswaR}{\@ifnextchar[{\basicssswaR}{\basicssswaR[71]}}%

\newcommand{\SSSEAR}[3]{\testdiagrammode%
\Y=#3%
\divide \Y by 2%
\Z=\Y%
\multiply \Z by 3%
\begin{picture}(0,0)%
\put(-\Y,\Z){\line(1,-3){#3}}%
\put(\Y,-\Z){\sssehead}%
\truex{200}\truez{600}%
\put(\value{x},\value{x}){\makebox(0,\value{z})[l]{${#1}$}}%
\put(-\value{x},-\value{z}){\makebox(0,\value{z})[r]{${#2}$}}%
\end{picture}}%

\def\basicsssear[#1]{\SSSEAR{}{}{#100}}%

\newcommand{\sssear}{\@ifnextchar[{\basicsssear}{\basicsssear[71]}}%

\def\basicSssear[#1]#2{\SSSEAR{#2}{}{#100}}%

\newcommand{\Sssear}{\@ifnextchar[{\basicSssear}{\basicSssear[71]}}%

\def\basicssseaR[#1]#2{\SSSEAR{}{#2}{#100}}%

\newcommand{\ssseaR}{\@ifnextchar[{\basicssseaR}{\basicssseaR[71]}}%

\newcommand{\NNNWAR}[3]{\testdiagrammode%
\Y=#3%
\divide \Y by 2%
\Z=\Y%
\multiply \Z by 3%
\begin{picture}(0,0)%
\put(\Y,-\Z){\line(-1,3){#3}}%
\put(-\Y,\Z){\nnnwhead}%
\truex{200}\truez{600}%
\put(\value{x},\value{x}){\makebox(0,\value{z})[l]{${#1}$}}%
\put(-\value{x},-\value{z}){\makebox(0,\value{z})[r]{${#2}$}}%
\end{picture}}%

\def\basicnnnwar[#1]{\NNNWAR{}{}{#100}}%

\newcommand{\nnnwar}{\@ifnextchar[{\basicnnnwar}{\basicnnnwar[71]}}%

\def\basicNnnwar[#1]#2{\NNNWAR{#2}{}{#100}}%

\newcommand{\Nnnwar}{\@ifnextchar[{\basicNnnwar}{\basicNnnwar[71]}}%

\def\basicnnnwaR[#1]#2{\NNNWAR{}{#2}{#100}}%

\newcommand{\nnnwaR}{\@ifnextchar[{\basicnnnwaR}{\basicnnnwaR[71]}}%

\newcommand{\NEENEAR}[3]{\testdiagrammode%
\Y=#3%
\divide \Y by 2%
\Z=#3%
\divide \Z by 3%
\begin{picture}(0,0)%
\put(-\Y,-\Z){\line(3,2){#3}}%
\put(\Y,\Z){\neenehead}%
\truex{200}\truey{800}\truez{600}%
\put(-\value{x},\value{x}){\makebox(0,\value{z})[r]{${#1}$}}%
\put(\value{x},-\value{y}){\makebox(0,\value{z})[l]{${#2}$}}%
\end{picture}}%

\def\basicneenear[#1]{\NEENEAR{}{}{#100}}%

\newcommand{\neenear}{\@ifnextchar[{\basicneenear}{\basicneenear[215]}}%

\def\basicNeenear[#1]#2{\NEENEAR{#2}{}{#100}}%

\newcommand{\Neenear}{\@ifnextchar[{\basicNeenear}{\basicNeenear[215]}}%

\def\basicneeneaR[#1]#2{\NEENEAR{}{#2}{#100}}%

\newcommand{\neeneaR}{\@ifnextchar[{\basicneeneaR}{\basicneeneaR[215]}}%

\newcommand{\SEESEAR}[3]{\testdiagrammode%
\Y=#3%
\divide \Y by 2%
\Z=#3%
\divide \Z by 3%
\begin{picture}(0,0)%
\put(-\Y,\Z){\line(3,-2){#3}}%
\put(\Y,-\Z){\seesehead}%
\truex{200}\truey{800}\truez{600}%
\put(\value{x},\value{x}){\makebox(0,\value{z})[l]{${#1}$}}%
\put(-\value{x},-\value{y}){\makebox(0,\value{z})[r]{${#2}$}}%
\end{picture}}%

\def\basicseesear[#1]{\SEESEAR{}{}{#100}}%

\newcommand{\seesear}{\@ifnextchar[{\basicseesear}{\basicseesear[215]}}%

\def\basicSeesear[#1]#2{\SEESEAR{#2}{}{#100}}%

\newcommand{\Seesear}{\@ifnextchar[{\basicSeesear}{\basicSeesear[215]}}%

\def\basicseeseaR[#1]#2{\SEESEAR{}{#2}{#100}}%

\newcommand{\seeseaR}{\@ifnextchar[{\basicseeseaR}{\basicseeseaR[215]}}%

\newcommand{\NWWNWAR}[3]{\testdiagrammode%
\Y=#3%
\divide \Y by 2%
\Z=#3%
\divide \Z by 3%
\begin{picture}(0,0)%
\put(\Y,-\Z){\line(-3,2){#3}}%
\put(-\Y,\Z){\nwwnwhead}%
\truex{200}\truey{800}\truez{600}%
\put(\value{x},\value{x}){\makebox(0,\value{z})[l]{${#1}$}}%
\put(-\value{x},-\value{y}){\makebox(0,\value{z})[r]{${#2}$}}%
\end{picture}}%

\def\basicnwwnwar[#1]{\NWWNWAR{}{}{#100}}%

\newcommand{\nwwnwar}{\@ifnextchar[{\basicnwwnwar}{\basicnwwnwar[215]}}%

\def\basicNwwnwar[#1]#2{\NWWNWAR{#2}{}{#100}}%

\newcommand{\Nwwnwar}{\@ifnextchar[{\basicNwwnwar}{\basicNwwnwar[215]}}%

\def\basicnwwnwaR[#1]#2{\NWWNWAR{}{#2}{#100}}%

\newcommand{\nwwnwaR}{\@ifnextchar[{\basicnwwnwaR}{\basicnwwnwaR[215]}}%

\newcommand{\SWWSWAR}[3]{\testdiagrammode%
\Y=#3%
\divide \Y by 2%
\Z=#3%
\divide \Z by 3%
\begin{picture}(0,0)%
\put(\Y,\Z){\line(-3,-2){#3}}%
\put(-\Y,-\Z){\swwswhead}%
\truex{200}\truey{800}\truez{600}%
\put(-\value{x},\value{x}){\makebox(0,\value{z})[r]{${#1}$}}%
\put(\value{x},-\value{y}){\makebox(0,\value{z})[l]{${#2}$}}%
\end{picture}}%

\def\basicswwswar[#1]{\SWWSWAR{}{}{#100}}%

\newcommand{\swwswar}{\@ifnextchar[{\basicswwswar}{\basicswwswar[215]}}%

\def\basicSwwswar[#1]#2{\SWWSWAR{#2}{}{#100}}%

\newcommand{\Swwswar}{\@ifnextchar[{\basicSwwswar}{\basicSwwswar[215]}}%

\def\basicswwswaR[#1]#2{\SWWSWAR{}{#2}{#100}}%

\newcommand{\swwswaR}{\@ifnextchar[{\basicswwswaR}{\basicswwswaR[215]}}%

\newcommand{\NENNEAR}[3]{\testdiagrammode%
\Y=#3%
\divide \Y by 2%
\Z=#3%
\multiply \Z by 3%
\divide \Z by 4%
\begin{picture}(0,0)%
\put(-\Y,-\Z){\line(2,3){#3}}%
\put(\Y,\Z){\nennehead}%
\truex{100}\truez{600}%
\put(-\value{x},\value{x}){\makebox(0,\value{z})[r]{${#1}$}}%
\put(\value{x},-\value{z}){\makebox(0,\value{z})[l]{${#2}$}}%
\end{picture}}%

\def\basicnennear[#1]{\NENNEAR{}{}{#100}}%

\newcommand{\nennear}{\@ifnextchar[{\basicnennear}{\basicnennear[143]}}%

\def\basicNennear[#1]#2{\NENNEAR{#2}{}{#100}}%

\newcommand{\Nennear}{\@ifnextchar[{\basicNennear}{\basicNennear[143]}}%

\def\basicnenneaR[#1]#2{\NENNEAR{}{#2}{#100}}%

\newcommand{\nenneaR}{\@ifnextchar[{\basicnenneaR}{\basicnenneaR[143]}}%

\newcommand{\SWSSWAR}[3]{\testdiagrammode%
\Y=#3%
\divide \Y by 2%
\Z=#3%
\multiply \Z by 3%
\divide \Z by 4%
\begin{picture}(0,0)%
\put(\Y,\Z){\line(-2,-3){#3}}%
\put(-\Y,-\Z){\swsswhead}%
\truex{100}\truez{600}%
\put(-\value{x},\value{x}){\makebox(0,\value{z})[r]{${#1}$}}%
\put(\value{x},-\value{z}){\makebox(0,\value{z})[l]{${#2}$}}%
\end{picture}}%

\def\basicswsswar[#1]{\SWSSWAR{}{}{#100}}%

\newcommand{\swsswar}{\@ifnextchar[{\basicswsswar}{\basicswsswar[143]}}%

\def\basicSwsswar[#1]#2{\SWSSWAR{#2}{}{#100}}%

\newcommand{\Swsswar}{\@ifnextchar[{\basicSwsswar}{\basicSwsswar[143]}}%

\def\basicswsswaR[#1]#2{\SWSSWAR{}{#2}{#100}}%

\newcommand{\swsswaR}{\@ifnextchar[{\basicswsswaR}{\basicswsswaR[143]}}%

\newcommand{\SESSEAR}[3]{\testdiagrammode%
\Y=#3%
\divide \Y by 2%
\Z=#3%
\multiply \Z by 3%
\divide \Z by 4%
\begin{picture}(0,0)%
\put(-\Y,\Z){\line(2,-3){#3}}%
\put(\Y,-\Z){\sessehead}%
\truex{200}\truez{600}%
\put(\value{x},\value{x}){\makebox(0,\value{z})[l]{${#1}$}}%
\put(-\value{x},-\value{z}){\makebox(0,\value{z})[r]{${#2}$}}%
\end{picture}}%

\def\basicsessear[#1]{\SESSEAR{}{}{#100}}%

\newcommand{\sessear}{\@ifnextchar[{\basicsessear}{\basicsessear[143]}}%

\def\basicSessear[#1]#2{\SESSEAR{#2}{}{#100}}%

\newcommand{\Sessear}{\@ifnextchar[{\basicSessear}{\basicSessear[143]}}%

\def\basicsesseaR[#1]#2{\SESSEAR{}{#2}{#100}}%

\newcommand{\sesseaR}{\@ifnextchar[{\basicsesseaR}{\basicsesseaR[143]}}%

\newcommand{\NWNNWAR}[3]{\testdiagrammode%
\Y=#3%
\divide \Y by 2%
\Z=#3%
\multiply \Z by 3%
\divide \Z by 4%
\begin{picture}(0,0)%
\put(\Y,-\Z){\line(-2,3){#3}}%
\put(-\Y,\Z){\nwnnwhead}%
\truex{200}\truez{600}%
\put(\value{x},\value{x}){\makebox(0,\value{z})[l]{${#1}$}}%
\put(-\value{x},-\value{z}){\makebox(0,\value{z})[r]{${#2}$}}%
\end{picture}}%

\def\basicnwnnwar[#1]{\NWNNWAR{}{}{#100}}%

\newcommand{\nwnnwar}{\@ifnextchar[{\basicnwnnwar}{\basicnwnnwar[143]}}%

\def\basicNwnnwar[#1]#2{\NWNNWAR{#2}{}{#100}}%

\newcommand{\Nwnnwar}{\@ifnextchar[{\basicNwnnwar}{\basicNwnnwar[143]}}%

\def\basicnwnnwaR[#1]#2{\NWNNWAR{}{#2}{#100}}%

\newcommand{\nwnnwaR}{\@ifnextchar[{\basicnwnnwaR}{\basicnwnnwaR[143]}}%

\newcommand{\Necurve}[2]%
{\testdiagrammode\begin{picture}(0,0)%
\truex{1300}\truey{2000}\truez{200}%
\put(0,\value{x}){\oval(#200,\value{y})[t]}%
\put(0,\value{x}){\makebox(0,0){\begin{picture}(#200,0)%
\put(#200,0){\line(0,-1){\value{z}}}%
\put(#200,-\value{z}){\shead}%
\put(0,0){\line(0,-1){\value{z}}}\end{picture}}}%
\truex{2500}%
\put(0,\value{x}){\makebox(0,0)[b]{${#1}$}}%
\end{picture}}%

\def\basicnecurvar[#1]{\Necurve{}{#1}}

\newcommand{\necurvar}{\@ifnextchar[{\basicnecurvar}{\basicnecurvar[160]}}%

\def\basicNecurvar[#1]#2{\Necurve{#2}{#1}}%

\newcommand{\Necurvar}{\@ifnextchar[{\basicNecurvar}{\basicNecurvar[160]}}%

\newcommand{\Nwcurve}[2]%
{\testdiagrammode\begin{picture}(0,0)%
\truex{1300}\truey{2000}\truez{200}%
\put(0,\value{x}){\oval(#200,\value{y})[t]}%
\put(0,\value{x}){\makebox(0,0){\begin{picture}(#200,0)%
\put(#200,0){\line(0,-1){\value{z}}}%
\put(0,0){\line(0,-1){\value{z}}}%
\put(0,-\value{z}){\shead}%
\end{picture}}}%
\truex{2500}%
\put(0,\value{x}){\makebox(0,0)[b]{${#1}$}}%
\end{picture}}%

\def\basicnwcurvar[#1]{\Nwcurve{}{#1}}

\newcommand{\nwcurvar}{\@ifnextchar[{\basicnwcurvar}{\basicnwcurvar[160]}}%

\def\basicNwcurvar[#1]#2{\Nwcurve{#2}{#1}}%

\newcommand{\Nwcurvar}{\@ifnextchar[{\basicNwcurvar}{\basicNwcurvar[160]}}%

\newcommand{\Securve}[2]%
{\testdiagrammode\begin{picture}(0,0)%
\truex{1300}\truey{2000}\truez{200}%
\put(0,-\value{x}){\oval(#200,\value{y})[b]}%
\put(0,-\value{x}){\makebox(0,0){\begin{picture}(#200,0)%
\put(#200,0){\line(0,1){\value{z}}}%
\put(0,0){\line(0,1){\value{z}}}%
\put(#200,\value{z}){\nhead}%
\end{picture}}}%
\truex{2500}%
\put(0,-\value{x}){\makebox(0,0)[t]{${#1}$}}%
\end{picture}}%

\def\basicsecurvar[#1]{\Securve{}{#1}}

\newcommand{\securvar}{\@ifnextchar[{\basicsecurvar}{\basicsecurvar[160]}}%

\def\basicSecurvar[#1]#2{\Securve{#2}{#1}}%

\newcommand{\Securvar}{\@ifnextchar[{\basicSecurvar}{\basicSecurvar[160]}}%

\newcommand{\Swcurve}[2]%
{\testdiagrammode\begin{picture}(0,0)%
\truex{1300}\truey{2000}\truez{200}%
\put(0,-\value{x}){\oval(#200,\value{y})[b]}%
\put(0,-\value{x}){\makebox(0,0){\begin{picture}(#200,0)%
\put(#200,0){\line(0,1){\value{z}}}%
\put(0,0){\line(0,1){\value{z}}}%
\put(0,\value{z}){\nhead}%
\end{picture}}}%
\truex{2500}%
\put(0,-\value{x}){\makebox(0,0)[t]{${#1}$}}%
\end{picture}}%

\def\basicswcurvar[#1]{\Swcurve{}{#1}}

\newcommand{\swcurvar}{\@ifnextchar[{\basicswcurvar}{\basicswcurvar[160]}}%

\def\basicSwcurvar[#1]#2{\Swcurve{#2}{#1}}%

\newcommand{\Swcurvar}{\@ifnextchar[{\basicSwcurvar}{\basicSwcurvar[160]}}%

\newcommand{\Escurve}[2]%
{\testdiagrammode\begin{picture}(0,0)%
\truex{1400}\truey{2000}\truez{200}%
\put(\value{x},0){\oval(\value{y},#200)[r]}%
\put(\value{x},0){\makebox(0,0){\begin{picture}(0,#200)%
\put(0,0){\line(-1,0){\value{z}}}%
\put(0,#200){\line(-1,0){\value{z}}}%
\put(-\value{z},0){\whead}%
\end{picture}}}%
\truex{2500}%
\put(\value{x},0){\makebox(0,0)[l]{${#1}$}}%
\end{picture}}%

\def\basicescurvar[#1]{\Escurve{}{#1}}

\newcommand{\escurvar}{\@ifnextchar[{\basicescurvar}{\basicescurvar[160]}}%

\def\basicEscurvar[#1]#2{\Escurve{#2}{#1}}%

\newcommand{\Escurvar}{\@ifnextchar[{\basicEscurvar}{\basicEscurvar[160]}}%

\newcommand{\Encurve}[2]%
{\testdiagrammode\begin{picture}(0,0)%
\truex{1400}\truey{2000}\truez{200}%
\put(\value{x},0){\oval(\value{y},#200)[r]}%
\put(\value{x},0){\makebox(0,0){\begin{picture}(0,#200)%
\put(0,0){\line(-1,0){\value{z}}}%
\put(0,#200){\line(-1,0){\value{z}}}%
\put(-\value{z},#200){\whead}%
\end{picture}}}%
\truex{2500}%
\put(\value{x},0){\makebox(0,0)[l]{${#1}$}}%
\end{picture}}%

\def\basicencurvar[#1]{\Encurve{}{#1}}

\newcommand{\encurvar}{\@ifnextchar[{\basicencurvar}{\basicencurvar[160]}}%

\def\basicEncurvar[#1]#2{\Encurve{#2}{#1}}%

\newcommand{\Encurvar}{\@ifnextchar[{\basicEncurvar}{\basicEncurvar[160]}}%

\newcommand{\Wscurve}[2]%
{\testdiagrammode\begin{picture}(0,0)%
\truex{1300}\truey{2000}\truez{200}%
\put(-\value{x},0){\oval(\value{y},#200)[l]}%
\put(-\value{x},0){\makebox(0,0){\begin{picture}(0,#200)%
\put(0,0){\line(1,0){\value{z}}}%
\put(0,#200){\line(1,0){\value{z}}}%
\put(\value{z},0){\ehead}%
\end{picture}}}%
\truex{2400}%
\put(-\value{x},0){\makebox(0,0)[r]{${#1}$}}%
\end{picture}}%

\def\basicwscurvar[#1]{\Wscurve{}{#1}}

\newcommand{\wscurvar}{\@ifnextchar[{\basicwscurvar}{\basicwscurvar[160]}}%

\def\basicWscurvar[#1]#2{\Wscurve{#2}{#1}}%

\newcommand{\Wscurvar}{\@ifnextchar[{\basicWscurvar}{\basicWscurvar[160]}}%

\newcommand{\Wncurve}[2]%
{\testdiagrammode\begin{picture}(0,0)%
\truex{1300}\truey{2000}\truez{200}%
\put(-\value{x},0){\oval(\value{y},#200)[l]}%
\put(-\value{x},0){\makebox(0,0){\begin{picture}(0,#200)%
\put(0,0){\line(1,0){\value{z}}}%
\put(\value{z},#200){\ehead}%
\put(0,#200){\line(1,0){\value{z}}}%
\end{picture}}}%
\truex{2400}%
\put(-\value{x},0){\makebox(0,0)[r]{${#1}$}}%
\end{picture}}%

\def\basicwncurvar[#1]{\Wncurve{}{#1}}

\newcommand{\wncurvar}{\@ifnextchar[{\basicwncurvar}{\basicwncurvar[160]}}%

\def\basicWncurvar[#1]#2{\Wncurve{#2}{#1}}%

\newcommand{\Wncurvar}{\@ifnextchar[{\basicWncurvar}{\basicWncurvar[160]}}%

\catcode`\@=12

\begin{document}

\begin{center}
{\Large \bf On the Structure and Function of Scientific Perspectivism in Quantum Mechanics}
\end{center}

\vspace{.5mm}

\begin{center}
{\large \bf Vassilios Karakostas and Elias Zafiris}
\end{center}

\footnotetext{To contact the authors, please write to: Vassilios Karakostas, {\it Department of Philosophy and History of Science, Faculty of Science, University of Athens, Athens 157 71, Greece; e-mail: karakost@phs.uoa.gr}, or to Elias Zafiris, {\it Department of Logic, E\"{o}tv\"{o}s University, Budapest, Hungary; e-mail: ezafiris@math.uoa.gr}.}

\begin{abstract}
\hspace{-45.2pt}
\begin{minipage}{1\textwidth}

\noindent
Contemporary scientific perspectivism is re-evaluated and extended to a comprehensive perspectivist methodology and `mediated' realistic epistemology, especially, with reference to quantum mechanics. In the present study, this is realized by representing categorically the global structure of a quantum algebra of events in terms of structured multitudes of interrelated local Boolean frames, realized as suitable perspectives or contexts for measuring physical quantities. The philosophical meaning of the proposed approach implies that the quantum world can be consistently approached and comprehended through a multilevel structure of locally variable perspectives, which interlock, in a category-theoretical environment, to form a coherent picture of the whole in a nontrivial way. Thus, in contrast to a panoptical ``view from nowhere'' of the classical paradigm, quantum theory acknowledges in an essential way a perspectival/contextual character of scientific knowledge.

\end{minipage}
\end{abstract}

\vspace{3.5mm}

\renewcommand{\baselinestretch}{1.15}

\section{Scientific Perspectivism as an Epistemological Methodological Framework of Theory-Building}
Contemporary scientific perspectivism, invigorated initially in the work of Ronald Giere (2006), is re-evaluated and extended to a comprehensive perspectivist methodology and `mediated' realistic epistemology, especially, with reference to quantum mechanics. In the present study, scientific perspectivism is primarily viewed as an epistemological methodological framework of how we obtain and form scientific knowledge of nature, through a broadly perspectivist process, constituting arguably an indispensable part in theory-building.

In relation to philosophical matters, scientific perspectivism occupies a middle ground between the extremes of the context-free universals of metaphysical objectivism, the rigid reductive methodology favored by positivist considerations of science, and the inherent relativism entailed by certain sociological accounts of science (e.g., Massimi 2012; Buzzoni 2016). Especially, in the proposed form, scientific perspectivism amply recognizes the existence of a mind-independent world as being logically {\it prior} to experience and knowledge, constituting the overarching condition for the possibility of knowledge. It emphasizes, however, that scientific knowledge of the world can never be pure, direct or unmediated, since it requires pre-conceptualization or structural organization; it requires the adoption of a perspective. Any knowledge and comprehension of something, either through a process of perception and classification or of normative structuring or purposeful acting, any attempt of combining individual experimental data or incorporating phenomena under generic patterns, etc., necessitates the endorsement of a conceptual-interpretative scheme and the selection of specific perspectives. In this respect, a context-free and interpretation-free access to reality as such seems an illusion. The `book of nature' proves too subtle and complex to be determined by just reading off reality.

In relation to methodological matters and, in contrast, to the usual understanding of a `perspective' as a visual metaphor involving a viewing projection, which, by itself, depicts the process of knowing as a passive activity, the concept of perspective is conceived in the present study in an {\it interactive} manner as the primary vehicle of tracing and investigating the world, as the principal unit of probing the world. A perspective is characterized endo-theoretically, namely, within a specific discipline, by a set of variables that are used to describe systems or to partition objects into parts, which together give a systematic account of a domain of phenomena.

It is worth noting that for perspectivism, the separation between the knowing subject and the object to be known, the partition between the observer and the observed, required for an objective description of phenomena, is {\it neither absolute nor catholic} as Cartesian-like epistemological approaches advocate, thus promoting an allegedly context-free account of the world (Karakostas 2012). The subject-object partition is accomplished upon the condition of the adopted perspective. The choice to adopt a particular perspective signifies also the approval of a conceptual scheme on the basis of which one may isolate which of the many available properties do, and which do not count for the purposes of description, since the world does not come with one preferred system of description. Consequently, scientific observation may be regarded as perspectival in the sense that claims about what is observed cannot be completely detached, in all circumstances, from the context of observation. The significance of this point is particularly pertinent to quantum theory due to the existence of {\it incompatible} physical quantities, represented by corresponding non-commuting self-adjoint operators, pertaining to any nontrivial quantum system; measuring apparatuses of such quantities cannot be held simultaneously in quantum mechanics (Sect. 2).

Especially, in relation to empirical testing of theories in contemporary physics, theoretical and methodological considerations specify the perspective from which we articulate the elementary yes-no experimental propositions or questions associated with properties of physical systems, in the sense that, on the one hand, they supply with a well-defined meaning the question that is put to nature, and, on the other hand, they specify the {\it kind} of the operations to be performed in order to ascertain particular answers to them (Karakostas 2014). In this sense, it is legitimate to say that the perspectival nature of experimental/empirical knowledge is an essential characteristic of acquiring scientific knowledge.

A single perspective provides, by definition, a partial/local and, thus, an incomplete description of the system to which it applies. Yet, the systematization of knowledge requires that perspectives associated with all aspects of a system can be correlated forming a {\it synthesized unity}, but they cannot be simply combined as independent integral parts of a third perspective (Sect. 3). Hence, a perspective of all perspectives or, equivalently, a panoptical perspective from nowhere does not exist. It is crucial, however, that a full-fledged analysis of a successful framework of perspectivism in science ought to provide a syntax of perspectives, illustrating how {\it locally} shared perspectives can (or cannot) be meaningfully combined at a higher theoretical level. Consequently, nature can be grasped scientifically, through structured multitudes of local variable perspectives, forming a {\it coherent} {\it multilevel} theoretical structure, exemplified by experimental procedures that render possible specific access to specific aspects of physical reality. Precisely this demanding task is accomplished by our category theoretic, perspectivist approach to quantum mechanics (Sect. 4).

\section{The Affinity of Perspectivist/Contextual Reasoning to Quantum Mechanics}
Standard quantum mechanics is formulated on a separable, complex Hilbert space associated to a physical system. In this framework, {\it quantum events} or {\it elementary propositions}, that is, true/false questions concerning values of physical quantities, are represented by orthogonal projection operators $\{\hat {P}_i\}$ on the system's Hilbert space $H$ or, equivalently, by the closed linear subspace $H_{\hat {P}_i}$ of $H$ upon which the projection operator $\hat {P}_i$ projects. The one-to-one correspondence between the {\it set of all closed linear subspaces} of $H$ and the {\it set of all projection operators}, denoted by $L_H$, allows a translation of the lattice structure of the subspaces of Hilbert space into the algebra of projections with the appropriate lattice theoretic characterizations (e.g., Varadarajan 2007).
Then, a {\it quantum algebra of events} is identified with the algebraic structure of all projection operators on Hilbert space, ordered by inclusion and carrying an orthocomplementation operation, thus forming a complete, atomic, orthomodular lattice. In effect, a non-classical, non-Boolean logical structure is induced which has its origin in quantum theory.

An immediate path for revealing the affinity of the perspectivist/contextual reasoning to quantum mechanics is provided through Kochen-Specker's celebrated theorem.
According to the latter, for any quantum system associated to a Hilbert space of dimension greater than two, {\it there does not exist a two-valued homomorphism} or, equivalently, {\it a truth-functional assignment} $h: {L}_H \rightarrow \{0, 1\}$ on the set of projection operators, ${L}_H$, interpretable as quantum mechanical propositions, preserving the lattice operations and the orthocomplement.
The essence of the theorem, when interpreted semantically, asserts the impossibility of assigning definite truth values to {\it all} propositions pertaining to a physical system at any one time, for any of its quantum states, without generating a contradiction.

The Kochen-Specker result shows, in physical terms, that in a system represented by a Hilbert space of three or more dimensions, there exist projection operators $\{\hat {P}_i\}$ such that it is not always possible to assign truth values 0 and 1 to all corresponding propositions pertaining to the system, so that the following conditions are fulfilled:
\begin{enumerate}
\item[(i)] For any orthogonal $i-$tuple of projection operators, $\{\hat {P}_i\}$, the assignment satisfies $\sum_{i}\hat {P}_{i} = \hat 1$, that is, one projection operator is mapped onto 1 (`true') and the remaining $i-1$ projection operators are mapped onto 0 (`false') (completeness of the basis condition).
\item[(ii)] If a projection operator, $\hat {P}_k$, belongs to multiple complete orthogonal bases, then, it is consistently assigned the {\it same} value in all bases (noncontextuality conditionn).
\end{enumerate}

The initial proof of Kochen and Specker establishes that no such assignment of 1's and 0's is possible for a special case restricted to a finite sublattice of projection operators on a three-dimensional Hilbert space, associated to a spin-1 quantum system, in a way that preserves the noncontextuality condition. The ingenuity of the proof, essentially of a geometrical nature, and its far reaching consequences have gradually generated an overwhelming production of theoretical and experimental research on foundational issues in quantum mechanics related to the contextual character of the theory as a structural feature of the quantum mechanical formalism itself (e.g., Cabello et al. 2010; Howard et al. 2014).

A failure of the noncontextuality condition means that the value assigned to a quantum mechanical observable $A$, whose representing self-adjoint operator $\hat{A}$ is analyzed in terms of spectral projections $\hat {P}_i$, depends on the {\it context} in which it is considered.
An equivalent way of expressing the above is to say that the value of $A$ depends on what {\it other} compatible observables are assigned values at the same time; i.e., it depends on a choice that concerns operators that commute with $\hat A$.
This dependence captures the endemic feature of {\it quantum contextuality} and may be highlighted by using an explicit example.  Let us consider, for reasons of simplicity, a triad of observables $\{A, B, C \}$ representing physical quantities of a quantum system $S$. According to quantum theory, it is possible to simultaneously measure a set of observables reliably if and only if the corresponding operators are commutative. Let us, then, assume that the operator $\hat A$ pertaining to system $S$ commutes with operators $\hat B$ and $\hat C$
($[\hat A, \hat B] = 0 = [\hat A, \hat C]$), not however the operators $\hat B$ and $\hat C$ with each other ($[\hat B, \hat C] \neq 0$).
Then, due to the non-commutativity of the last pair of operators, the result of a measurement of observable $A$ depends on whether the system had previously been subjected to a measurement of $B$ or a measurement of $C$ or in none of them.
Thus, the value of the observable $A$ depends upon the set of mutually compatible observables one may consider it with, that is, the value of $A$ depends upon the selected set of measurements. In other words, the value of the observable $A$ cannot be thought of as pre-fixed, as being independent of the experimental context actually chosen, as specified, in our example, by the $\{\hat A, \hat B \}$ or $\{\hat A, \hat C \}$ frame of mutually commuting operators.
It is worth noting that the formalism of quantum theory does not imply how such a contextual valuation might be obtained on the set ${L}_H$ of all projection operators on a Hilbert space or what properties it should possess.

To this end, we resort to the powerful methods of category theory, which directly captures the idea of structures varying over contexts, thus providing a natural setting for investigating multilevel structures, studying contextual phenomena, and providing context-dependent operations in logic (e.g., Zafiris 2004; D\"{o}ring \& Isham 2008; Abramsky \& Brandenburger 2011; Zafiris \& Karakostas 2013).
In the proposed category theoretic, perspectival representation of a quantum algebra of events, developed in Sect. 4, the notion of {\it perspective} that is applied on a quantum system is tantamount to a set of mutually compatible physical quantities, as in the preceding example, or, more precisely, to a complete Boolean algebra of commuting projection operators generated by such a set.
In other words, in our approach a Boolean algebra in the lattice of quantum events, picked by a self-adjoint operator, serves as a {\it Boolean probing frame} relative to which corresponding results of measurement are being coordinatized (cf., Svozil 2009).
This naturally leads, by extension, to a {\it horizon of perspectives} on the structure of quantum events with respect to various Boolean frames realized as experimental contexts for measuring physical quantities. Be sure, no single context or perspective can deliver a complete picture of the quantum system, but, by applying category theoretic reasoning, it is possible to combine them suitably in an overall structure that will capture the entire system. It is also of great importance how the various contexts or contextual perspectives relate to each other.
Categorically speaking, this consideration is naturally incorporated into our scheme, since the category theoretic representation of quantum event algebras in terms of Boolean probing contexts can be described by means of a multilevel structure, mathematically known as a {\it topos}, which stands for a {\it category of sheaves of variable, overlapping and interconnected families of local Boolean frames}, capable of carrying all the information encoded in the former.

\section{The Categorical Imperative: A Novel Mode of Object Specification in Perspectivist Terms}
Category theory provides a general theoretical framework for the study of structured systems in terms of their {\it mutual relations} and {\it admissible transformations}. Contrary to the atomistic approach of set theory, which crucially depends on the concept of elements-points and the membership relationship of a variable $x$ in a set $X$, $x \in X$, in category theory the notion of {\it morphism} or {\it arrow} undertakes primary role.
A morphism, for instance, $f: A \rightarrow B$ in a category $\mathcal C$ expresses one of the many possible ways in which the object $A$ relates to the object $B$ within the context of category $\mathcal C$. Thus, the incoming morphism to $B$ from any other object $A$ in category $\mathcal C$ may naturally be considered as a {\it local perspective} targeting $B$ whose source is $A$ in the same category.
The category theoretic mode of thinking incorporates internally the very nature of ``pointing at" or ``viewing from" (e.g., Lawvere \& Schanuel 2009).

Accordingly, the categorical framework signifies a remarkable conceptual change in the way of conceiving the form and function of mathematical objects, departing from the axiomatic set-theoretic approach.
The emphasis now is put on the specification of objects in terms of the relations they bear with other objects of the same category.
However counterintuitive it may initially appear, in category theory the nature of the objects is a {\it derivative} aspect of the patterns described by the morphisms or mappings that connect the objects.
In fact, by virtue of Yoneda's fundamental lemma, an object can be {\it completely classified} and {\it retrieved} by the network of all morphisms -- the structure preserving relations -- thought of in the present study as local perspectives, targeting this object within the same category (e.g., Awodey 2010, 187-189).
The qualification of complete classification refers to the fact that the whole network of targeting morphisms specifies the object uniquely up to canonical isomorphism.
The ground breaking consequence of this proposition is that the investigated object can be legitimately subjugated or even conceptually substituted by the network of all structure-preserving relations targeting it within the same category. In this manner, the object constitutes a {\it perspectival representation} of the whole network of relations directed to it, and, inversely, this network becomes {\it uniquely representable} up to equivalence by the targeted object.

Most significantly, the preceding considerations do not exclude the inter-level relational determination of objects belonging to {\it distinct} categorical species of structure, under the condition that there exists a {\it bi-directional} functorial correlation between them, formulated in the language of {\it adjunctions}. It is precisely the latter development that gradually introduced into category theory a paradigm change in understanding structures of general types and paved the way for forming bridges between seemingly unrelated mathematical disciplines, revealing, at the same time, the philosophical significance of category theory (e.g., Marquis 2009).
As analyzed in Sect. 4, it is indeed the categorical notion of adjunction, consisting of a pair of adjoined functors, that allows us to produce a perspectival representation of a quantum event algebra by linking appropriately the Boolean and the quantum structural levels.

In this general setting, the notion of a perspective on an object of a category is conceived as the principal means of probing or resolving this object, independently of any a priori requirement of analysis of the object in its set-theoretic elements.
This is the case because the notion of a perspective is not subordinate to a set-theoretic function between a probe and an object, but on the contrary, subsumes a well-defined structural characterization derived from the internalization of a probing relation within the category deciphering the species of the object under investigation (Sect. 4.1).
A probing relation, therefore, is qualified as a potential perspective on an object of a category, if and only if, it can be {\it internalized} within this category so that it can be expressed as a structure preserving morphism targeting the object of inquiry within this category.

It is worthy to underline that a single perspective, although incomplete in its capacity to resolve the investigated object globally or in its entirety, shapes the target locally or partially in a structurally adaptable and congruent manner, so that it can be internally extended under the proviso that overlaps compatibly with some other perspective deciphering another local cover of the investigated object (Sect. 4.2).
A crucial feature of this local perspectival schematism of an object in a category is that it does not assume or require the existence of an all-encompassing perspective, meaning that local perspectives should not be thought of as independent parts of an overall fixed perspective. In contradistinction, the perspectival schematism or formation of an object is based on the idea of a {\it multiplicity} of possible local perspectives, covering the object entirely only under their {\it joint} action, which is only constrained by the normative requirement of compatible interconnection on their pairwise overlaps whenever this is the case (Sect. 4.3).

Category theory is suitably equipped to deal successfully with such demanding structural problems in the natural sciences, since, besides the consideration of morphisms between objects in a category, there also exist, at the next higher level, structure preserving mappings between categories, namely {\it functors}, and, further on, mappings between functors called {\it natural transformations}. It is apparent that the theory of categories proceeds inherently in an hierarchical manner, encompassing structures of increasing abstraction and complexity. Thus, essential categorical notions and constructions present themselves in ascending levels of generality and depth: category, functor, natural transformation, adjointness, etc. The basic categorical principles that we adopt in the subsequent analysis are summarized as follows:

(i) To each kind of mathematical structure used to represent a system, there corresponds a {\it category} whose objects have that structure and whose arrows or morphisms preserve it.

(ii) To any natural construction on structures of one kind, yielding structures of another kind, there corresponds a {\it functor} from the category of the first specified kind to the category of the second. Specifically, given categories $\mathcal C$ and $\mathcal D$, a functor $\mathbb F: \mathcal C \rightarrow \mathcal D$ is a type of mapping that associates to every object $A$ of category
$\mathcal C$ an object ${\mathbb F}(A)$ of category $\mathcal D$, and to every morphism $f: A \rightarrow B$ in $\mathcal C$ a morphism ${\mathbb F}(f): {\mathbb F}(A) \rightarrow {\mathbb F}(B)$ in $\mathcal D$, by preserving the essential structural relationships among objects, that is, identity morphisms and composition of morphisms. Thus, a functor $\mathbb F: \mathcal C \rightarrow \mathcal D$ provides a sort of `picture' of category $\mathcal C$ in $\mathcal D$ by preserving the structure of $\mathcal C$.

(iii) To each natural translation between two functors having identical domains and codomains, there corresponds a {\it natural transformation}. Natural transformations define structure preserving mappings of functors, namely, they provide a way of transforming one functor into another while respecting the internal structure of the categories involved.
Given a pair of functors $\mathbb F : \mathcal C \rightarrow \mathcal D$ and $\mathbb G : \mathcal C \rightarrow \mathcal D$, having both the same domain and codomain categories, one may think of each of them as producing a `picture' of category $\mathcal C$ inside $\mathcal D$.
Then, intuitively, a natural transformation is a way to transform globally or systematically the `picture' defined by functor $\mathbb F$ onto the `picture' defined by $\mathbb G$.
The specification ``natural", particularly, in the notion of natural transformations refers to the comparison of two functorial processes, sharing the same source and target categories, in a way that captures the {\it shared structure} or {\it generic common properties} existing in different categorical contexts.
No doubt, the key concept of natural transformations acquires in category theory the status of a principle, analogous to general covariance in physics, that penetrates deeper than is initially discernible.

(iv) To any canonical {\it bi-directional functorial correlation} between two kinds of mathematical structures, there corresponds an {\it adjunction} between the corresponding categories. Thus, an adjunction between categories $\mathcal C$ and $\mathcal D$ is established, if and only if, there exist a pair of functors $\mathbb F: \mathcal C \rightarrow \mathcal D$ and
$\mathbb G: \mathcal D \rightarrow \mathcal C$ which are bijectively related, i.e., $\mathbb F :  {\mathcal C}  \adjar  { {\mathcal D}} : \mathbb G$. Then, $\mathbb F$ is called {\it left adjoint} to $\mathbb G$, correspondingly, $\mathbb G$ is designated {\it right adjoint} to $\mathbb F$.
Importantly, this means that the objects of the categories $\mathcal C$ and $\mathcal D$ are related with each other through natural transformations. It should be remarked that adjointness, and the related pair of adjoint functors, is a concept of fundamental logical and mathematical importance contributed to mathematics by category theory (e.g., Goldblatt 2006, 438).

\section{Perspectival Representation of a Quantum Event Structure via Adjunction of Boolean Frames}
The conceptual basis of the proposed perspectival representation of a quantum structure of events in terms of interconnected families of Boolean probing frames relies on the physically significant fact that it is possible to analyze or `coordinatize' the information contained in a quantum event algebra $L$ by means of structure preserving morphisms $B \rightarrow L$, having as their domains locally defined Boolean event algebras $B$. As alluded to in Sect. 2, any single map from a Boolean domain to a quantum event algebra does not suffice for a complete determination of the latter's information content.
In fact, it contains only the amount of information related to a particular Boolean frame, prepared for a specific kind of measurement, and inevitably is constrained to represent exclusively the abstractions associated with it.
This problem is confronted by employing a sufficient amount of maps, organized in terms of {\it covering sieves}, from the coordinatizing Boolean domains to a quantum event algebra so as to cover it completely. These maps furnish the role of local {\it Boolean covers} for the filtration of the information associated with a quantum structure of events, in that, their domains $B$ provide Boolean coefficients associated with typical measurement situations of quantum observables.
The local Boolean covers capture, in essence, separate, complementary features of the quantum system under investigation, thus providing a structural decomposition of a quantum event algebra in the descriptive terms of Boolean probing frames.
In turn, the incomplete and complementary local Boolean descriptions or perspectives can be smoothly {\it pasted} or {\it glued} together, by demanding the satisfaction of {\it partial compatibility} between {\it overlapping} local Boolean covers, so that one may arrive at the synthesis and actual determination of the global quantum event algebra itself.
The implementation of the {\it perspectival representation of a quantum event algebra in terms of structured multitudes of interconnected Boolean probing frames} requires distinct notions of Boolean and quantum categorical event structures, respectively.
The categorical methodology involved in the realization of the suggested approach necessitates the application of categorical sheaf theory\renewcommand{\baselinestretch}{1}\footnote{It is not possible to provide here a concise account of categorical sheaf theory. For a systematic introduction to this well-developed mathematical framework, the reader may consult, for instance, Mac Lane \& Moerdijk (1992) or Borceux (1994/2008).}
\renewcommand{\baselinestretch}{1.15}
to quantum structures, which, in relation to foundations of quantum logic, including also the fundamental problem of truth valuation in quantum mechanics, has been formulated and further expanded in recent publications of the authors (e.g., Zafiris 2006; Zafiris \& Karakostas 2013; Karakostas \& Zafiris 2017).

\subsection{Categories of Boolean and Quantum Event Structures}

\theoremstyle{definition}
\begin{definition}
A {\it Boolean categorical event structure} is a small category, denoted by
$\mathcal B$, which is called the category of Boolean event
algebras. The objects of $\mathcal B$ are complete Boolean
algebras of events and the morphisms are the corresponding Boolean
algebraic homomorphisms.
\end{definition}

\theoremstyle{definition}
\begin{definition}
A {\it quantum categorical event structure} is a locally small
co-complete category, denoted by $\mathcal L$, which is called the
category of quantum event algebras. The objects of $\mathcal L$
are quantum event algebras and the morphisms are quantum algebraic
homomorphisms.
\end{definition}

It is important to note that any arbitrary pair of events $l$ and ${\acute l}$  belonging to a quantum event algebra $L$ in $\mathcal L$, $l, {\acute l} \in L$, are compatible if the sublattice generated by
$\{{l,l^{\ast},{\acute l},{{\acute l}^{\ast}}}\}$, where $[-]^{\ast} : L \rightarrow L$ denotes the orthocomplementation operation, is a Boolean algebra, namely, if it is a Boolean sublattice of $L$. In perspectivist terms, this indicates that a Boolean event algebra is {\it structurally adaptable} to a quantum event algebra since it encodes a structurally invariant context of co-measurable observables by means of their joint compatible spectral resolution. It is natural, therefore, to consider a Boolean categorical event structure $\mathcal B$ as a {\it category of probes} for the category of quantum categorical event structure $\mathcal L$. The structural adaptability of the category of Boolean event algebras $\mathcal B$ to the category of quantum event algebras $\mathcal L$ gives rise to the Boolean probing or shaping functor of $\mathcal L$ by $\mathcal B$.

\theoremstyle{definition}
\begin{definition}
A {\it Boolean probing or shaping functor} of a quantum categorical event structure ${\mathcal L}$,  ${\mathbb M}:{\mathcal B} \rightarrow {\mathcal L}$, assigns to each Boolean event algebra in ${\mathcal B}$ the underlying quantum event algebra from ${\mathcal L}$, and to each Boolean homomorphism the underlying quantum algebraic homomorphism.
\end{definition}

\theoremstyle{corollary}
\begin{corollary}
The quantum algebraic homomorphism, $\psi_B: {\mathbb M} (B) \rightarrow L$, constitutes a {\it Boolean frame} of a quantum event algebra $L$, or, equivalently, a {\it Boolean perspective} on $L$ whose source is the Boolean probe $B$.
\end{corollary}

\theoremstyle{proposition}
\begin{proposition}
The Boolean shaping functor of a quantum categorical event structure $\mathcal L$,  ${\mathbb M}:{\mathcal B} \rightarrow {\mathcal L}$, is not invertible, i.e., there is no
opposite-directing functor from ${\mathcal L}$ to ${\mathcal B}$.
\end{proposition}

The proof of the above proposition follows immediately from Kochen-Specker's theorem, analyzed in Sect. 2, and, according to which, there does not exist a global Boolean two-valued truth-functional assignment pertaining to a quantum event algebra.

The shaping functor  ${\mathbb M}:{\mathcal B} \rightarrow {\mathcal L}$ is technically a forgetful functor that fulfills the requirement of structural adaptability of the category ${\mathcal B}$ of Boolean probes to the category of quantum event algebras. Because of the fact that an opposite-directing functor from ${\mathcal L}$ to ${\mathcal B}$ is not feasible, since a quantum event algebra cannot be realized within any Boolean event algebra, we seek for an extension of ${\mathcal B}$ into a larger categorical environment where such a realization becomes possible.
This extension is expected to conform with the intended perspectivist semantics of adjoining a multiplicity of Boolean probing frames to a quantum event algebra, understood equivalently as Boolean perspectives on the latter.
For this reason, it is necessary to extend the categorical level of $\mathcal B$ to the categorical level of diagrams in $\mathcal B$, such that the global information encoded in a quantum event algebra may be recovered in a structure preserving way by an appropriate sheaf-theoretic construction gluing together categorical diagrams of locally variable
Boolean frames.\renewcommand{\baselinestretch}{1}\footnote{In general, a diagram
$\mathbb X=(\{X_i \}_{i \in I}, \{F_{ij}\}_{i,j \in I})$ in a category $\mathcal C$ is defined as an indexed family of objects $\{X_i \}_{i \in I}$ and a family of
morphisms sets $\{F_{ij}\}_{i,j \in I} \subseteq {Hom_{\mathcal C}}({X_i},{X_j})$.}
\renewcommand{\baselinestretch}{1.15}
This is accomplished by means of the categorical technique of Yoneda's embedding ${\mathbb y} : \mathcal B \rightarrow {{\mathbb S}ets^{{\mathcal B}^{op}}}$, which is a full and faithful functor (e.g., Mac Lane \& Moerdijk 1992, 26).

\subsection{Functor of Boolean Frames and Pasting Maps}
It is apparent, therefore, that the realization of this extension process requires the construction of the functor category, ${{\mathbb S}ets^{{\mathcal B}^{op}}}$, called the
{\it category of presheaves of sets on Boolean event algebras}, where ${\mathcal B}^{op}$ is the opposite category of ${\mathcal B}$.

\theoremstyle{definition}
\begin{definition}
The functor category ${{\mathbb S}ets^{{\mathcal B}^{op}}}$ has
objects all contravariant functors $ {\mathbb P}: {\mathcal B}^{op} \rightarrow {{\mathbb S}ets}$ and morphisms all natural transformations between such functors.
\end{definition}
Each object ${\mathbb P}$ in the functor category ${{\mathbb S}ets^{{\mathcal B}^{op}}}$ is a contravariant
set-valued functor on ${\mathcal B}$,  called a {\it presheaf of sets} on ${\mathcal B}$ (e.g., Borceux 1994/2008, 195).
In order to obtain a clear understanding of the structure of the functor category ${{\mathbb S}ets^{{\mathcal B}^{op}}}$, it is useful to think of a presheaf of sets ${\mathbb P}$ in ${{\mathbb S}ets^{{\mathcal B}^{op}}}$ as a right action of the category $\mathcal B$ on
a set of events, which is partitioned into a variety of Boolean spectral kinds parameterized  by
the Boolean event algebras $B$ in $\mathcal B$. Such an action $\mathbb P$ is
equivalent to the specification of a {\it diagram} in $\mathcal B$,
to be thought of as a $\mathcal B$-variable set forming a presheaf ${\mathbb P}(B)$ on ${\mathcal B}$.

For each Boolean algebra $B$ of ${\mathcal B}$,  ${\mathbb P}$(B) is
a set, and for each Boolean homomorphism $f: C \rightarrow B$, ${\mathbb P} (f) :
{\mathbb P}(B) \rightarrow   {\mathbb P}$(C) is a set-theoretic function, such that, if $p \in {\mathbb
P}$(B), the value ${\mathbb P}(f) (p)$ for an arrow $f : C \rightarrow B$
in ${\mathcal B}$ is called the {\it restriction} of $p$ along $f$ and
is denoted by ${\mathbb P}(f) (p)=p \cdot f$.
From a physical viewpoint, the purpose of introducing the notion of a presheaf ${\mathbb P}$ on $\mathcal B$, in the environment of the functor category ${{\mathbb S}ets^{{\mathcal B}^{op}}}$, is to identify an element of ${\mathbb P}$(B), that is, $p \in {\mathbb P}$(B), with an event observed by means of a physical procedure over a Boolean domain cover for a quantum event algebra.
As demonstrated in Proposition 4.3, this identification forces the interrelation of observed events, over all Boolean probing frames of the base category $\mathcal B$, to fulfil the requirements of a uniform and homologous fibred structure.

\theoremstyle{definition}
\begin{definition}
The {\it Boolean realization functor} {\it of a quantum categorical event structure} ${\mathcal L}$ in ${{\mathbb {S} ets}^{{\mathcal B}^{op}}}$, namely, the functor of generalized elements of $\mathcal L$ in the environment of the category of presheaves on Boolean event algebras, is defined as
\begin{equation}
{\mathbb {R}}: { {\mathcal L}} \rightarrow {{\mathbb {S} ets}}^{{{\mathcal{B}}}^{op}} \\,
\end{equation}
where the action on a Boolean algebra $B$ in $\mathcal B$ is given by
\begin{equation}
{\mathbb {R}}(L)(B)={Hom_{\mathcal L}}({\mathbb M}(B),L) \\.
\end{equation}
\end{definition}
The presheaf functor ${\mathbb {R}}(L)(-)={Hom_{\mathcal L}}({\mathbb M}(-),L)$
constitutes the image of ${\mathbb {R}}$ in ${{\mathbb {S} ets}}^{{{\mathcal{B}}}^{op}}$ and is called the {\it functor of Boolean frames} or {\it functor of Boolean perspectives} on a quantum event algebra $L$ in ${\mathcal L}$.
Since the physical interpretation of the presheaf functor ${\mathbb {R}}(L)(-)$ refers to the functorial realization of a quantum event algebra $L$ in $\mathcal L$ in terms of structured multitudes of local Boolean frames adjoined to it, intuitively, it is natural to think of ${\mathbb {R}}(L)(-)$ as comprising the {\it network of relationships} that $L$ has with all admissible Boolean frames-perspectives on $L$.

\theoremstyle{proposition}
\begin{proposition}
The Boolean frames pertaining to a quantum event algebra $L$,
\begin{equation}
{\psi_B}: {\mathbb M}(B) \rightarrow L,
\end{equation}
being instantiated by the evaluation of the functor ${\mathbb {R}}(L)(-)$ at each $B$ in $\mathcal B$, are interrelated by the operation of presheaf restriction.
\end{proposition}
It is sufficient to observe that for each Boolean homomorphism $f : C \rightarrow B$, ${{\mathbb {R}}(L)} (f) :
{{\mathbb {R}}(L)}(B) \rightarrow   {{\mathbb {R}}(L)}(C)$ is a function between sets of Boolean frames of $L$ in the opposite direction, such that, if $\psi_B
\in {{\mathbb {R}}(L)}(B)$ is a Boolean frame of $L$, the value of ${{\mathbb {R}}(L)}(f) (\psi_B)$, or, equivalently, the corresponding Boolean frame $\psi_C: {\mathbb M}(C) \rightarrow L$ is given by the restriction or pullback of $\psi_B$ along $f$, denoted by ${{\mathbb {R}}(L)}(f) (\psi_B)=\psi_B \cdot f=\psi_C.$

\theoremstyle{corollary}
\begin{corollary}
For a fixed quantum event algebra $L$ in $\mathcal L$, the set of all pairs $(B,\psi_B)$, where $B$ is a Boolean event algebra and $\psi_B: \mathbb M (B) \rightarrow L$ a Boolean probing frame of $L$ defined over $B$, has the structure of a category.
\end{corollary}
On the basis of Definition 4.5, the functor of Boolean frames ${\mathbb {R}}(L)$ of a quantum event algebra $L$ in $\mathcal L$ forms a presheaf of sets on Boolean event algebras $B$ in $\mathcal B$.
Thus, we can legitimately consider the {\it category of elements} corresponding to the functor ${\mathbb {R}}(L)$, denoted by ${{\int}}({\mathbb {R}}(L),{\mathcal B})$, and specified as follows: it has objects all pairs $(B,\psi_B)$ and morphisms ${(\acute{B},{\psi_{\acute B}})
{\rightarrow}(B,\psi_B)}$ are those Boolean homomorphisms $u : \acute{B}
{\rightarrow} B$ of category ${\mathcal B}$ for which $\psi_B \cdot u={\psi_
{\acute B}}$, that is, the restriction or pullback of the Boolean frame $\psi_B$ along $u$ is $\psi_{\acute B}$.
This category is naturally called the {\it category of Boolean frames} of $L$, or, equivalently, the {\it category of Boolean perspectives} on $L$.

\theoremstyle{proposition}
\begin{proposition}
For a fixed quantum event algebra $L$ in $\mathcal L$, the category of  Boolean frames of $L$ induces a split, discrete and uniform fibration of $L$ over its Boolean probes,  where $\mathcal B$ is the base category of the fibration.
\end{proposition}
By projecting on the second coordinate of the category of Boolean frames of $L$, ${{\int}}({{\mathbb {R}}(L)},{\mathcal B})$, we obtain a functor, ${{{\int}}}_{{\mathbb {R}}(L)} :
{{\int}}({{\mathbb {R}}(L)},{\mathcal B})  {\rightarrow}  {\mathcal B}$, as in the diagram below:
\begin{diagram}
€{\mathbb {\int}}({{\mathbb {R}}(L)}, {\mathcal B} )€€ €\Sar {{\mathbb
{\int}}_{{{\mathbb {R}}(L)}}}€€ €{\mathcal B} €\Ear {{{\mathbb {R}}(L)}} € {{\mathbb {S} ets}}€€
\end{diagram}
We first note that, in the case under study, the fibers are categories in which the only arrows
are identity arrows and, thus, the fibration induced by the presheaf functor ${\mathbb {R}}(L)(-)$ of Boolean frames of $L$ is discrete.
If $B$ is a Boolean probe in ${\mathcal B}$, the
inverse image under ${{\int}}({\mathbb {R}}(L),{\mathcal B})$ of $B$ is simply
the set of Boolean frames of $L$, i.e. ${\mathbb {R}}(L)(B)$, although its elements are written as
pairs $(B,\psi_B)$ so as to form a disjoint union.
The emergence of a measurement event $q \in \mathbb M(B)$ with respect to the Boolean frame $\psi_B: \mathbb M (B) \rightarrow L$ amounts to the choice of a projection $q \in B$. In this sense, the Boolean frame $\psi_B$ becomes a pointed one.
Therefore,  choice of projections effected by measurement procedures with respect to Boolean frames of $L$ make the fibration split.
Finally, the fibration is uniform over the base category $\mathcal B$ because for any two measurement events over
the same Boolean event algebra, the structure of all Boolean frames that relate to the first event cannot be distinguished in any
possible way from the structure of Boolean frames relating to the second. Henceforth, all possible events with respect to any particular Boolean frame
are uniformly equivalent to each other. Accordingly, the fibration ${{{\int}}}_{{\mathbb {R}}(L)} :
{{\int}}({{\mathbb {R}}(L)},{\mathcal B})  {\rightarrow}  {\mathcal B}$ amounts to a {\it partitioning} of a quantum event algebra $L$ into {\it partially congruent Boolean perspectives} parameterized by the Boolean probes of the base category $\mathcal B$ of the fibration.

The explicit representation of a quantum event algebra $L$ in $\mathcal L$ in terms of coherently interconnected families of Boolean probing frames, capable of carrying all the information encoded in the former, requires the satisfaction of partial compatibility between overlapping pairs of Boolean frames.
\theoremstyle{definition}
\begin{definition}
The {\it categorical overlap or pullback} of any pair of Boolean frames ${\psi}_B :
{\mathbb M}(B) \rightarrow L$, $B$ in $\mathcal B$, and ${\psi}_{\acute
B} : {\mathbb M}({\acute B}) \rightarrow L$, ${\acute B}$ in $\mathcal B$, with common codomain a quantum event algebra $L$,
consists of the event algebra  ${\mathbb M}(B) {\times}_L {\mathbb
M}({\acute B})$ together with the two arrows $\psi_{B \acute B}$ and
$\psi_{\acute B B}$, called projections, as shown in the diagram:
\begin{diagram}
{\mathbb M (\acute{\acute B})}€€
 €\Sear u€\Esear h €€
 €\sseaR g €{{\mathbb M}(B)} {\times}_L {{\mathbb M}(\acute B)}       €\Ear {{\psi}_{B,{\acute B}}}  €{{\mathbb M}(B)} €€
 €          €\saR {{\psi}_{{\acute B},B}}   €         €\saR  {{\psi}_B}  €€
 €          €{{\mathbb M}({\acute B})}       €\Ear  {{\psi}_{\acute B}} €L€€
\end{diagram}
\end{definition}

\theoremstyle{proposition}
\begin{proposition}
If the Boolean frames ${\psi}_B$ and ${\psi}_{\acute B}$ of $L$ are injective, then their
pullback is given by their intersection.
\end{proposition}
Note that the square in the preceding diagram {\it commutes} and for any Boolean domain object ${\mathbb M}(\acute{\acute B})$ or event algebra $\acute{\acute B}$ in ${\mathcal B}$ and arrows $h$ and $g$ that make the outer square commute, there is a unique $u : {\mathbb M}(\acute{\acute B}) \ar {\mathbb M}(B) {\times}_L {\mathbb M}({\acute B})$ that makes the whole diagram commutative. Hence, we obtain the compatibility condition: ${\psi}_{\acute B} \circ g = {\psi}_B \circ h$. If, therefore, ${\psi}_B$ and ${\psi}_{\acute B}$ are injective morphisms, then their pullback is isomorphic with the intersection ${\mathbb M}(B) \cap {\mathbb M}({\acute B})$. Accordingly, we can define the {\it gluing or pasting map} between Boolean frames on their overlap, which is an isomorphism.

\theoremstyle{definition}
\begin{definition}
The {\it pairwise gluing isomorphism} of the Boolean frames ${\psi}_B$ and ${\psi}_{\acute B}$ of $L$ is defined as follows:
\begin{equation}
{\Omega}_{B, \acute B} : \psi_{\acute B
B}({\mathbb M}(B) {\times}_L {\mathbb M}({\acute B})) \ar \psi_{B \acute B}({\mathbb M}(B) {\times}_L {\mathbb M}({\acute B})),
\end{equation}
\begin{equation}
{\Omega}_{B, \acute B}=\psi_{B \acute B} \circ  {\psi_{\acute B
B}}^{-1}.
\end{equation}
\end{definition}

\theoremstyle{proposition}
\begin{proposition}
The Boolean coordinatizing maps $\psi_{\acute B B} : ({\mathbb M}(B) {\times}_L {\mathbb
M}({\acute B})) \rightarrow L$ and $ \psi_{B \acute B} : ({\mathbb M}(B) {\times}_L
{\mathbb M}({\acute B})) \rightarrow L$ cover $L$ in a compatible way on their intersection.
\end{proposition}
An immediate consequence of Definition 4.7 is the satisfaction of the following {\it cocycle conditions}:
\begin{align}
{\Omega}_{B, B}&=1_B                                                                               & 1_{B}&: \text{identity of}\,\, B \\
{\Omega}_{B, \acute B} \circ {\Omega}_{\acute B, \acute{\acute B}}&={\Omega}_{B, \acute{\acute B}} & \text{if}\,\,\,& {\mathbb M}(B)\cap {\mathbb M}(\acute B)\cap {\mathbb M}(\acute{\acute B}) \neq 0 \\
{\Omega}_{B, \acute B}&={{\Omega}^{-1}}_{\acute B, B}                                              & \text{if}\,\,\,& {\mathbb M}(B)\cap {\mathbb M}(\acute B) \neq 0.
\end{align}
Thus, the pairwise gluing isomorphism ${\Omega}_{B, \acute B}$ between any two injective Boolean frames of $L$ assures that $\psi_{\acute B B} : ({\mathbb M}(B) {\times}_L {\mathbb
M}({\acute B}))$ and $ \psi_{B \acute B} : ({\mathbb M}(B) {\times}_L
{\mathbb M}({\acute B}))$ probe $L$ on their common refinement in a compatible way.

\subsection{The Boolean Frames -- Quantum Adjunction}
We constructed in Sect. 4.2 the Boolean realization functor of a quantum categorical event structure ${\mathcal L}$,
${\mathbb {R}}: { {\mathcal L}} \rightarrow {{\mathcal {S} ets}}^{{{\mathcal{B}}}^{op}}$,
realized for each $L$ in $\mathcal L$ by its presheaf functor of Boolean frames,
${\mathbb {R}}(L)(B)={Hom_{\mathcal L}}({\mathbb M}(B),L)$, where the shaping functor ${\mathbb M}:{\mathcal B} \rightarrow {\mathcal L}$
fulfills the requirement of structural adaptability of the category of Boolean event algebras to the category of quantum event algebras.
In this categorical setting, the problem of establishing a perspectival representation of a quantum event algebra via adjunction of Boolean frames is solved exactly
by considering the opposite-directing, left adjoint functor $\mathbb L : {{\mathbb {S} ets}^{{{\mathcal B}}^{op}}} \rightarrow \mathcal L$,
to the Boolean realization functor $\mathbb R: { {\mathcal L}} \rightarrow {{\mathbb {S} ets}^{{{\mathcal B}}^{op}}}$.
If, therefore, for a fixed quantum event algebra $L$ in $\mathcal L$, the right adjoint functor $\mathbb R$ partitions or decomposes $L$ in an orderly manner via the action of Boolean probing frames $\psi_B: {\mathbb M}(B) \rightarrow L$, functioning as suitable perspectives or contexts for measurement of observables, then, inversely,
the left adjoint functor $\mathbb L$ provides a {\it perspectival synthesis} of a quantum event algebra, in a structure preserving manner, by gluing together structured families or diagrams of variable local Boolean frames.

Thus, the existence of the functor $\mathbb L$, being the left adjoint to $\mathbb R$, gives rise to a {\it categorical adjunction} that has been recently proved to exist between the category of quantum event algebras ${\mathcal L}$ and the category of presheaves ${{\mathbb {S} ets}^{{{\mathcal B}}^{op}}}$
on Boolean event algebras (Zafiris 2006; see also Zafiris \& Karakostas 2013 for a more detailed treatment including in addition the involved logical aspects).
Since the proposed perspectivist interpretation of a quantum event structure is based on this pair of adjoint functors, it is useful to express their established bi-directional correspondence in the form of the following theorem.

\theoremstyle{theorem}
\begin{theorem}
There exists a categorical adjunction between the categories ${{\mathbb {S} ets}^{{\mathcal B}^{op}}}$ and ${\mathcal L}$, called the Boolean frames--quantum adjunction, established by the pair of adjoint functors ${\mathbb L}$ and ${\mathbb R}$, as follows:
\begin{equation}
\mathbb L : {{\mathbb {S} ets}^{{\mathcal B}^{op}}} \adjar {\mathcal L} : \mathbb R \\.
\end{equation}
Equivalently, there exists a bijection, which is natural in both $\mathbb P$ in ${{\mathbb {S} ets}^{{\mathcal B}^{op}}}$ and $L$ in ${\mathcal L}$,
\begin{equation}
{Hom}_{{{\mathbb {S} ets}^{{\mathcal B}^{op}}}} ({\mathbb P},{\mathbb R}(L)) \cong {{Hom}_{\mathcal
L}({\mathbb L \mathbb P}, L)} \\,
\end{equation}
abbreviated as
\begin{equation}
Nat({\mathbb P},{\mathbb R}(L)) \cong {Hom}_{\mathcal
L}({\mathbb L \mathbb P}, L) \\.
\end{equation}
\end{theorem}

\theoremstyle{corollary}
\begin{corollary}
The left adjoint functor of the Boolean frames-quantum adjunction, $\mathbb L : {{\mathbb {S} ets}^{{{\mathcal B}}^{op}}} \rightarrow \mathcal L$, is defined for each presheaf $\mathbb P$ in
${{\mathbb {S} ets}^{{\mathcal B}^{op}}}$ as the colimit ${\mathbb L}(\mathbb P)$:\renewcommand{\baselinestretch}{1}\footnote{For readers not familiar with the categorical construction of colimits we note that their existence expresses in category theoretical language the basic intuition that a complex object may be conceived as arising from the interconnection of partially or locally defined informational units within a category. In a nutshell, colimits may be viewed as binding factors `gluing' parts together. The colimit
${\mathbb L}({\mathbb P})$ of Eq. (4.12) is explicitly constructed in Zafiris \& Karakostas (2013).}
\renewcommand{\baselinestretch}{1.15}
\begin{equation}
{\mathbb
L}({\mathbb P})= {\it Colim} \{ \bf{{\int}}({\mathbb P},{\mathcal
B}) \Ar {{\mathbf {\int}}_{\mathbb P}} {\mathcal B} \Ar {\mathbb
M} {\mathcal L} \} \;.
\end{equation}
\end{corollary}

The essential functioning of the Boolean frames - quantum adjunction, specified by the pair of adjoint functors ${\mathbb L} \dashv {\mathbb R}$,
is made transparent if we consider that it provides a bi-directional mechanism of encoding and decoding information between diagrams of Boolean event algebras $B$ and quantum event algebras $L$ via the action of Boolean probing frames or perspectives $\psi_B: {\mathbb M}(B) \rightarrow L$.
Thus, if we think of ${{\mathbb {S} ets}^{{\mathcal B}^{op}}}$ as the categorical universe of variable local Boolean frames modeled in $\mathbb {S} ets$, and of ${\mathcal L}$ as the categorical universe of quantum event structures, then the left adjoint functor $\mathbb L :  {{\mathbb {S} ets}^{{\mathcal B}^{op}}}  \to {\mathcal L}$ signifies a translational code of information from the level of local Boolean algebras to the level of global quantum event algebras, whereas the Boolean realization functor ${\mathbb {R}}: { {\mathcal L}} \rightarrow {{\mathbb {S} ets}^{{{\mathcal B}}^{op}}}$ signifies a translational code in the inverse direction.
In general, the content of the information cannot remain completely invariant with respect to translating from one categorical universe to another, and conversely.
However, as suggested by Theorem 4.1,
there remain two alternatives for a variable set over local Boolean frames $\mathbb P$, standing for a presheaf functor $\mathbb P$ in
${{\mathbb {S} ets}^{{\mathcal B}^{op}}}$, to exchange information with a quantum algebra $L$.
Either the content of information is transferred in quantum terms with the colimit in the category of elements of $\mathbb P$ translating, represented as the quantum morphism
${\mathbb L \mathbb P} \rightarrow {L}$, or the content of information is transferred in Boolean terms with the functor of Boolean frames of $L$ translating, represented correspondingly as the natural transformation ${\mathbb P} \rightarrow {{\mathbb R}({L})}$.
In the first case, from the setting of $L$, information is being received in quantum terms, while in the second, from the setting of $\mathbb P$, information is being sent in Boolean terms. Then, the natural bijection of Eq. (4.11) corresponds to the assertion that these two distinct ways of information transfer are equivalent. Thus, the fact that these two functors are adjoint underlines an amphidromous dependent variation, safeguarding that the global information encoded in a quantum kind of event structure is retrievable in a structure-preserving manner by all possible partial structural congruences with the Boolean kind of event structure.

Importantly, by virtue of the existence of the Boolean frames - quantum adjunction, $\mathbb L :{{{\mathbb {S} ets}}^{{\mathcal B}^{op}}} \adjar {\mathcal L} : \mathbb R$,
every probing relation from a Boolean event algebra $B$ in $\mathcal B$ to a quantum event algebra $L$ in $\mathcal L$ shaped by the functor $\mathbb M: \mathcal B \rightarrow \mathcal L$, or, equivalently, every Boolean frame-perspective on $L$, factors uniquely through the category of presheaves of sets ${{{\mathbb {S} ets}}^{{\mathcal B}^{op}}}$, as revealed by the following commutative diagram:
\begin{diagram}
€{\mathcal B}€€ €\Smono {\mathbb y} €\Sear {\mathbb M}€€ €{\mathbb
Sets}^{{\mathcal B}^{op}}€\Edotar {\mathbb L}  \wdotaR {\mathbb R}€{\mathcal L}€€
\end{diagram}

This epitomizes the fact that there exists an exact solution to the problem of specifying a quantum event algebra $L$ in perspectivist terms, by means of Boolean probing frames acting on it, which is provided by the left adjoint colimit functor $\mathbb L: {{{\mathcal {S} ets}}^{{\mathcal B}^{op}}} \rightarrow \mathcal L$ of the Boolean frames-quantum adjunction.
Henceforth, the specification of a quantum event algebra $L$ in perspectivist terms, firstly, is not subordinate to ad hoc choices of Boolean frames adjoined to it, and, secondly, it is synthesized in the limit of the joint compatible action of all Boolean frames-perspectives acting on it.

\section{Concluding Remarks}
The key philosophical meaning of this approach implies, therefore, the view that the quantum world can be consistently approached and comprehended through a multilevel structure of overlapping Boolean frames, understood as locally variable perspectives applied on a quantum system, which interlock, in a category-theoretical environment, to form a coherent picture of the whole in a nontrivial way.
In quantum mechanics the relation between the global theoretical structure and its various empirical sub-structures is indeed such that, depending on the type of experimental context a quantum system is brought to interact, different manifested aspects of the system are disclosed, impossible to be combined into a single picture as in classical physics, although only one type of system is concerned.
Thus, by virtue of the proposed category-theoretic perspectivist approach to quantum mechanics, a quantum event structure can only be unfolded through structured interconnected families of Boolean probing frames capable of carrying all the information encoded in the former, and inversely.

In closing, and in view of the preceding considerations, it is only natural to assert that, in contrast to an Archimedean panoptical ``view from nowhere" of the classical paradigm, the general epistemological implication of quantum theory acknowledges in an essential way a perspectival/contextual character of knowledge.
Furthermore, the suggested perspectivist approach to quantum mechanics provides, from the viewpoint of theory construction, the appropriate mathematical substratum for developing a post-classical, structured view of scientific theorizing in the sense of comprehending a theory not just as a class of empirical models simpliciter, as a structureless set of ``models of the data", but also establishing mappings between these models allowing thereby their coherent embedding in a global theoretical structure.

\vspace{1cm}

\begin{center}{ {\bf \large References

}

\vspace{.5mm} }\end{center}

\begin{enumerate}
\renewcommand{\baselinestretch}{0}
{\small
\item {Abramsky, S., \& Brandenburger, A. (2011). The sheaf-theoretic structure of non-locality and contextuality. {\it New Journal of Physics}, {\it 13}, 113036.}
\item {Awodey, S. (2010). {\it Category theory} (2nd ed.). Oxford: Oxford University Press.}
\item {Borceux, F. (1994/2008). {\it Categories of sheaves}. Cambridge: Cambridge University Press.}
\item {Buzzoni, M. (2016). Mechanisms, experiments, and theory-ladeness: A realist perspectivalist view. {\it Axiomathes}, {\it 26}, 411-427.}
\item {Cabello, A., Severini, S., \& Winter, A. (2010). (Non-)Contextuality of physical theories as an axiom. {\it arXiv:1010.2163}.}
\item {D\"{o}ring, A., Isham, C.J. (2008). A topos foundation for theories of physics: I. Formal languages for physics. {\it Journal of Mathematical Physics}, {\it 49}, 053515.}
\item {Giere, R.N. (2006). {\it Scientific perspectivism}. Chicago: The University of Chicago Press.}
\item {Goldblatt, R. (2006). {\it Topoi: The categorial analysis of logic} (2nd ed.). New York: Dover.}
\item {Howard, M., Wallman, J., Veitch, V., \& Emerson, J. (2014). Contextuality supplies the `magic' for quantum computation. {\it Nature}, {\it 510}, 351-355.}
\item {Karakostas, V. (2012). Realism and objectivism in quantum mechanics. {\it Journal for General Philosophy of Science}, {\it 43}, 45-65.}
\item {Karakostas, V. (2014). Correspondence truth and quantum mechanics. {\it Axiomathes}, {\it 24}, 343-358.}
\item {Karakostas, V., \& Zafiris, E. (2017). Contextual semantics in quantum mechanics from a categorical point of view. {\it Synthese}, {\it 194}, 847-886.}
\item {Lawvere, W., \& Schanuel, S. (2009). {\it Conceptual mathematics} (2nd ed.). Cambridge: Cambridge University Press.}
\item {Mac Lane, S., \& Moerdijk, I. (1992). {\it Sheaves in geometry and logic: A first introduction to topos theory}. New York: Springer.}
\item {Marquis, J.P. (2009). {\it From a geometrical point of view: A study in the history and philosophy of category theory}. New York: Springer.}
\item {Massimi, M. (2012). Scientific perspectivism and its foes. {\it Philosophica}, {\it 84}, 25-52.}
\item {Svozil, K. (2009). Contexts in quantum, classical and partition logic. In K. Engesser, D. Gabbay, \& D. Lehmann (Eds.) {\it Handbook of quantum logic and quantum structures: Quantum logic} (pp. 551-586). Amsterdam: Elsevier.}
\item {Varadarajan, V.S. (2007). {\it Geometry of quantum theory} (2nd ed.). New York: Springer.}
\item {Zafiris, E. (2004). Boolean coverings of quantum observable structure: A setting for an abstract differential geometric mechanism. {\it Journal of Geometry and Physics}, {\it 50}, 99-114.}
\item {Zafiris, E. (2006). Sheaf-theoretic representation of quantum measure algebras. {\it Journal of Mathematical Physics}, {\it 47}, 092103.}
\item {Zafiris, E., \& Karakostas, V. (2013). A categorial semantic representation of quantum event structures. {\it Foundations of Physics}, {\it 43}, 1090-1123.}

}

\end{enumerate}

\end{document}